\documentclass[aps,prd,amsmath,floats,floatfix,twocolumn,superscriptaddress,
  nofootinbib,showpacs]{revtex4} 
\usepackage{graphicx} 
\usepackage{bm}

\renewcommand{\Re}{\mathrm{Re}}

\newcommand{\Mirr}{M_{\text{irr}}}
\newcommand{\Eadm}{M_{\text{ADM}}}
\newcommand{\Jadm}{J_{\text{ADM}}}

\newcommand{\Caltech}{\affiliation{Theoretical Astrophysics 130-33,
    California Institute of Technology, Pasadena, CA 91125}}
\newcommand{\LIGO}{\affiliation{LIGO Laboratory, California Institute
    of Technology, Pasadena, California 91125}}
\newcommand{\Cornell}{\affiliation{Center for Radiophysics and Space
    Research, Cornell University, Ithaca, New York, 14853}}
\newcommand{\WakeForest}{\affiliation{Department of Physics, Wake
    Forest University, Winston-Salem, North Carolina 27106}}
\newcommand{\Syracuse}{\affiliation{Department of Physics, Syracuse
    University, Syracuse, New York, 13244}}

\begin{document}
\vspace{-2.5cm} 

\title{High-accuracy comparison of numerical relativity simulations
with post-Newtonian expansions}

\author{Michael Boyle} \Caltech
\author{Duncan A. Brown} \Caltech \LIGO \Syracuse
\author{Lawrence E. Kidder} \Cornell
\author{Abdul H. Mrou\'e} \Cornell
\author{Harald P. Pfeiffer} \Caltech
\author{Mark A. Scheel} \Caltech
\author{Gregory B. Cook} \WakeForest
\author{Saul A. Teukolsky} \Cornell


\date{\today}

\begin{abstract}
Numerical simulations of 15 orbits of an equal-mass binary black hole
system are presented.  Gravitational waveforms from these simulations,
covering more than 30 cycles and ending about 1.5 cycles before
merger, are compared with those from quasi-circular zero-spin
post-Newtonian (PN) formulae.  The cumulative phase uncertainty of
these comparisons is about 0.05 radians, dominated by effects arising
from the small residual spins of the black holes and the small
residual orbital eccentricity in the simulations.  Matching numerical
results to PN waveforms early in the run yields excellent agreement
(within $0.05$ radians) over the first $\sim 15$ cycles, thus
validating the numerical simulation and establishing a regime where PN
theory is accurate.  In the last 15 cycles to merger, however, {\em
  generic} time-domain Taylor approximants build up phase differences of
several radians.  
But, apparently by coincidence, one specific
post-Newtonian approximant, TaylorT4 at 3.5PN order, agrees much
better with the numerical simulations, with accumulated phase
differences of less than 0.05 radians over the 30-cycle waveform.
Gravitational-wave amplitude comparisons are also done between
numerical simulations and post-Newtonian, and the agreement depends on
the post-Newtonian order of the amplitude expansion: the amplitude
difference is about 6--7\% for zeroth order and becomes smaller for
increasing order.  A newly derived 3.0PN amplitude correction improves
agreement significantly ($<1\%$ amplitude difference throughout most
of the run, increasing to $4\%$ near merger) over the previously known
2.5PN amplitude terms.
\end{abstract}


\maketitle


\section{Introduction}

The last two years have witnessed tremendous progress in simulations
of black hole binaries, starting with the first stable simulation of
orbiting and merging black holes~\cite{Pretorius2005a,Pretorius2006},
development of the moving puncture
method~\cite{Campanelli2006a,Baker2006a} and rapid progress by other
groups~\cite{Campanelli-Lousto-Zlochower:2006,Herrmann2007b,Diener2006,%
Scheel2006,Sperhake2006,%
Bruegmann2006,Marronetti2007,Etienne2007, Szilagyi2007}.  Since then, an
enormous amount of work has been done on the late inspiral and merger of
black hole binaries, among them studies of the universality of the
merger waveforms \cite{Baker2006b,Baker-Campanelli-etal:2007},
investigations into black hole kicks
\cite{Baker2006c,Gonzalez2007,Koppitz2007,Campanelli2007,Gonzalez2007b,%
Herrmann2007,Sopuerta-Yunes-Laguna:2007,Choi-Kelly-Boggs-etal:2007,%
Campanelli2007a,Bruegmann-Gonzalez-Hannam-etal:2007,Baker2007,Herrmann2007c,%
Herrmann2007b,Schnittman2007} and spin
dynamics~\cite{Campanelli2007b,Campanelli2006d, Campanelli2006c},
comparisons to post-Newtonian models
\cite{Buonanno-Cook-Pretorius:2007,Ajith-Babak-Chen-etal:2007,%
Berti-Cardoso-etal:2007}, and applications to gravitational wave
data analysis~\cite{Pan2007,Buonanno2007,Baumgarte:2006en}.

Compared to the intense activity focusing on simulations close to
merger, there have been relatively few simulations covering the inspiral
phase.  To date, only three simulations~\cite{Baker2006d,Baker2006e,%
  Pfeiffer-Brown-etal:2007,Hannam2007,Husa2007} cover more than five
orbits.  Long inspiral simulations are challenging for a variety of
reasons:  First, the orbital period increases rapidly with separation,
so that simulations must cover a significantly longer evolution
time. In addition, the gravitational waveform must be extracted at
larger radius (and the simulation must therefore cover a larger spatial volume) 
because the gravitational wavelength is longer.  Furthermore,
gravitational wave data analysis requires small {\em absolute} accumulated
phase uncertainties in the waveform, so the relative phase uncertainty of the
simulation must be smaller.

Gravitational wave detectors provide a major driving force for
numerical relativity (NR).  The first generation interferometric
gravitational wave detectors, such as
LIGO~\cite{Barish:1999,Waldman:2006}, GEO600~\cite{Hild:2006} and
VIRGO~\cite{Acernese:2002,Acernese-etal:2006}, are now operating at or
near their design sensitivities. Furthermore, the advanced generation
of detectors are entering their construction phases. This new
generation of interferometers will improve detector sensitivity by a
factor of $\sim 10$ and hence increase expected event rates by a
factor of $\sim 1000$~\cite{Fritschel2003}.  One of the most promising
sources for these detectors is the inspiral and merger of binary black
holes (BBHs) with masses $m_1 \sim m_2 \sim
10$--$20\,M_\odot$~\cite{Flanagan1998a}.  These systems are expected
to have circularized long before their gravitational waves enter the
sensitive frequency band of ground-based detectors~\cite{Peters1964}.

A detailed and accurate understanding of the gravitational waves
radiated as the black holes spiral towards each other will be crucial
not only to the initial detection of such sources, but also to
maximize the information that can be obtained from signals once they
are observed.  When the black holes are far apart, the gravitational
waveform can be accurately computed using a post-Newtonian (PN)
expansion. As the holes approach each other and their velocities
increase, the post-Newtonian expansion is expected to diverge from the
true waveform.  It is important to quantify any differences between
theoretical waveforms and the true signals, as discrepancies will
cause a reduction of search sensitivity.  Several techniques have been
proposed to address the problem of the breakdown of the post-Newtonian
approximation~\cite{Damour98,Buonanno99,Buonanno:2002ft}, but
ultimately, the accuracy of the post-Newtonian waveforms used in
binary black hole gravitational wave searches can only be established
through comparisons with full numerical simulations.

Unfortunately, comparing post-Newtonian approximations to numerical
simulations is not straightforward, the most obvious problem being the
difficulty of producing long and sufficiently accurate numerical
simulations as explained above.  In addition, post-Newtonian waveforms
typically assume circular orbits, and most astrophysical binaries are
expected to be on circular orbits late in their inspiral, so the
orbital eccentricity within the numerical simulation must be
sufficiently small\footnote{Unfortunately, this circularization occurs
on extremely long time scales~\cite{Peters1964}, thousands of orbits,
making it impossible to run the numerical simulation long enough to
radiate the eccentricity away.}.  Another factor that complicates
comparisons is the variety of post-Newtonian approximants available,
from several straightforward Taylor expansions to more sophisticated
Pad\'e resummation techniques and the effective one-body approach (see
e.g.~\cite{Damour2001,Damour02,Damour98,Buonanno99,
Buonanno00,2000PhRvD..62h4011D,Damour01c,Damour03,Buonanno06}, as well
as Section~\ref{sec:taylor-approximants} below).  While all
post-Newtonian approximants of the same order should agree
sufficiently early in the inspiral (when neglected higher-order terms
are small), they begin to disagree with each other during the late
inspiral when the post-Newtonian approximation starts to break
down---exactly the regime in which NR waveforms are becoming
available.

Finally, agreement (or disagreement) between NR and PN waveforms will
also depend very sensitively on the precise protocol used to compare
the waveforms.  Are PN and NR waveforms matched early or late in the inspiral?
Is the matching done at a particular time,
or is a least-squares fit performed over part (or all) of the
waveform?  Does one compare frequencies $\omega(t)$ or phases
$\phi(t)$?  Are comparisons presented as functions of time or of
frequency?  Up to which cutoff frequency does one compare PN with NR?

Despite these difficulties, several comparisons between NR and PN have
been done for the last few orbits of an equal-mass, non-spinning black
hole binary.  The first such study was done by Buonanno {\em et
  al}~\cite{Buonanno-Cook-Pretorius:2007} based on simulations
performed by Pretorius~\cite{Pretorius2005a} lasting somewhat more
than 4 orbits ($\sim 8$ gravitational wave cycles).  This comparison
performs a least-squares fit over the full waveform, finds 
agreement between the numerical evolution and a
particular post-Newtonian approximant (in our language TaylorT3
3.0/0.0\footnote{We identify post-Newtonian approximants with three
  pieces of information: the label introduced by~\cite{Damour2001} for
  how the orbital phase is evolved; the PN order to which the orbital
  phase is computed; and the PN order that the amplitude of the
  waveform is computed.  See Sec.~\ref{sec:taylor-approximants} for
  more details.}) and notes that another approximant (TaylorT4 3.5/0.0) 
will give similarly good agreement.  However, as the authors note, 
this study is
severely limited by numerical resolution, sizable initial eccentricity
($\sim 0.015$), close initial separation of the black holes, and
coordinate artifacts; for these reasons,
the authors do not quantify the level of agreement.
  
More recently, Baker et al.~\cite{Baker2006d,Baker2006e}
performed simulations covering the last $\sim 14$ cycles before
merger.  These simulations have an orbital eccentricity $\sim
0.008$~\cite{Baker2006d}, forcing the authors to use
a fitted smooth (``de-eccentrized'') gravitational wave phase to 
obtain a monotonically increasing gravitational wave frequency.   
Comparing to TaylorT4
3.5/2.5, they find agreement between numerical and post-Newtonian
gravitational wave phase to within their numerical errors, which are
about 2 radians.  The authors also indicate that other post-Newtonian
approximants do not match their simulation as well as TaylorT4,
but unfortunately, they do not mention whether any disagreement is
significant (i.e., exceeding their numerical errors).
Pan {\em et. al}~\cite{Pan2007} performed a more comprehensive
analysis of the numerical waveforms computed by
Pretorius~\cite{Buonanno-Cook-Pretorius:2007} and the Goddard
group~\cite{Baker2006d,Baker2006e}, confirming that TaylorT4 3.5/0.0
matches the numerical results best.  

The most accurate inspiral simulation to date was performed by the
Jena group and presented in Husa et al.~\cite{Husa2007} and
Hannam et al.~\cite{Hannam2007}.  This simulation covers 18
cycles before merger and has an orbital eccentricity of $\sim
0.0018$~\cite{Husa-Hannam-etal:2007}.  Discarding the first two cycles
which are contaminated by numerical noise, and terminating the
comparison at a gravitational-wave frequency $m\omega=0.1$ (see
Eq.~(\ref{eq:omega-definition}) for the precise definition) their
comparison extends over 13 cycles.  We discuss the results of
Ref.~\cite{Hannam2007} in more detail in
Sec.~\ref{sec:ComparisonWithPN-TaylorT1}.

This paper presents a new inspiral simulation of a non-spinning equal
mass black hole binary.  This new simulation more than doubles the
evolution time of the simulations 
in Refs.~\cite{Baker2006d,Baker2006e,Hannam2007,Husa2007}, resulting in a
waveform with 30 gravitational wave cycles, ending $\sim 1.5$ cycles
before merger, and improves numerical truncation errors by one to two
orders of magnitude
over those in Refs.~\cite{Baker2006d,Baker2006e,Hannam2007,Husa2007}.  
The orbital eccentricity of our simulations is $\sim 6\times 10^{-5}$;
this low eccentricity is achieved using refinements of techniques described
in~\cite{Pfeiffer-Brown-etal:2007}.  We present a detailed analysis of
various effects which might influence our comparisons to
post-Newtonian waveforms for non-spinning black hole binaries on
circular orbits. These effects result in an uncertainty of
$\sim 0.05$ radians out of the accumulated $\sim 200$ radians.
Perhaps surprisingly, the largest uncertainty arises from the residual
orbital eccentricity, despite its tiny value.  The second largest
effect arises due to a potential residual spin on the black holes,
which we bound by $|S|/\Mirr^2<5\times 10^{-4}$.

We compare the numerical waveforms with four different time-domain
post-Newtonian
Taylor-approximants~\cite{Damour2001,Damour02,Buonanno:2002ft} and we
match PN and NR waveforms at a specific time during the inspiral.  We
explore the effects of varying this matching time.  When matching
$\sim 9$ cycles after the start of our evolution, all post-Newtonian
approximants of 3.0PN and 3.5PN order in orbital phase agree with our
simulation to within $\sim 0.03$ radians over the first 15 cycles.
This agreement is better than the combined uncertainties of the
comparison, thus validating our simulations in a regime where the
3.5PN truncation error of post-Newtonian theory is comparable to the
accuracy of our simulations.  Lower order post-Newtonian approximants
(2.0PN and 2.5PN order), however, accumulate a significant
phase difference of $\sim 0.2$ radians over this region.

Extending the comparison toward merger (as well as when matching
closer to merger), we find, not surprisingly, that the agreement
between PN and NR at late times depends strongly on exactly what
post-Newtonian approximant we use~\cite{Damour2001,Damour02}. Typical
accumulated phase differences are on the order of radians at frequency
$m\omega=0.1$.
One particular post-Newtonian approximant, TaylorT4 at 3.5PN order in
phase, agrees with our NR waveforms far better than the other
approximants, the agreement being within the phase uncertainty of the
comparison (0.05 radians) until after the gravitational wave frequency passes
$m\omega=0.1$ (about 3.5 cycles before merger).  It remains to be seen
whether this agreement is fundamental or accidental, and whether it
applies to more complicated situations (e.g. unequal masses,
nontrivial spins).

We also compare the post-Newtonian gravitational wave amplitude to the
numerical amplitude, where we estimate the uncertainty of this
comparison to be about $0.5\%$.  Restricted waveforms (i.e., 0PN order
in the amplitude expansion) are found to disagree with the numerical
amplitudes by 6--7\%.  An amplitude expansion of order 2PN shows
significantly better agreement than the expansion at order 2.5PN.  A
newly derived 3PN amplitude~\cite{Kidder07a} is found to give much
better agreement than the 2.0PN amplitude.

This paper is organized as follows: Section~\ref{sec:NRGeneration}
discusses our numerical techniques.  In particular, we describe how we
construct binary black hole initial data, evolve these data for 15
orbits, extract gravitational wave information from the evolution, and
produce a gravitational waveform as seen by an observer at infinity.
Section~\ref{sec:PN} details the generation of post-Newtonian
waveforms, including details of how we produce the four approximants
that we compare against NR.  We describe our procedure for comparing
NR and PN waveforms in Sec.~\ref{sec:PNComparison}, and present a
detailed study of various sources of uncertainty in
Sec.~\ref{sec:SummaryErrors}.  The comparisons between NR and PN are
presented in Section~\ref{sec:Results}.  This section is split into
two parts: First, we compare each PN approximant separately with the
numerical simulation.  Subsequently, we show some additional figures
which facilitate cross-comparisons between the different PN
approximants.  Finally, we present some concluding remarks in
Section~\ref{sec:conclusions}.  The impatient reader primarily
interested in NR-PN comparisons may wish to proceed directly to
Table~\ref{tab:Errors} summarizing the uncertainties of our
comparisons, and then continue to Sec.~\ref{sec:Results}, starting
with Fig.~\ref{fig:NR-TaylorT1}.

\section{Generation of numerical waveforms}
\label{sec:NRGeneration}

In order to do a quantitative comparison between numerical and
post-Newtonian waveforms, it is important to have a code capable of
starting the black holes far enough apart to be in a regime where we
strongly believe the post-Newtonian approximation is valid, track the
orbital phase extremely accurately, and do so efficiently so the
simulation can be completed in a reasonable amount of time.
Furthermore, the gravitational waves from such a simulation must be
extracted in such a manner that preserves the accuracy of the
simulation and predicts the waveform as seen by a distant observer,
so a comparison with the post-Newtonian waveform can be made.  In this
section we describe the techniques we use to do this, as well as the
results of a simulation starting more than 15 orbits prior to merger.

When discussing numerical solutions of Einstein's equations, we write
all dimensioned quantities in terms of some mass scale $m$, which we
choose to be the sum of the irreducible masses of the two black holes
in the initial data:
\begin{equation}
\label{eq:m}
m=M_{{\rm irr},1} + M_{{\rm irr},2}.
\end{equation}
The irreducible mass of a single hole is defined as 
\begin{equation}
\label{eq:MirrDefinition}
\Mirr \equiv \sqrt{A/16\pi}, 
\end{equation}
where $A$ is the surface area of the event horizon; in practice we
take $A$ to be the surface area of the apparent horizon.  More
generally, it is more appropriate to use the Christodoulou mass of
each black hole,
\begin{equation}\label{eq:Christoudoulou-mass}
M_{\rm BH}^2 =\Mirr^2+\frac{S^2}{4\Mirr^2},
\end{equation}
instead of the irreducible mass. Here $S$ is the spin of the hole.
However, for the case considered in this paper, the spins are
sufficiently small that there is little difference between $M_{\rm BH}$
and $\Mirr$.


\subsection{Initial data}

Initial data are constructed within the conformal thin sandwich
formalism~\cite{York1999,Pfeiffer2003b} using a pseudo-spectral
elliptic solver~\cite{Pfeiffer2003}.  We employ quasi-equilibrium
boundary conditions~\cite{Cook2002,Cook2004} on spherical excision
boundaries, choose conformal flatness and maximal
slicing, and use Eq.~(33a) of Ref.~\cite{Caudill-etal:2006} as the
lapse boundary condition.  The spins of the black holes are made very
small via an appropriate choice of the tangential shift at the
excision surfaces, as described in~\cite{Caudill-etal:2006}.  

\begin{table*}
\caption{\label{tab:ID}Summary of the initial data sets used in
  this paper.  The first block of numbers ($d$, $\Omega_0$, $f_r$, and
  $v_r$) represent raw parameters entering the construction of the
  initial data.  The second block gives some properties of each
  initial data set: $m$ denotes the sum of the irreducible
  masses, $\Eadm$ and $\Jadm$ the ADM energy and angular momentum, and
  $s_0$ the initial proper separation between the horizons.  The
  last column lists the eccentricity computed from
  Eq.~(\ref{eq:e-ds/dt}).  The initial data set 30c is used for
  all evolutions (except for consistency checks) described in this paper.}
\begin{tabular}{c|cccr|cccc|c}
Name & $d$ & $\Omega_0$ & $f_r$ & $v_r\times 10^4$ 
   & $m\Omega_0$ & $\Eadm/m$  &  $\Jadm/m^2$  
   & $s_0/m$ & $e_{ds/dt}$\\\hline

30a
 & 30  & 0.0080108 & 0.939561 &  0.00\;\; & 0.01664793 & 0.992333 & 1.0857 
   & 17.37 & $1.0\times 10^{-2}$   \\

30b & 30  & 0.0080389 & 0.939561 & -4.90\;\; & 0.0167054 &0.992400 & 1.0897
   & 17.37 & $6.5\times 10^{-4}$   \\

{\bf 30c} & \;{\bf 30}\; & {\bf 0.0080401} & {\bf 0.939561} &{\bf  -4.26\;} 
  & \;{\bf 0.0167081}\; & {\bf 0.992402 } & {\bf 1.0898} 
   & \;{\bf 17.37}\;&  \;$\mathbf{ 5\times 10^{-5}}$\; \\ 

\hline

24a & 24  & 0.0110496 & 0.92373 &  -8.29\;\; & 0.0231947 & 0.990759 & 1.0045
   & 14.15 & $1.1\times 10^{-3}$ \\

24b & 24  & 0.0110506 & 0.923739 & -8.44\;\; & 0.0231967 & 0.990767 & 1.0049
   & 14.15 &  $1.5\times 10^{-4}$\\

\end{tabular}
\end{table*}

\begin{figure}
\includegraphics[scale=0.49]{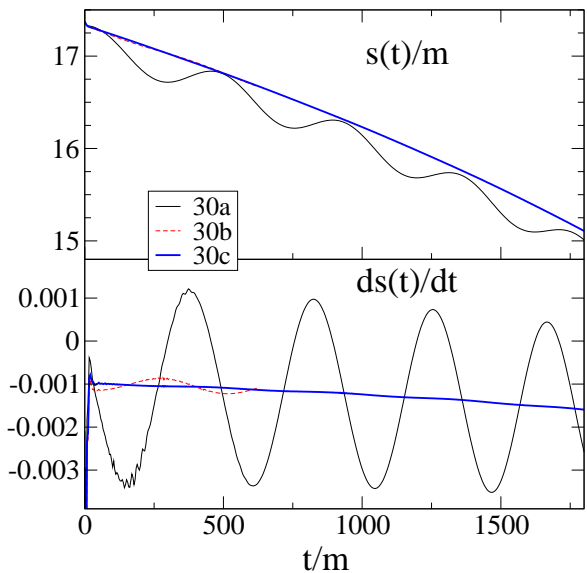}
\caption{\label{fig:ProperSep} Proper separation (top panel) and its
  time derivative (lower panel) versus time for short evolutions of
  the $d=30$ initial data sets 30a, 30b, and 30c (see
  Table~\ref{tab:ID}). These three data sets represent zero through
  two iterations of our eccentricity-reduction procedure.  The
  orbital eccentricity is reduced significantly by each iteration.  }
\end{figure}

As the most accurate post-Newtonian waveforms available assume
adiabatic inspiral of quasi-circular orbits, it is desirable to reduce
the eccentricity of the numerical data as much as possible. Using
techniques developed in~\cite{Pfeiffer-Brown-etal:2007}, each black
hole is allowed to have a nonzero initial velocity component towards
the other hole. This small velocity component $v_r$ and the initial
orbital angular velocity $\Omega_0$ are then fine-tuned in order to
produce an orbit with very small orbital eccentricity\footnote{An
  alternative method of producing low-eccentricity initial data, based
  on post-Newtonian ideas, is developed
  in~\cite{Husa-Hannam-etal:2007}.  While that technique is
  computationally more efficient than ours, it merely reduces orbital
  eccentricity by a factor of $\sim 5$ relative to quasi-circular
  initial data, which is insufficient for the comparisons
  presented here. (cf. Sec.~\ref{sec:EccentricityMatchingTime}).}.  
We have improved our
eccentricity-reduction procedure since the version described
in~\cite{Pfeiffer-Brown-etal:2007}, so we summarize our new iterative
procedure here:

We start with a quasi-circular (i.e., $v_r = 0$) initial data set at
coordinate separation $d=30$, where $\Omega_0$ is determined by
equating Komar mass with Arnowitt-Deser-Misner (ADM) 
mass~\cite{Caudill-etal:2006}.  We then
evolve these data for about 1.5 orbits, corresponding to a time
$t/m\approx 600$.  From this short evolution, we measure the proper
separation $s$ between the horizons by integration along the
coordinate axis connecting the centers of the black holes.  We fit the
time derivative $ds/dt$ in the interval $100\lesssim t/m\lesssim 600$
to the function
\begin{equation}\label{eq:fit}
\frac{ds}{dt}=A_0 + A_1 t + B \cos(\omega t+\varphi),
\end{equation}
where we vary all five parameters $A_0, A_1, B, \omega$ and $\varphi$
to achieve the best fit.
The desired smooth inspiral is represented by
the part $A_0+A_1t$; the term $B\cos(\omega t+\varphi)$ corresponds to
oscillations caused by orbital eccentricity.  

For a {\em Newtonian}
orbit with radial velocity $B\cos(\omega t+\varphi)$ at initial 
separation $s_0$, it is straightforward to determine the changes
to the orbital frequency and the radial velocity which make the 
orbit perfectly circular, namely 
\begin{align}
\label{eq:Omega0-update}
\Omega_0 &\to \Omega_0 + \frac{B\sin\varphi}{2 s_0},\\
\label{eq:vr-update}
v_r &\to v_r-\frac{B\cos\varphi}{2}.
\end{align}
For Newtonian gravity, Eq.~(\ref{eq:vr-update}) will of course result
in a circular orbit with $v_r=0$.  In {\em General Relativity},
$\Omega_0$ and $v_r$ will be different from their Newtonian values,
for instance $v_r<0$ to account for the inspiral of the two black
holes.  Nevertheless, we assume that small perturbations around the
zero-eccentricity inspiral trajectory behave similarly to small
perturbations around a Newtonian circular orbit.  Therefore, we
apply the same formulae, Eqs.~(\ref{eq:Omega0-update})
and~(\ref{eq:vr-update}), to obtain improved values for $\Omega_0$ and
$v_r$ for the black hole binary, where $s_0$ is the initial proper
separation between the horizons.  
We
then use the new values of $\Omega_0$ and $v_r$ to construct a new
initial data set, again evolve for two orbits, fit to
Eq.~(\ref{eq:fit}), and update $\Omega_0$ and $v_r$.  We
continue iterating this procedure until the eccentricity is
sufficiently small.  

We estimate the eccentricity for each iteration
from the fit to Eq.~(\ref{eq:fit}) using the formula
\begin{equation}\label{eq:e-ds/dt}
e_{ds/dt}=\frac{B}{s_0\omega},
\end{equation}
which is valid in Newtonian gravity for small eccentricities.
Successive iterations of this procedure are illustrated 
in Fig.~\ref{fig:ProperSep} and yield the initial data sets
30a, 30b, and 30c summarized in Table~\ref{tab:ID}.
Eccentricity decreases by roughly a factor of 10 in each iteration,
with 30c having $e_{ds/dt}\approx 5\times 10^{-5}$.  The
evolutions used during eccentricity reduction need not be very accurate
and need to run only for a short time, $t\sim 600m$.  
One iteration of this procedure at our second lowest
resolution requires about 250 CPU-hours.  For
completeness, Table~\ref{tab:ID} also lists parameters for initial
data at smaller separation; these data will be used for consistency
checks below.  Apart from these consistency checks, the remainder of
this paper will focus exclusively on evolutions of the
low-eccentricity initial data set 30c.

\subsection{Evolution of the inspiral phase}
\label{sec:InspiralEvolution}

The Einstein evolution equations are solved with the pseudo-spectral
evolution code described in Ref.~\cite{Scheel2006}.  This code evolves
a first-order representation~\cite{Lindblom2006} of the generalized
harmonic system~\cite{Friedrich1985,Garfinkle2002,Pretorius2005c}.  
We handle the singularities by excising the black hole
interiors from our grid. Our outer boundary
conditions~\cite{Lindblom2006,Rinne2006,Rinne2007} are designed to
prevent the influx of unphysical constraint
violations~\cite{Stewart1998,FriedrichNagy1999,Bardeen2002,Szilagyi2002,%
Calabrese2003,Szilagyi2003,Kidder2005}
and undesired incoming gravitational radiation~\cite{Buchman2006},
while allowing the outgoing gravitational radiation to pass freely
through the boundary.

The code uses a fairly complicated domain decomposition to achieve
maximum efficiency.  Each black hole is surrounded by several
(typically six) concentric spherical shells, with the inner boundary
of the innermost shell (the excision boundary) just inside the
horizon. A structure of touching cylinders (typically 34 of them)
surrounds these shells, with axes along the line between
the two black holes. The outermost shell around each black hole
overlaps the cylinders.  The outermost cylinders overlap a set of
outer spherical shells, centered at the origin, which extend to large
outer radius.  External boundary conditions are imposed only on the
outer surface of the largest outer spherical shell.  We vary the
location of the outer boundary by adding more shells at the outer
edge.  Since all outer shells have the same angular resolution, the
cost of placing the outer boundary farther away (at full resolution)
increases only linearly with the radius of the boundary.  External
boundary conditions are enforced using the method of
Bjorhus~\cite{Bjorhus1995}, while inter-domain boundary conditions are
enforced with a penalty method~\cite{Gottlieb2001,Hesthaven2000}.

We employ the dual-frame method described in Ref.~\cite{Scheel2006}:
we solve the equations in an 'inertial frame' that is asymptotically
Minkowski, but our domain decomposition is fixed in a 'comoving frame'
that rotates with respect to the inertial frame and also shrinks with
respect to the inertial frame as the holes approach each other. The
positions of the holes are fixed in the comoving frame; we account for
the motion of the holes by dynamically adjusting the coordinate
mapping between the two frames.  Note that the comoving frame is
referenced only internally in the code as a means of treating moving
holes with a fixed domain. Therefore all coordinate quantities
(e.g. black hole trajectories, wave-extraction radii) mentioned in
this paper are inertial-frame values unless explicitly stated otherwise.

One side effect of our dual frame system is that the outer boundary of
our domain (which is fixed in the comoving frame) moves inward with
time as observed in the inertial frame. This is because the comoving
frame shrinks with respect to the inertial frame to follow the motion
of the holes.  In Refs.~\cite{Scheel2006,Pfeiffer-Brown-etal:2007} the
inertial frame coordinate radius $r$ (with respect to the center of mass)
and the comoving coordinate radius $r'$ are related by a simple scaling
\begin{equation}\label{eq:SpatialCoordMapLinear}
r = a(t) r'.
\end{equation}
The expansion parameter $a(t)$ is initially set to unity and decreases
dynamically as the holes approach each other, so that the
comoving-frame coordinate distance between the holes remains constant.
The outer boundary of the computational grid is at a fixed comoving
radius $R'_{\rm bdry}$, which is mapped to the inertial coordinate
radius $R_{\rm bdry}(t)=a(t)R'_{\rm bdry}$.  Because we wish to
accurately compute the gravitational radiation as measured far from
the holes, it is desirable to have a moderately large outer boundary
($R_{\rm bdry}(t)\gtrsim 200m$) throughout the run.  For the linear
mapping, Eq.~(\ref{eq:SpatialCoordMapLinear}), this requires a very
distant outer boundary early in the run, $R_{\rm bdry}(0)\simeq
1000m$.  Computationally this is not very expensive.  However, the
initial junk radiation contaminates the evolutions for a time interval
proportional to the light-crossing time to the outer boundary, and for
$R_{\rm bdry}(0)\simeq 1000m$ it would be necessary to discard a
significant portion of the evolution.

We therefore use the mapping
\begin{equation}\label{eq:CubicScaleMap}
r = \left[a(t) + \left(1-a(t)\right) \frac{r'^2}{R_0'^2} \right] r', 
\end{equation}
for some constant $R_0'$ which is chosen to be roughly the radius of
the outer boundary in comoving coordinates.  This mapping has the
following properties: (1) At the initial time $t=0$, the map reduces
to the identity map because $a(0)=1$. Thus we do not need to re-map
our initial data before evolving.  (2) For small radii (i.e., at the
locations of the black holes), the map reduces to the linear map,
$r=a(t)r'+{\cal O}(r'^3)$.  This allows use of the control system
without modifications.  (3) The moving radius $r'=R_0'$ is mapped to a
{\em constant} inertial radius: $r(R_0')=R_0'$.  This allows us to
keep the inertial radius of the outer boundary constant (or nearly
constant\footnote{In practice, we choose $R_0'$ somewhat larger than
the outer boundary, so that the outer boundary of the computational
domain slowly contracts in inertial coordinates. This makes the
zero-speed characteristic fields {\em outgoing} there, avoiding the
need to impose boundary conditions on those fields.}) in time rather
than shrinking rapidly.

\begin{table}
\caption{\label{tab:Evolutions}Overview of low-eccentricity
  simulations discussed in this paper.  $R_{\rm bdry}$ is the initial
  coordinate radius of the outer boundary; this radius changes during the
  evolution according to the choice of ``radial map'' between inertial
  and comoving coordinates.  The last column lists the different
  resolutions run for each evolution, N6 being highest resolution.
  Evolution 30c-1/N6 forms the basis of our post-Newtonian comparisons,
  and is used in all figures unless noted otherwise.  }
\begin{tabular}{c|c|c|cc|c}
Name & ID & $N_{\rm orbits}$ & $R_{\rm bdry}$ & radial map & resolutions \\\hline
{\bf 30c-1} & {\bf 30c} & {\bf 15.6} & $\bm{462m}$ & {\bf Eq.~(\ref{eq:CubicScaleMap})} & 
{\bf N1, N2, \ldots, N6}  
\\ 
30c-2 & 30c & 15.6 & $722m$ & Eq.~(\ref{eq:SpatialCoordMapLinear}) & 
N2, N4, N6
\\ 
30c-3 & 30c & 15.6 & $202m$ & Eq.~(\ref{eq:SpatialCoordMapLinear}) & 
N2, N3, \ldots, N6
\\\hline 
24b-1 & 24b & 8.3 & $160m$ & Eq.~(\ref{eq:SpatialCoordMapLinear}) & 
N2, N3, N4
\end{tabular}
\end{table}

\begin{figure}
\includegraphics[scale=0.49]{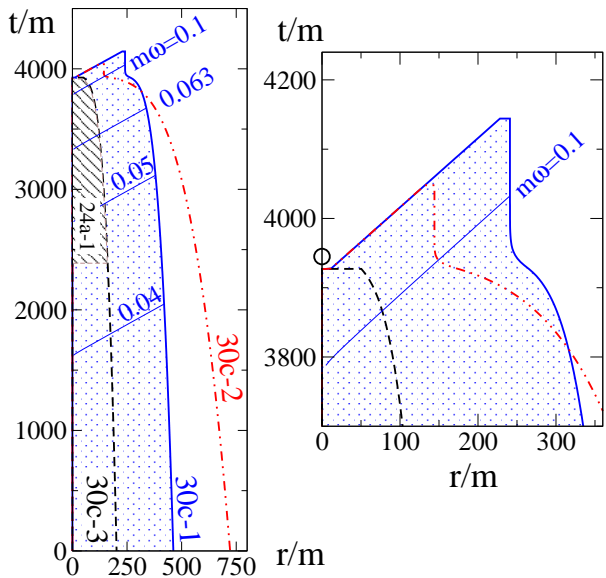}
\caption{\label{fig:SpacetimeDiagram}Spacetime diagram showing the
  spacetime volume simulated by the numerical evolutions listed in
  Tab.~\ref{tab:Evolutions}.  The magnified view in the right panel
  shows how the gravitational waves are escorted to our extraction
  radii (see Sec.~\ref{sec:Escorting}) after the simulation in the
  center has already crashed at $t\sim 3930m$, and after the estimated
  time of the black hole merger, which is indicated by the circle.
  The thin diagonal lines are lines of constant $t-r^*$; each
  corresponds to a retarded time at which the gravitational wave
  frequency $\omega$ at infinity assumes a particular value. }
\end{figure}

In total, we have run three evolutions of the 30c initial data set;
these use different combinations of outer boundary radius and radial
mapping between inertial and moving coordinates.  Some properties of
these evolutions are summarized in Table~\ref{tab:Evolutions}.  We
also performed extensive convergence testing, running the same
evolution on up to six distinct resolutions, N1 to N6.  The coarsest
resolution 30c-1/N1 uses approximately $41^3$ grid points (summing all
grid points in all the subdomains), while the most accurate evolution,
30c-1/N6, uses about $67^3$ grid points.  The run 30c-1/N2 required
about 2,500 CPU-hours and run 30c-1/N6 about 19,000, where our
simulations do not take advantage of symmetries.  
The distance to the outer boundary is adjusted by adding or removing
outer spherical shells to an otherwise unmodified domain-decomposition.
Run 30c-1 has 20 such outer spherical shells, while 30c-2 utilizes 32
and 30c-3 only 8.  Thus, the total number of grid points varies
slightly between runs, e.g. about $71^3$ for 30c-2/N6.
Figure~\ref{fig:SpacetimeDiagram} indicates the different
behavior of the outer boundary location for these three evolutions.

\begin{figure}
\includegraphics[scale=0.49]{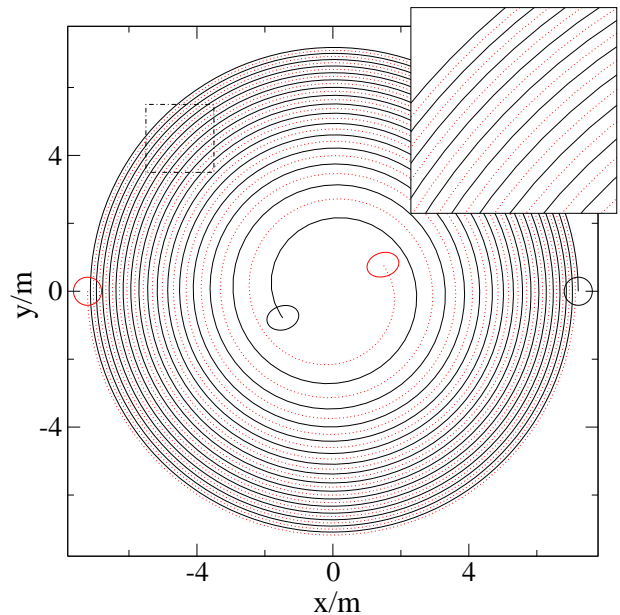}
\caption{\label{fig:Trajectories}Coordinate trajectories of the
  centers of the black holes.  The small circles/ellipsoids show the
  apparent horizons at the initial time and at the time when the
  simulation ends and wave escorting begins. The inset shows an
  enlargement of the dashed box.}
\end{figure}

For all of the evolutions 30c-1/2/3, the coordinate trajectories of
the centers of the apparent horizons appear as in
Fig.~\ref{fig:Trajectories}.  The regular inspiral pattern without
noticeable oscillations once again indicates that our evolutions
indeed have very low eccentricity.

\begin{figure}
\includegraphics[scale=0.49]{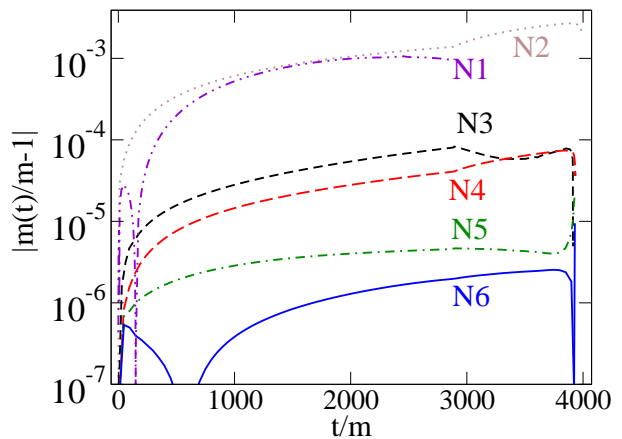}
\caption{\label{fig:AhMass}Deviation of total irreducible mass
  $m(t)=2\Mirr(t)$ from its value in the initial data.  Plotted are
  the six different resolutions of run 30a-1.}
\end{figure}

Figure~\ref{fig:AhMass} demonstrates the convergence of the black
hole mass $m(t)$ with spatial resolution for run 30c-1. The mass
$m(t)$ is computed as the sum of the irreducible masses of both black
holes, as defined in Eq.~(\ref{eq:MirrDefinition}). 
At the highest resolution,
$m(t)$ deviates by only a few parts in $10^{6}$ from its initial
value $m$.

Our apparent horizon finder works by expanding the radius of the
apparent horizon
as a series in spherical harmonics up to some order $L$.  We utilize
the fast flow methods developed by Gundlach~\cite{Gundlach1998} to
determine the expansion coefficients; these are significantly faster
than our earlier minimization
algorithms~\cite{baumgarte_etal96,Pfeiffer2000}.  The apparent horizon
is almost spherical during the inspiral, so that the expansion in $L$
converges exceedingly fast: $L=8$ results in a relative error of the
irreducible mass of better than $10^{-8}$.  The distortion of the
horizons becomes more pronounced toward the end of the evolution when
the black holes approach each other rapidly.  This results in an error
of $10^{-6}$ in the $L=8$ apparent horizon expansion for the last
$10m$ of the evolution.

We also measure the quasi-local spin using coordinate rotation vectors
projected into the apparent horizon 
surfaces~\cite{BrownYork1993, Ashtekar2001, Ashtekar2003}.  Only
the z-component of the spin is non-zero (i.e., the spins are aligned
with the orbital angular momentum). The spin starts at $S_z/M_{\rm
  irr}^2\approx -6\times 10^{-5}$ and increases slowly to $-5\times
10^{-4}$ during the evolution, where the minus sign indicates that the
black hole spin is anti-aligned with the orbital angular momentum.
Thus it appears the black hole's spins move further away from the
corotational state. We believe
this effect is caused by the use of coordinate rotation vectors when
calculating the quasi-local spin, rather than more sophisticated
approximate Killing vectors~\cite{Dreyer2003, Cook2007, OwenThesis}.  
Preliminary results with approximate
Killing vectors find the initial spin to be less than $10^{-6}$, and
slowly increasing during the evolution to a final value of $2\times
10^{-5}$ at the end of the comparison interval to post-Newtonian
theory.  Given the preliminary character of these results, we will
take here the conservative bound $|\mathbf{S}|/\Mirr^2\le 5\times
10^{-4}$ obtained from coordinate rotation vectors.

\subsection{Escorting gravitational waves}
\label{sec:Escorting}

The simulation presented in Fig.~\ref{fig:Trajectories} stops when the
horizons of the black holes become too distorted just before merger.
%
At this time, most of the domain (all regions
except for the immediate vicinity of the two holes) is still
well resolved, and the spacetime contains gravitational radiation that
has not yet propagated out to the large radii where we perform wave
extraction.  So instead of losing this information, which consists of
several gravitational-wave cycles, we evolve only the outer portions
of our grid beyond the time at which the code crashes in the center,
effectively 'escorting' the radiation out to the extraction radii.

To do this, we first stop the evolution shortly before it crashes, and
we introduce a new spherical excision boundary that surrounds both
black holes and has a radius of roughly three times the black hole
separation.  This new excision boundary moves radially outward at slightly
faster than the speed of light so that it is causally disconnected
from the interior region where the code is crashing, and so
that no boundary
conditions are required on this boundary.  We then continue the
evolution on the truncated spherical-shell domain that extends from the new
excision boundary to the outer boundary.
To move both boundaries appropriately, we employ a
new radial coordinate mapping
\begin{equation}
r = A(t) r(r') + B(t),
\end{equation}
where $r(r')$ is given by Eq.~(\ref{eq:CubicScaleMap}). The
functions $A(t)$ and $B(t)$ are chosen to satisfy three criteria:
First, the inner boundary of the spherical shell moves outward with
coordinate speed of unity, which turns out to be slightly
superluminal.  Second, the outer boundary location $R_{\rm bdry}(t)$
has continuous first and second time derivatives at the time we transition
to the truncated domain. And finally, the outer boundary location 
$R_{\rm bdry}(t)$ approaches some fixed value at late times.  The right panel
of Fig.~\ref{fig:SpacetimeDiagram} shows the motion of the inner and
outer radii for evolutions 30c-1 and 30c-2 (we did not perform wave
escorting for 30c-3).  For 30c-1, wave escorting extends the evolution
for an additional time $220m$ beyond the
point at which the simulation stops in the center.

\begin{figure}
\includegraphics[scale=0.49]{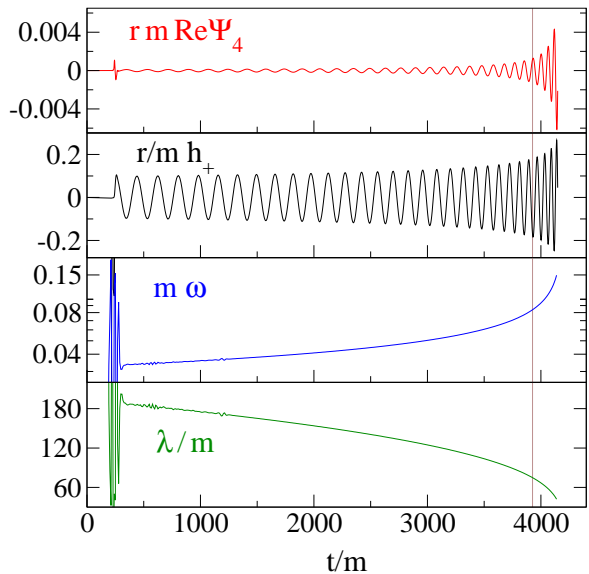}
\caption{\label{fig:0093g_Lev6_R0240m}Gravitational waveform extracted
  at $r=240m$.
  From top panel to bottom: The real part of the $(2,2)$
  component of $r\Psi_4$; the gravitational wave strain, obtained by
  two time integrals of $\Re(r\Psi_4)$; the frequency of the
  gravitational wave, Eq.~(\ref{eq:omega-definition}); the
  gravitational wavelength, $\lambda=2\pi/\omega$.  The vertical brown
  line at $t\approx 3930m$ indicates the time when ``wave escorting''
  starts.}
\end{figure}

Figure~\ref{fig:0093g_Lev6_R0240m} shows the gravitational waveform
extracted at inertial coordinate radius $R=240m$ for the run 30c-1.
The brown vertical line indicates the time when wave escorting
starts. Wave escorting allows us to extract another 4 cycles of
gravitational waves.  
When computing the gravitational wave strain $h(t)$ from the
Newman-Penrose scalar $\Psi_4$ (see Eq.~(\ref{eq:Psi4Definition})
below), one must choose integration constants during the
time integration.  These integration constants were chosen such that
$h(t)$ has zero average and first
moment~\cite{Pfeiffer-Brown-etal:2007}, which is is sufficiently
accurate for the illustrative Fig.~\ref{fig:0093g_Lev6_R0240m}.  To
avoid errors caused by the choice of integration constants, the comparison to
post-Newtonian waveforms below is based entirely on $\Psi_4$.

In the lower two panels of Fig.~\ref{fig:0093g_Lev6_R0240m} there is a
significant amount of noise near the beginning of the run, at
$t<250m$. This noise is barely evident in the top panel of
Fig.~\ref{fig:0093g_Lev6_R0240m} as well.  The noise is a
manifestation of `junk radiation', a pulse of radiation often seen at
the beginning of numerical relativity simulations, and is caused by the
initial data not being precisely a snapshot of an evolution that has
been running for a long time.  Among
the effects that produce junk radiation are incorrect initial
distortions of the individual holes, so that each hole radiates as
it relaxes to its correct quasi-equilibrium shape.

\begin{figure}
\includegraphics[scale=0.49]{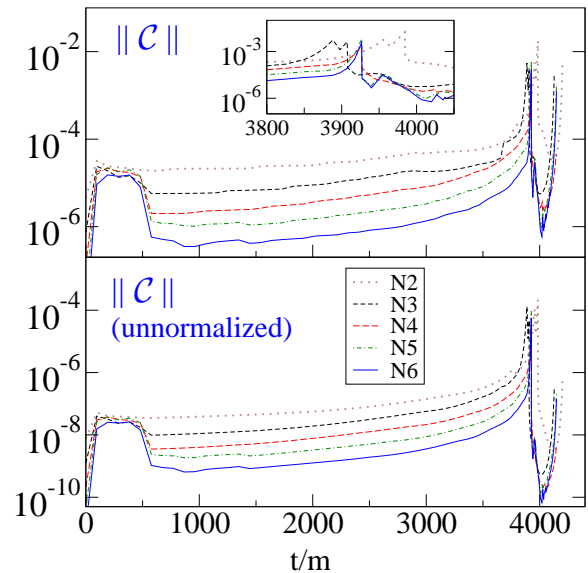}
\caption{\label{fig:Constraints}
  Constraint violations of run 30c-1.  The top panel shows the
  $L^2$ norm of all constraints, normalized by the $L^2$ norm of
  the spatial gradients of all dynamical fields. 
  The bottom panel shows the same data, but without the normalization factor. 
  Norms are taken only in the regions outside apparent horizons.
}
\end{figure}

Our evolution code does not explicitly enforce either the Einstein
constraints or the secondary constraints that arise from writing the
system in first-order form.  Therefore, examining how well these
constraints are satisfied provides a useful consistency check.  
Figure~\ref{fig:Constraints} shows the constraint
violations for run 30c-1.  The top panel shows the $L^2$ norm of all
the constraint fields of our first order generalized harmonic system,
normalized by the $L^2$ norm of the spatial gradients of the dynamical
fields (see Eq.~(71) of Ref.~\cite{Lindblom2006}).  The bottom panel
shows the same quantity, but without the normalization factor ({\em
  i.e.,} just the numerator of Eq.~(71) of Ref.~\cite{Lindblom2006}).
The $L^2$ norms are taken over the entire computational volume that
lies outside of apparent horizons.  At early times, $t<500m$,
the constraints converge rather slowly with resolution because the
junk radiation contains high frequencies.  Convergence is more rapid
during the smooth inspiral phase, after the junk radiation has exited
through the outer boundary.  The constraints increase around
$t\sim 3900m$ as the code begins to fail near
the two merging holes, but then the constraints decrease again after
the failing region is excised for wave escorting.  The normalized
constraint violations are less than $10^{-4}$ until just before the peak
(which occurs at $t=3930m$ for all but
the lowest resolutions).  The size of the peak causes some concern
that the waveforms at late times may be contaminated by constraint
violations to a non-negligible degree.  However, near the peak,
the constraint
violations are large only in the inner regions of the domain near
the black holes (note
that the curves in Fig.~\ref{fig:Constraints} decrease by
two orders of magnitude immediately after these inner
regions are excised at $t=3930m$).  Because all constraint quantities
propagate at the speed of light or slower for the formulation of
Einstein's equations that we use, any influence that the constraint
peak has on the extracted waveform occurs after the constraint violations
have had time to propagate out to the wave extraction zone.
This is very late in the waveform, well after
the gravitational wave frequency reaches $m\omega=0.1$, as
can be seen from the right panel of the spacetime diagram in
Fig.~\ref{fig:SpacetimeDiagram}.

\subsection{Waveform extraction}
\label{sec:waveform-extraction}
Gravitational waves are extracted using the Newman-Penrose scalar
$\Psi_4$, using the same procedure as
in~\cite{Pfeiffer-Brown-etal:2007}.  To summarize, given a spatial
hypersurface with timelike unit normal $n^\mu$, and given a spatial
unit vector $r^\mu$ in the direction of wave propagation, the standard
definition of $\Psi_4$ is the following component of the Weyl
curvature tensor,
\begin{equation}
\Psi_4 = - C_{\alpha\mu\beta\nu} \ell^\mu \ell^\nu \bar{m}^\alpha\bar{m}^\beta,
\label{eq:Psi4Definition}
\end{equation}
where $\ell^\mu \equiv \frac{1}{\sqrt{2}}(n^\mu - r^\mu)$,
and $m^\mu$ is a complex null vector (satisfying $m^\mu \bar{m}_\mu = 1$)
that is orthogonal to $r^\mu$ and $n^\mu$. Here an overbar denotes complex
conjugation.  

For (perturbations of) flat spacetime, $\Psi_4$ is typically evaluated
on coordinate spheres, and in this case the usual
choices for $n^\mu$, $r^\mu$ and $m^\mu$ are
\begin{subequations}
\begin{eqnarray}
  n^\mu &=& \left(\frac{\partial}{\partial t}\right)^\mu,\\
  r^\mu &=& \left(\frac{\partial}{\partial r}\right)^\mu,
            \label{eq:FlatspaceRadialTetrad}\\
  m^\mu &=& \frac{1}{\sqrt{2} r}
            \left(\frac{\partial}{\partial \theta} 
            +    i\frac{1}{\sin\theta}\frac{\partial}{\partial \phi}\right)^\mu,
  \label{eq:FlatspaceMTetrad}
\end{eqnarray}
\end{subequations}
where $(r,\theta,\phi)$ denote the standard spherical coordinates.
With this choice, $\Psi_4$
can be expanded in terms of spin-weighted spherical harmonics of weight $-2$:
\begin{equation}
\Psi_4(t,r,\theta,\phi) 
= \sum_{l m} \Psi_4^{l m}(t,r)\, {}_{-2}Y_{l m}(\theta,\phi),
\label{eq:Psi4Ylm}
\end{equation}
where the $\Psi_4^{l m}$ are expansion coefficients defined by this equation.

For curved spacetime, there is considerable freedom in the choice of
the vectors $r^\mu$ and $m^\mu$, and different researchers have made
different choices~\cite{Buonanno-Cook-Pretorius:2007,Fiske2005,%
  Beetle2005,Nerozzi2005,Burko2006,Campanelli2006,Bruegmann2006} that
are all equivalent in the $r\to\infty$ limit.  We choose these vectors
by first picking an extraction two-surface $\cal E$ that is a
coordinate sphere ($r^2=x^2+y^2+z^2$ using the global asymptotically
Cartesian coordinates employed in our code) centered on the center of
mass of the binary system, i.e. the point of symmetry.  We choose
$r^\mu$ to be the outward-pointing spatial unit normal to $\cal E$
(that is, we choose $r_i$ proportional to $\nabla_i r$ and raise the
index with the
spatial metric).  Then we choose $m^\mu$ according to
Eq.~(\ref{eq:FlatspaceMTetrad}), using the standard spherical
coordinates $\theta$ and $\phi$ defined on these coordinate spheres.
Finally we use Eqs.~(\ref{eq:Psi4Definition}) and~(\ref{eq:Psi4Ylm})
to define the $\Psi_4^{l m}$ coefficients.

Note that the $m^\mu$ vector used here is not exactly null nor
exactly of unit magnitude at finite $r$.  The resulting $\Psi_4^{l m}$
at finite $r$ will disagree with the waveforms observed at infinity.
Our definition does, however, agree with the standard definition given
in Eqs.~(\ref{eq:Psi4Definition})--(\ref{eq:Psi4Ylm}) as $r\to\infty$.
Because we extrapolate the extracted waves to find the asymptotic
radiation field (see Section~\ref{sec:Extrapolation}), these effects
should not play a role in our PN comparisons:  Relative errors in $\Psi_4^{lm}$
introduced by using
the simple coordinate tetrad fall off like $1/r$, 
and thus should vanish after extrapolating to obtain
the asymptotic
behavior.  While more careful treatment of the extraction
method---such as those discussed in \cite{Nerozzi2006, Pazos2006,
  Lehner2007}---may improve the quality of extrapolation and would be
interesting to explore in the future, the naive choice made here
should be sufficient to ensure that the waveform after extrapolation
is correct to the accuracy needed for these simulations.

In this paper, we focus on the $(l,m)=(2,2)$ mode.  Following
common practice (see e.g.~\cite{Baker2006b,Bruegmann2006}), we split
the extracted waveform into real phase $\phi$ and real amplitude $A$,
defined by
\begin{equation}\label{eq:A-phi-definition}
\Psi^{22}_4(r,t) = A(r,t)e^{-i\phi(r,t)}.
\end{equation}
The gravitational-wave frequency is given by
\begin{equation}\label{eq:omega-definition}
\omega=\frac{d\phi}{dt}
\end{equation}
The minus sign in the definition of $\phi$ is chosen so that
the phase increases in time and $\omega$ is positive.
Equation~(\ref{eq:A-phi-definition}) defines $\phi$ only up to
multiples of $2\pi$.  These multiples of $2\pi$ are chosen to make
$\phi$ continuous through each evolution, still leaving an overall
multiple of $2\pi$ undetermined.  We will consider
only phase differences in this paper, so the choice of this overall
phase offset is irrelevant.

\subsection{Convergence of extracted waveforms}

In this section we examine the convergence of the gravitational
waveforms extracted at fixed radius, without extrapolation to
infinity.  This allows us to study the behavior of our code without
the complications of extrapolation.  The extrapolation process and the
resulting extrapolated waveforms are discussed in
Sec.~\ref{sec:Extrapolation}.

\begin{figure}
\includegraphics[scale=0.49]{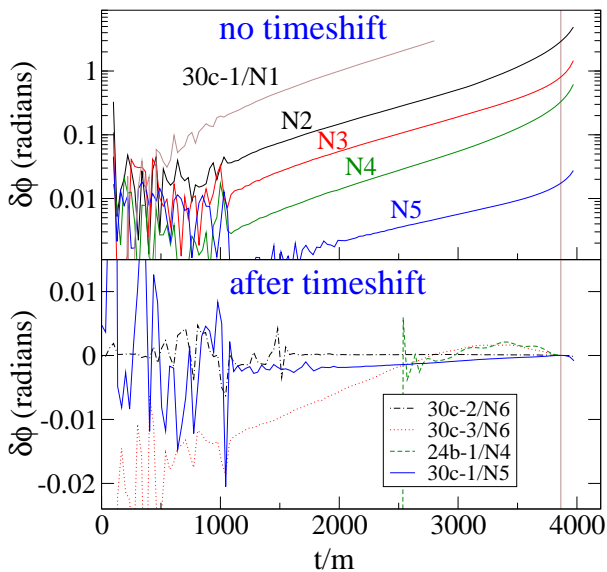}
\caption{\label{fig:PhaseConvergence}Convergence of the gravitational
  wave phase extracted at radius $R=77m$.  All lines show differences
  with respect to our highest resolution run, 30c-1/N6.  The top panel
  shows different resolutions of the same run 30c-1; no time or
  phase shifts have been performed.  The bottom panel compares
  different runs, aligning the runs at $m\omega=0.1$ by a time and
  phase shift.  The thin vertical line 
  indicates the time at which $m\omega=0.1$ for 30c-1/N6. 
} 
\end{figure}

The top panel of Fig.~\ref{fig:PhaseConvergence} shows the convergence
of the gravitational wave phase $\phi$ with numerical
resolution for the run 30c-1.  
For this plot, the waveform is extracted at a fixed radius $R=77m$.
Each line shows the difference between $\phi$ computed at some
particular resolution and $\phi$ computed from our highest-resolution
run 30c-1/N6. When subtracting results at different resolutions, no
time or phase adjustment has been performed.  The difference in $\phi$
between the two highest-resolution runs is smaller than $0.03$ radians
throughout the run, and it is smaller than $0.02$ radians between
$t=1000m$ and the point at which $m\omega=0.1$. 

At times before $1000m$, the phase convergence of our simulation is
limited to about $0.05$ radians because of effects of junk radiation
(described at the end of Section~\ref{sec:Escorting}).  The sharp
pulse of junk radiation has comparatively large numerical truncation
error, and excites all characteristic modes at truncation-error level,
including waves that propagate back toward the origin.  Generation of
these secondary waves stops when the pulse of junk radiation leaves
through the outer boundary ({\em i.e.,} after one light-crossing
time).  Because we use the improved outer boundary conditions of Rinne
et al.~\cite{Rinne2007}, there are no significant reflections
when the junk radiation passes through the outer boundary.  However,
the waves produced before the junk radiation leaves remain in the
computational domain for two additional light-crossing times, until
they eventually leave through the outer boundary.

The bottom panel of Fig.~\ref{fig:PhaseConvergence} shows phase
comparisons between different waveforms after we perform a time shift
and phase shift so that the waveforms agree at $m\omega=0.1$.  Our
procedure for time shifting and phase shifting is the same as the
shifting procedure we use to compare NR with PN waveforms (see
Sec.~\ref{sec:matching-procedure}), so that the error estimates we
extract from the bottom panel of Fig.~\ref{fig:PhaseConvergence}
are relevant for our later NR-PN comparison.

There are three different types of comparisons shown in the bottom
panel of Fig.~\ref{fig:PhaseConvergence}: Phase differences between
runs with the same initial data but with different outer boundary
locations, phase differences between runs with different initial data,
and phase differences between different numerical resolutions of the
same run (this last comparison is the same as what is shown in the top
panel, except in the bottom panel the waveforms are time and phase
shifted).  We will discuss all three of these in turn.

First, we compare the phase difference of 30c-1/N6 with
runs that have different outer
boundary locations.  Run 30c-2 (with more distant outer boundary)
agrees to within $0.002$ radians with run 30c-1, but run 30c-3 (with
closer outer boundary), has a much larger phase difference with 30c-1.
We believe that this is because run 30c-3 has a very small ratio of
outer boundary location to gravitational wavelength: $R/\lambda$ is about 1.1
for the first two-thirds of the run, and remains less than 2 for the entire
run.

We can explain the order of magnitude of these phase differences using
the analysis of Buchman \& Sarbach~\cite{Buchman2006}.
Our outer boundary conditions are not perfectly absorbing, but instead
they reflect some fraction of the outgoing radiation.\footnote{However,
in a comparison of various boundary conditions~\cite{Rinne2007},
the boundary conditions we use produced smaller reflections than
other boundary conditions commonly used in numerical relativity.} 
The ratio of the
amplitude of curvature perturbations (i.e. $\Psi_4$) of the
reflected wave to that of the outgoing wave is
\begin{equation}
q\approx \frac{3}{2(2\pi)^4}\left(\frac{\lambda}{R}\right)^4.
\end{equation}
The incoming reflected waves grow like $1/r$ as they travel inward just
like the outgoing waves decrease by $1/r$ as they propagate outward.
Therefore, the ratio of amplitudes of incoming and outgoing waves will
have approximately the same value, $q$, at smaller radii, 
and we assume for the sake of this rough argument that this ratio 
remains equal to $q$ even in the vicinity of the black holes (where it
is no longer technically meaningful to talk about 'radiation').
Now consider the second time derivative of the
gravitational wave phase, $\ddot \phi$; this is nonzero only because
of gravitational wave emission, so $\ddot \phi$ is proportional to
some power of the outgoing wave amplitude.  
To get the correct power, we can use
Eq.~(\ref{eq:xdot}) to find $\dot x\sim x^5$, so
Eq.~(\ref{eq:dPhidt}) yields $\ddot\phi \sim x^{11/2}$ (we assume
gravitational wave phase is twice the orbital phase).  The
amplitude of $\Psi_4$ scales like $x^4$, so
$\ddot\phi \sim A^{11/8}$.  Let us assume for the sake
of this rough error estimate that the change in $\ddot \phi$ due to
the {\it ingoing} reflected wave scales similarly with amplitude,
$\ddot \phi\sim \bar{A}^{11/8}$, 
where $\bar{A}=q A$ is the amplitude of the reflected ingoing wave.
Therefore the unphysical gravitational-wave force
acting back on the system due to boundary reflections will cause
fractional errors in the second derivative of the phase of about $q^{11/8}$. 
That is, the error $\delta\phi$ caused by the
improper boundary condition will be given by
\begin{equation}
\frac{d^2 \delta\phi}{dt^2} = q^{11/8} \frac{d^2\phi}{dt^2}.
\end{equation}
Integrating this yields $\delta\phi=q^{11/8}\phi$, where $\phi$ is the
total gravitational wave phase accumulated during the evolution.  For
30c-3, $\lambda/R\sim 0.9$, so $q\sim 6\times 10^{-4}$, which yields
$\delta\phi\sim 0.08$ radians for an accumulated gravitational wave phase of
about $200$ radians. This rough estimate agrees in order of magnitude
with the phase difference between 30c-3 and 30c-1 as shown in the bottom
panel of Figure~\ref{fig:PhaseConvergence}.  The run 30c-1 has an outer
boundary about 2.5 farther away, reducing the reflection coefficient
by a factor $2.5^4\approx 40$, so for 30c-1 this estimate of the
phase error gives $\delta\phi=5\times 10^{-4}$ radians.  Therefore, we
expect reflection of the outgoing radiation at the outer boundary to
be insignificant for 30c-1. 
This is confirmed by the excellent agreement between runs
30c-1 and 30c-2 (the latter having even larger outer boundary).

The second comparison shown in the lower panel of
Fig.~\ref{fig:PhaseConvergence} is the phase difference between
30c-1/N6 and 24b-1/N4, a shorter 8-orbit evolution started from a
separate initial data set (set 24b in Table~\ref{tab:ID})
with a separate eccentricity-reduction
procedure.  The phase agreement between these two runs (including an
overall time shift and phase shift) is better than $0.01$ radians for
a total accumulated phase of $\sim 100$ radians of the 8-orbit run, i.e. better
than one part in $10^4$.  Run 24b-1 has a similar outer boundary
location as run 30c-3, and indeed both of these runs show similar
phase differences from 30c-1.

Finally, the third comparison shown in the lower panel of
Fig.~\ref{fig:PhaseConvergence} is the phase difference between
the two highest resolutions of the run 30c-1 when a time shift is
applied.  For $t\gtrsim 1000m$
the agreement is much better than without the time shift
(see upper panel), indicating that the dominant error is a small
difference in the overall evolution time.  For the post-Newtonian
comparisons we perform in the second part of this paper, waveforms are
always aligned at specific frequencies by applying time and phase
shifts.  Therefore, the time-shifted phase difference as displayed in
the lower panel is the most appropriate measure of numerical
truncation error for these PN comparisons. This difference is less
than $0.003$ radians after $t=1000m$ but is larger, about $0.02$
radians, at early times where the waveforms are noisy because of junk
radiation.

\begin{figure}
\includegraphics[scale=0.49]{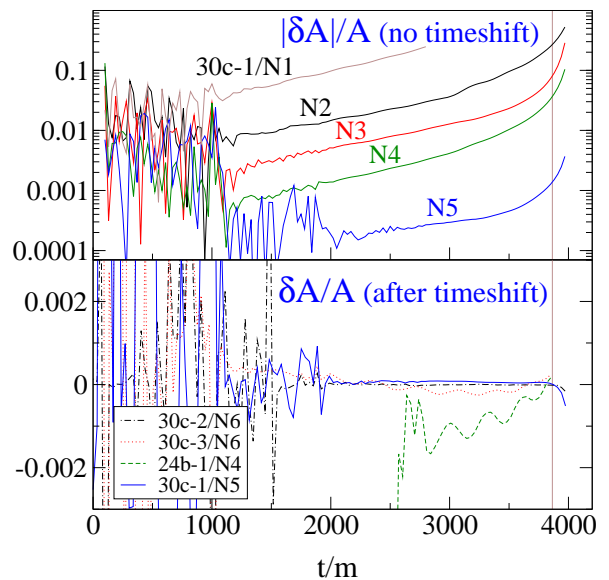}
\caption{\label{fig:AmplitudeConvergence}Convergence of the
  gravitational wave amplitude extracted at radius $R=77m$.  This plot
  corresponds to Fig.~(\ref{fig:PhaseConvergence}), except that
  relative amplitude differences are shown.  The thin vertical line
  indicates the time at which $m\omega=0.1$ for 30c-1/N6.}
\end{figure}

We now compare the gravitational wave amplitudes of different runs in the
same manner as we compared the gravitational wave phases.
Figure~\ref{fig:AmplitudeConvergence} presents convergence data for
the amplitude of the gravitational waves for the same runs as shown
in Fig.~\ref{fig:PhaseConvergence}.  Spatial truncation error for the
amplitude is less then $0.1$ percent for $t/m>1000$, and earlier than
this it is limited by residual noise from the junk radiation.
Differences (including time shifts) between runs of different lengths
are shown in the lower panel of Fig.~\ref{fig:AmplitudeConvergence}.
These differences are even smaller,
but because of their small size, they are dominated by noise for about
the first half of the run.  The oscillations apparent in the comparison
to 24b-1 are caused by the larger orbital eccentricity of 24b-1
(cf. Tab.~\ref{tab:ID}).

\subsection{Extrapolation to infinity}
\label{sec:Extrapolation}

The quantity of interest to gravitational wave detectors is the
gravitational waveform as seen by an observer effectively infinitely
far from the source.  Our numerical simulations, in contrast, cover
only a region of finite volume around the source, and our numerical
waveforms are extracted at a finite radius.  Waveforms extracted at a
finite radius can differ from those extracted at infinity because of
effects discussed in Section~\ref{sec:waveform-extraction}; these
effects can lead to phase errors of several tenths of a radian and
relative amplitude errors of several percent.  To avoid such errors we
extrapolate to infinite extraction radius as follows.

\begin{figure}
\includegraphics[scale=0.49]{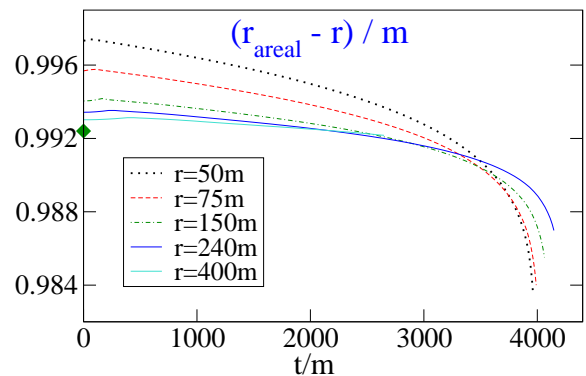}
\caption{\label{fig:ArealRadius}Difference between areal radius
  $r_{\rm areal}$ and coordinate radius $r$ of selected extraction surfaces.
  $r_{\rm areal}$ remains constant to within $~0.01m$ during the
  evolution.  The diamond indicates $\Eadm/m$ of the initial data.  }
\end{figure}

We extract data for $\Psi_4$ on coordinate spheres of coordinate radii
$r/m=75, 80, 85, \ldots, 240$, as described in
Section~\ref{sec:waveform-extraction}.
These extracted waveforms are shifted in time relative
to one another because of the finite light-travel time between these
extraction surfaces.  We correct for this by shifting each waveform by
the tortoise-coordinate radius at that extraction
point~\cite{Fiske2005}
\begin{equation}
\label{eq:Rstar}
  r^\ast = r_{\rm areal} + 
2\Eadm\ln\left( \frac{r_{\rm areal}}{2\Eadm} - 1 \right)\ .
\end{equation}
Here $\Eadm$ is the ADM mass
of the initial data, and
$r_{\rm areal}=\sqrt{A/4\pi}$, where $A$ is the area of the extraction
sphere.
This is not the only possible choice for the
retarded time---for example, the waveforms could be shifted
so that the maxima of the amplitude align~\cite{Hannam2007}.  It has also
been suggested~\cite{Kocsis2007} that the time shift should change with the
amount of radiated energy---essentially, that the factor of $\Eadm$
should be replaced by the amount of mass interior to the
extraction radius at each time.  We leave investigation
of other choices of retarded time for future work.

Figure~\ref{fig:ArealRadius} presents the areal radius during the
evolution at several typical extraction radii.
The areal radius of these extraction surfaces is constant to within
about $0.01m$, and to the same precision, $r_{\rm areal}=r+\Eadm$.
This relationship is not surprising, because the initial data is
conformally flat, so that for coordinate spheres $r_{\rm
  areal}=r+\Eadm+{\cal O}(\Eadm/r)$.  For convenience, we simply set
$r_{\rm areal}=r+\Eadm$ in Eq.~(\ref{eq:Rstar}), rather than
explicitly integrating to find the area of each extraction sphere.

\begin{figure}
  \includegraphics[scale=0.49]{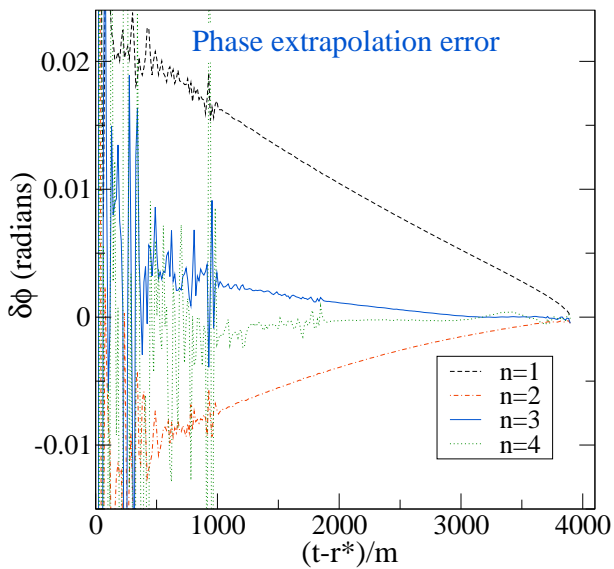}
 \caption{\label{fig:ExtrapolationConvergence_Phi} Error of phase
   extrapolation to infinity for extrapolation of order
   $n$, cf. Eq.~(\ref{eq:phi-Extrapolation}).  Plotted are absolute
   differences between extrapolation with order $n$ and $n+1$.
   Increasing the order of the polynomial increases accuracy, but also
   amplifies noise.}
\end{figure}

After the time shift, each waveform is a function of retarded time,
$t-r^\ast$.  At a given value of retarded time, we have a series of
data points---one for each extraction radius.  We fit phase and
amplitude of these data separately to a polynomial in $1/r$,
\begin{align}\label{eq:phi-Extrapolation}
  \phi(t-r^\ast, r)&=\phi_{(0)}(t-r^\ast)+
\sum_{k=1}^n \frac{\phi_{(k)}(t-r^\ast) }{r^k},\\
\label{eq:A-Extrapolation}
  r A(t-r^\ast, r)&=A_{(0)}(t-r^\ast)+
\sum_{k=1}^n \frac{A_{(k)}(t-r^\ast) }{r^k}.
\end{align}
The leading-order term of each polynomial, as a function of
retarded time, is then the desired asymptotic waveform:
\begin{align}
\phi(t-r^\ast)&=\phi_{(0)}(t-r^\ast),\\
 r A(t-r^\ast)&=A_{(0)}(t-r^\ast).
\end{align}

We find good convergence of this method as we increase the order $n$ of
the extrapolating polynomial.
Figure~\ref{fig:ExtrapolationConvergence_Phi} shows the difference in
phase between waveforms extrapolated using successively higher-order
polynomials.  We see a broad improvement in the accuracy of the phase
with increasing order, but unfortunately, higher order extrapolations
tend to amplify the noise.  Our preferred choice is $n=3$
extrapolation, resulting in extrapolation errors of $\lesssim 0.003$
radians for $t-r^\ast\gtrsim 1000m$.

\begin{figure}
  \includegraphics[scale=0.49]{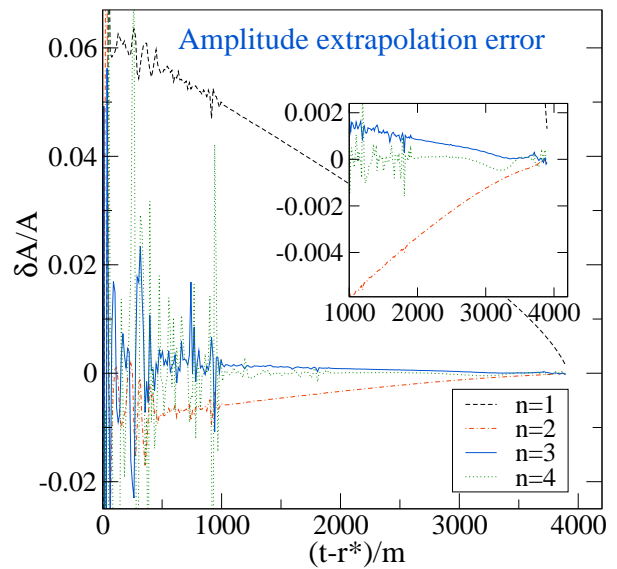}
  \caption{Error of amplitude extrapolation to infinity for
    extrapolation with order $n$, cf. Eq.~(\ref{eq:A-Extrapolation}).
    Plotted are relative amplitude differences between extrapolation
    with orders $n$ and $n+1$.  The inset is an enlargement for
    $t-r^\ast\ge 1000m$.
    \label{fig:ExtrapolationConvergence_Amp}}
\end{figure}

Figure~\ref{fig:ExtrapolationConvergence_Amp} is analogous to
Fig.~\ref{fig:ExtrapolationConvergence_Phi}, except that it shows
relative differences in the extrapolated amplitudes.  The basic
picture agrees with the phase extrapolation: Higher order
extrapolation reduces the errors, but amplifies noise.  Our preferred
choice $n=3$ gives a relative amplitude error of $\lesssim 0.002$ for
$t-r^\ast\gtrsim 1000m$, dropping to less than $0.001$ for
$t-r^\ast\gtrsim 2000m$.

Phase and amplitude extrapolation become increasingly more accurate
at late times.  The main obstacle to accuracy seems to be near-zone
effects scaling with powers of $(\lambda/r)$, where $\lambda$ is the
wavelength of the gravitational wave.  The wavelength is quite large
at the beginning of the simulation ($\lambda\approx 180m$,
cf. Fig.~\ref{fig:0093g_Lev6_R0240m}), but becomes shorter during the
evolution, so that even low-order extrapolation is quite accurate at
late times.  Alternatively, near-zone effects can be mitigated by
using data extracted at large values of $r$.  It is precisely because
of these near-zone effects that we have chosen to ignore
data extracted at $r<75m$ when we extrapolate to infinity.

\begin{figure}
  \includegraphics[scale=0.49]{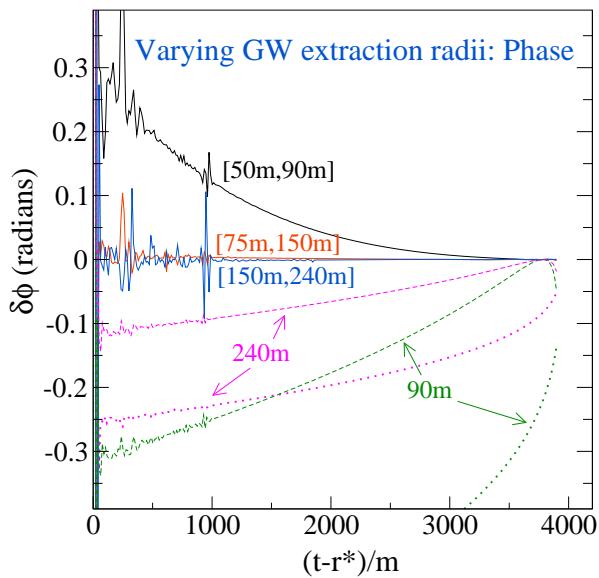}
  \caption{ \label{fig:ExtrapolationRadii_Phi}Effect of wave
    extraction radii on extrapolated phase.  Each curve represents the
    difference from our preferred wave extrapolation using $r\in[75m,
      240m]$.  The three solid curves represent extrapolation from
    different intervals of extraction radii.  The curves labeled
    ``240m'' and ``90m'' represent differences from
    waves extracted at these two radii,
    without any extrapolation, for two cases: time and phase shifted
    so that $\phi$ and $\dot\phi$ match at $m\omega=0.1$ (dashed),
    and without these shifts (dotted).  }
\end{figure}

\begin{figure}
  \includegraphics[scale=0.49]{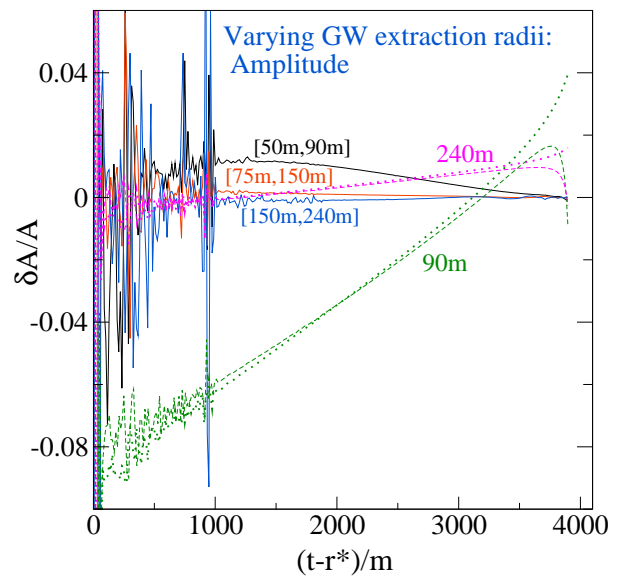}
  \caption{ \label{fig:ExtrapolationRadii_Amp} Effect of choice of
    wave extraction radii on extrapolated amplitude.  Each curve
    represents the (relative) amplitude difference to our preferred
    wave extrapolation using $r\in[75m, 240m]$.  The three solid curves
    represent extrapolation from different intervals of extraction
    radii.  
    The curves labeled
    ``240m'' and ``90m'' represent differences from
    waves extracted at these two radii,
    without any extrapolation, for two cases: time and phase shifted
    so that $\phi$ and $\dot\phi$ match at $m\omega=0.1$ (dashed),
    and without these shifts (dotted).  }
\end{figure}

In Figs.~\ref{fig:ExtrapolationRadii_Phi}
and~\ref{fig:ExtrapolationRadii_Amp}, we show the effects of
extrapolation using different ranges of extracted data.  Using data
extracted every $5m$ in the range $r\!=\!50m\text{--}90m$ results in
noticeable differences early in the run---though it is adequate later
in the run.  For ranges at higher radii (e.g. $[75m, 150m]$ or $[150m,
  240m]$), the accuracy is not highly variable, though we find that
noise is increased when using data from such a smaller range of
extraction radii.

To estimate the errors generated by not extrapolating waveforms to
infinity at all, Fig.~\ref{fig:ExtrapolationRadii_Phi} contains also
the phase difference between wave extraction at two finite radii
($90m$ and $240m$) and our preferred extrapolated phase at infinity.
The dotted lines show such phase differences when only a time shift by
the tortoise-coordinate radius of the extraction sphere is applied.
The errors are dramatic, tenths of radians or more, even very late in
the run.  When matching to post-Newtonian waveforms, we are free to
add an overall time and phase shift
(cf. Section~\ref{sec:matching-procedure}). Therefore, the dashed
lines in Fig.~\ref{fig:ExtrapolationRadii_Phi} show phase differences
with the same unextrapolated waveforms as shown by the dotted
lines, except that a phase and time shift has been applied so that the
$\phi$ and $\dot\phi$ agree with those of the extrapolated waveform
late in the run (where $m\omega=0.1$),
where the wavelengths are shortest and wave
extraction is expected to work best.  Even with such an adjustment,
the gravitational wave phase extracted at $r=90m$ differs by about 0.1
radian at $t\sim 1000m$ before coalescence, with this difference growing to
0.3 radians at the start of our simulation.

Figure~\ref{fig:ExtrapolationRadii_Amp} makes the same comparison for
the gravitational wave amplitude.  Wave extraction at $r=90m$ results in
relative amplitude errors of up to 8 per cent, and of about 2 per cent
even in the last $1000m$ of our simulation.  We also point out that the
errors due to finite extraction radius decay approximately as the
inverse of the extraction radius: For waves extracted at $r=240m$ the
errors are smaller than for waves extracted at $r=90m$
by about a factor of three, as can be seen in
Figs.~\ref{fig:ExtrapolationRadii_Phi}
and~\ref{fig:ExtrapolationRadii_Amp}; for wave extraction at $r=45m$,
the errors would be approximately twice as large as the $r=90m$ case.
The errors introduced by using a finite extraction radius are
significantly larger than our truncation error (even at extraction
radius $240m$).  Therefore extrapolation to infinity is essential to
realize the full accuracy of our simulations.

\subsection{Estimated time of merger}
\label{sec:EstimatedTimeOfMerger}

Since we have not yet been successful with simulating the merger,
we do not precisely know when merger occurs.  However, by comparing
the orbital and gravitational wave frequencies to already published
results, we can nevertheless estimate the time of merger. 

The simulation presented in Fig.~\ref{fig:Trajectories} stops at time
$t=3929m$ when the horizons of the black holes become too distorted
just before merger.  At that point, the proper separation between the
horizons is $\sim 4.0m$, and the orbital frequency has reached
$m\Omega_{\rm orbit}=0.125$; comparison
with~\cite{Buonanno-Cook-Pretorius:2007} suggests this is about $15m$
before formation of a common apparent horizon, i.e. the common horizon
should form in our simulations at $t_{\rm CAH}\approx 3945m$.

The waveform extrapolated to infinity ends at $t-r^\ast=3897m$ at a
gravitational wave frequency of $m\omega\approx 0.16$.  This places
the end of the waveform at about $50m$ (or $\sim 1.5$ cycles) before
formation of a common apparent horizon\footnote{The waveform ends
  somewhat further from merger than the orbital trajectory, because
  the artificial boundary is placed initially at a radius $\sim 15m$,
  and then moves outward somewhat faster than the speed of light, thus
  overtaking the very last part of the waveform as it travels to the
  wave-extraction radii.} (judged by comparison with
\cite{Buonanno-Cook-Pretorius:2007}).  Thus, we estimate the formation
of a common horizon to correspond to a retarded time of
approximately $(t-r^\ast)_{\rm CAH}\approx 3950m$.

\section{Generation of post-Newtonian waveforms}
\label{sec:PN}

It is not our intention to review all of post-Newtonian (PN) theory,
but to summarize the important points that go into the construction of
the post-Newtonian waveforms that we will compare to our numerical
simulation.  For a complete review of post-Newtonian methods applied to
inspiralling compact binaries, see the review article by
Blanchet\cite{Blanchet2006}.  

The post-Newtonian approximation is a slow-motion, weak-field
approximation to general relativity with an expansion parameter
$\epsilon \sim (v/c)^2 \sim (Gm/rc^2)$. For a binary system of two
point masses $m_1$ and $m_2$, $v$ is the magnitude of the
relative velocity, $m$ is the
total mass, and $r$ is the separation.  In order to produce a
post-Newtonian waveform, it is necessary to solve both the
post-Newtonian equations of motion describing the binary, and the
post-Newtonian equations describing the generation of gravitational
waves.

Solving the equations of motion yields explicit expressions for the
accelerations of each body in terms of the positions and velocities of
the two bodies\cite{Jaranowski98a, Jaranowski99a, Damour00a,
  Damour01a, Blanchet00a, Blanchet01a, Damour01b, Blanchet04, Itoh01,
  Itoh03, Itoh04}.  The two-body equations of motion can then be
reduced to relative equations of motion in the center-of-mass frame in
terms of the relative position and velocity\cite{Blanchet03a}.  The
relative acceleration is currently known through 3.5PN order, where
0PN order for the equations of motion corresponds to Newtonian
gravity.  The effects of radiation reaction (due to the emission of
gravitational waves) enters the relative acceleration starting at
2.5PN order.  The relativistic corrections to the relative
acceleration at 1PN, 2PN and 3PN order (ignoring the radiation
reaction terms at 2.5PN and 3.5PN order) admit a conserved center of
mass binding energy through 3PN order\cite{Andrade01}.  There is no
2.5PN or 3.5PN order contribution to the energy.

Solving the post-Newtonian wave generation problem yields expressions
for the gravitational waveform $h_{ij}$ and gravitational wave flux
${\cal L}$ in terms of radiative multipole moments\cite{thorne80}.
These radiative multipole moments are in turn related to the source
multipole moments, which can be given in terms of the relative
position and relative velocity of the binary\cite{Blanchet98}.  For
the gravitational wave generation problem, PN orders are named with
respect to the leading order waveform and flux, which are given by the
quadrupole formalism.  Thus, for example, 1.5PN order in the wave
generation problem represents terms of order $(v/c)^3$ beyond
quadrupole. Higher order effects enter both through post-Newtonian
corrections to the mass quadrupole, as well as effects due to higher
multipole moments.  Starting at 1.5PN order the radiative multipole
moments include non-linear and non-instantaneous (i.e. depend upon the
past history of the binary) interactions among the source multipole
moments (e.g. gravitational wave tails)\cite{Blanchet98, Blanchet92,
  Blanchet98a, Blanchet98b}.

It was recognized early that simply plugging in the orbital evolution
predicted by the equations of motion into the expressions for the
waveform would not generate templates accurate enough for matched
filtering in detecting gravitational waves\cite{cutler_etal93}.  This
is because radiation reaction enters the equations of motion only at
the 2.5PN order; hence computing a waveform to $k$ PN order beyond the
quadrupole formalism would require $2.5+k$ PN orders in the equations
of motion.  In order to obtain as accurate a post-Newtonian waveform
as possible it is thus necessary to introduce the assumption of an
adiabatic inspiral of a quasi-circular orbit, as well as the
assumption of energy
balance between the orbital binding energy and the energy emitted by
the gravitational waves.

\subsection{Adiabatic inspiral of quasi-circular orbits}

The emission of gravitational radiation causes the orbits of an
isolated binary system to circularize~\cite{Peters1964}.  Thus it is a
reasonable assumption to model the orbital evolution of the binary as
a slow adiabatic inspiral of a quasi-circular orbit.  With this
assumption, post-Newtonian expressions for the orbital energy $E$ and
gravitational energy flux ${\cal L}$ are currently known through
3.5PN order
\cite{Blanchet02,Blanchet02a,Blanchet05a,Blanchet04a,Blanchet05}.
These expressions can be given in terms of a parameter related to
either the harmonic coordinate separation $r$, or to the orbital
frequency $\Omega$.  We choose to use the expressions given in terms
of a frequency-related parameter
\begin{equation}
\label{eq:xDefinition}
x \equiv \left( \frac{G m \Omega}{c^3} \right)^{2/3}
\end{equation}
rather than a coordinate-related parameter, because the coordinate
relationship between the numerical simulation and the harmonic
coordinates used in post-Newtonian approximations is unknown.
The orbital energy for an equal mass system is given by\cite{Blanchet2006}
\begin{align}
\label{eq:PNEnergyDefinition}
E = - \frac{m c^2}{8} x \Bigg[ &1 - \frac{37}{48} x - \frac{1069}{384} x^2 
\nonumber \\
&+\left( \frac{1427365}{331776} -\frac{205}{384} \pi^2 \right) x^3 \Bigg],
\end{align}
and the gravitational wave flux for an equal mass system is given by
\cite{Blanchet2006}
\begin{eqnarray}
\label{eq:PNFluxDefinition}
{\cal L} &=& \frac{2 c^5}{5 G} x^5 \left\{
1 - \frac{373}{84} x + 4 \pi  x^{3/2} - \frac{59}{567} x^2 
- \frac{767}{42} \pi x^{5/2} \right. \nonumber \\ && \left. +
\left[ \frac{18608019757}{209563200} + \frac{355}{64} \pi^2
- \frac{1712}{105} \gamma \right. \right. \nonumber \\* && \left. \left.
- \frac{856}{105} \ln{(16 x)} \right] x^3 
+ \frac{16655}{6048} \pi x^{7/2} \right\},
\end{eqnarray}
where $\gamma = 0.577216 \ldots$ is Euler's constant.

\subsection{Polarization Waveforms}
\label{sec:PN-PolarizationWaveforms}

The gravitational polarization waveforms for a quasi-circular orbit in the
$x-y$ plane, as measured by an observer at spherical coordinates $(R,{\hat
\theta},{\hat \phi})$, are given by
\begin{eqnarray}
h_+ &=& \frac{2 G \mu}{c^2 R} x \left\{ -(1 + \cos{\hat \theta}) 
\cos{2 (\Phi - {\hat \phi})} + \cdots \right\} \\*
h_\times &=& \frac{2 G \mu}{c^2 R} x \left\{ -2 \cos{\hat \theta} 
\sin{2 (\Phi - {\hat \phi})} + \cdots \right\},
\end{eqnarray}
where $\Phi$ is the orbital phase (measured from the x-axis) and $\mu
= m_1 m_2 / m$ is the reduced mass.  The polarization waveforms
are currently known through 2.5PN order\cite{Arun2004,Kidder07}.

\subsubsection{Optimally oriented observer}

For an equal-mass binary the polarization waveforms along the $z$-axis
(i.e. the optimally oriented observer along the normal to the orbital plane)
are given by \cite{Arun2004,Kidder07}
\begin{widetext}
\begin{eqnarray}
h_+^{(z)} &=& \frac{G m}{2 c^2 R} x \left( \cos{2 \Phi} \left\{
-2 + \frac{17}{4} x - 4 \pi x^{3/2} + \frac{15917}{2880} x^2 + 9 \pi
x^{5/2} \right\} 
\right. \nonumber \\* && \qquad\qquad\left. 
+ \sin{2 \Phi} \left\{
- 12 \ln{\left(\frac{x}{x_0}\right)} x^{3/2} +
\left[\frac{59}{5} + 27 \ln{\left(\frac{x}{x_0}\right)}
\right] x^{5/2} \right\} \right) 
\label{eq:hZaxisplus}
\\
h_\times^{(z)} &=& \frac{G m}{2 c^2 R} x \left( \sin{2 \Phi}
\left\{ -2 + \frac{17}{4} x - 4 \pi x^{3/2} + \frac{15917}{2880} x^2 +
9 \pi x^{5/2} \right\} 
\right. \nonumber \\* && \qquad\qquad\left. 
+ \cos{2 \Phi}
\left\{ 12 \ln{\left(\frac{x}{x_0}\right)} x^{3/2} -
\left[\frac{59}{5} + 27 \ln{\left(\frac{x}{x_0}\right)}
\right] x^{5/2} \right\} \right),
\label{eq:hZaxiscross}
\end{eqnarray} 
\end{widetext}
where 
\begin{equation}
\label{eq:LogParameter}
\ln{x_0} \equiv \frac{11}{18} - \frac{2}{3}\gamma + 
\frac{2}{3}\ln{\left(\frac{Gm}{4bc^3}\right)}
\end{equation}
is a constant frequency scale that depends upon the constant time scale
$b$ entering the gravitational wave tail contribution to the
polarization waveforms~\cite{Wiseman93,Blanchet93}.
The freely-specifiable constant
$b$ corresponds to a choice of the origin of radiative time $T$ with
respect to harmonic time $t$, and enters the relation between the
retarded time $T_R = T - R/c$ in radiative coordinates (the
coordinates in which the waveform is given) and the retarded time $t -
r/c$ in harmonic coordinates (the coordinates in which the equations
of motion are given)~\cite{Wiseman93,Blanchet93}:
\begin{equation}
T_R = t - \frac{r}{c} - \frac{2 G \Eadm}{c^3} \ln{\left(\frac{r}{bc}\right)}.
\end{equation}
Here $\Eadm$ is the ADM mass (mass monopole) of the binary system.
 
\subsubsection{The (2,2) mode}
\label{sec:PN-22-mode}

When comparing a post-Newtonian waveform with data from a physical
gravitational wave detector, it is necessary to compare waves emitted
in a certain direction $(\hat{\theta},\hat{\phi})$ with respect to the
source.  However, comparing waveforms between PN and numerical
simulations can be done in all directions simultaneously by
decomposing the waveforms in terms of spherical harmonics and then
comparing different spherical harmonic modes.  Since the power in
each spherical harmonic mode decreases rapidly with spherical harmonic
index, with the $(2,2)$ mode dominating (for an equal-mass
non-spinning binary), it is possible to do a very accurate comparison
that is valid for all angles by using only a few modes.  In addition,
as pointed out by Kidder~\cite{Kidder07a}, the dominant (2,2) mode can
be computed to 3PN order.  For an equal-mass binary, the (2,2) mode is
\begin{widetext}
\begin{eqnarray}
h_{(2,2)} &=& -2 \sqrt{\frac{\pi}{5}} \frac{G m}{c^2 R} e^{-2i \Phi} 
x \left\{ 1 - \frac{373}{168} x + \left[ 2 \pi + 6 i
  \ln{\left(\frac{x}{x_0}\right)} \right] x^{3/2} -
\frac{62653}{24192} x^2 -  \left[
  \frac{197}{42} \pi + \frac{197 i}{14}
  \ln{\left(\frac{x}{x_0}\right)} + 6 i \right]
x^{5/2} \right. \nonumber \\* 
\label{eq:h22}
&+& \left.\!\!\!\!\left[
  \frac{43876092677}{1117670400} + \frac{99}{128} \pi ^2 -
  \frac{428}{105} \ln{x} - \frac{856}{105} \gamma - \frac{1712}{105}
  \ln{2} - 18 \left[ \ln{\left(\frac{x}{x_0}\right)}
    \right]^2 + \frac{428}{105} i \pi + 12 i \pi
  \ln{\left(\frac{x}{x_0}\right)} \right]\!\!x^3\!\!
\right\}.
\end{eqnarray}
\end{widetext}

Since the (2,2) mode of the numerical waveforms is less noisy than the
waveform measured along the $z$-axis, and since we have access to the 3PN
amplitude correction of the (2,2) mode, we will use the (2,2)
waveforms rather than the $z$-axis waveforms for our comparisons between
NR and PN in Sec.~\ref{sec:Results}.  We have verified (for all
comparisons using post-Newtonian waveforms of $\le 2.5$PN order in
amplitude) that our results do not change significantly when we use
$z$-axis waveforms instead of (2,2) waveforms.

\subsection{Absorbing amplitude terms into a redefinition of the phase}

The logarithms of the orbital frequency parameter $x$ (as well as the
constant frequency scale $x_0$) that appear in the amplitude
expressions~(\ref{eq:hZaxisplus}),~(\ref{eq:hZaxiscross}),
and~(\ref{eq:h22}) can be absorbed into a redefinition of the phase by
introducing an auxiliary phase variable $\Psi = \Phi + \delta$.
Noting that the $\ln{x}$ terms first enter at 1.5 PN order, it is
straightforward to show that
choosing~\cite{Blanchet96,Arun2004,Kidder07a}
\begin{equation}
\label{eq:PhaseShift}
\delta =  - 3 \frac{\Eadm}{m} x^{3/2} \ln{\left(\frac{x}{x_0}\right)},
\end{equation}
where $\Eadm /m = 1 - x/8 + O(x^2)$ for an equal mass system, will
eliminate the $\ln{x}$ terms from both the (2,2) mode as well as
for the polarization waveforms.  This follows from 
\begin{eqnarray*}
h_{(2,2)} &=& A e^{-2i \Psi} \\
&=& A e^{-2i \Phi} e^{-2i \delta} \\
&=& A e^{-2i \Phi} ( 1 - 2i \delta - 2\delta^2 + O(x^{9/2}) ),
\end{eqnarray*}
and similarly for the polarization waveforms.  Furthermore, since the
orbital phase as a function of frequency goes as $x^{-5/2}$ at leading
order (see Eq.~(\ref{eq:TaylorPhase}) below), the $\ln{x}$ terms,
which were 1.5PN, 2.5PN, and 3PN order in the original amplitude
expressions, now appear as phase corrections at relative order 4PN,
5PN, and 5.5PN.  As these terms are beyond the order to which the
orbital phase evolution is known (3.5PN order), it can be argued that
these terms can be ignored. Note that the choices of $x_0$ in
Eq.~(\ref{eq:LogParameter}) and $\delta$ in Eq.~(\ref{eq:PhaseShift})
are not unique; they were made to gather all logarithmic terms into
one term, as well as to simplify the waveform~\cite{Blanchet96}.

\subsection{Energy balance}

The second assumption that goes into making as accurate a
post-Newtonian waveform as possible is that of energy balance.  It is
assumed that the energy carried away by the emission of gravitational
waves is balanced by the change in the orbital binding energy of the
binary,
\begin{equation}
\label{eq:EnergyBalance}
\frac{dE}{dt} = - {\cal L}.
\end{equation}
While this is extremely plausible, it has only been confirmed through
1.5 PN order\cite{Blanchet97}.

Given the above expressions for the energy, flux, and waveform
amplitude, there is still a set of choices that must be made in order
to produce a post-Newtonian waveform that can be compared to our numerical
waveform. These include
\begin{enumerate}
\item The PN order through which terms in the orbital energy
  and luminosity are retained.
\item The procedure by which the energy balance equation
  is used to obtain $x(t)$ and $\Phi(t)$.
\item The PN order through which terms in the waveform amplitude
  are kept.
\item The treatment of the $\ln{x}$ terms. These terms can be included
      in the amplitude or included in the orbital phase via the auxiliary phase
      $\Psi \equiv \Phi + \delta$. If the latter is chosen, these terms can
      be retained or ignored; ignoring them can be justified because they
      occur at higher order than all known terms in the orbital phase.
\end{enumerate}

We always expand energy and luminosity to the same order, which may be
different from the order of the amplitude expansion; both of these
expansion orders are indicated explicitly in each of our comparisons.
We ignore
the $\ln{(x/x_0)}$ terms in the amplitude by absorbing them into the phase
and dropping them because of their high PN order.  In the next section we
describe several choices for obtaining $x(t)$ and $\Phi(t)$ from the
energy balance equation.

\subsection{Taylor approximants: Computing $\Phi(t)$}
\label{sec:taylor-approximants}
In this section we describe how to obtain the orbital phase as a
function of time, $\Phi(t)$, using the energy balance
equation~(\ref{eq:EnergyBalance}).  Different methods of doing this
exist; here we follow the naming convention of~\cite{Damour2001}.  
These methods, and variations
of them, are called Taylor approximants, and all formally agree to a
given PN order but differ in how higher-order terms are truncated.  We
discuss four time-domain approximants here, but more can be defined.

\subsubsection{TaylorT1}

The TaylorT1 approximant is obtained by numerically integrating the
ODEs
\begin{eqnarray}
\frac{dx}{dt} &=& - \frac{{\cal L}}{(dE/dx)} \label{eq:fluxenergy}\\*
\frac{d\Phi}{dt} &=& \frac{c^3}{G m} x^{3/2}\label{eq:dPhidt},
\end{eqnarray}
to produce $\Phi(t)$.  The fraction on the right side of
Eq~(\ref{eq:fluxenergy}) is retained as a ratio of post-Newtonian
expansions, and is not expanded further before numerical integration.
This is the approximant used in the NR-PN comparisons
in~\cite{Hannam2007,Pan2007}.

\subsubsection{TaylorT2}

The TaylorT2 approximant is obtained by starting with the parametric
solution of the energy balance equation:
\begin{eqnarray}
t(x) &=& t_0 + \int_x^{x_0} dx \frac{(dE/dx)}{{\cal L}} \\*
\Phi(x) &=& \Phi_0 + \int_x^{x_0} dx \frac{x^{3/2} c^3}{G m}
\label{eq:T2phaseIntegral}
\frac{(dE/dx)}{{\cal L}}.
\end{eqnarray}
The integrand of each expression is re-expanded as a single post-Newtonian
expansion in $x$ and truncated at the appropriate PN-order; 
these integrals are then evaluated analytically to obtain
for an equal-mass binary \cite{Damour2001,Damour02}:
\begin{eqnarray}
t &=& t_0 - \frac{5 G m}{64 c^3}  x^{-4} \left\{
1 + \frac{487}{126} x - \frac{32}{5} \pi x^{3/2} + \frac{2349439}{254016} x^2 
\right. \nonumber \\* && \left.
- \frac{1864}{63} \pi x^{5/2}  +
\left[ -\frac{999777207379}{5867769600} + \frac{1597}{48} \pi^2
\right. \right. \nonumber \\ && \left. \left.
+ \frac{6848}{105} \gamma 
+ \frac{3424}{105} \ln{(16 x)} \right] x^3 
- \frac{571496}{3969} \pi x^{7/2} \right\} \\* 
\Phi &=& \Phi_0 - \frac{1}{8} x^{-5/2} \left\{
1 + \frac{2435}{504} x - 10 \pi  x^{3/2} + \frac{11747195}{508032} x^2
\right. \nonumber \\* && \left.
+ \frac{1165}{42} \pi  x^{5/2} \ln{x}
+ \left[ \frac{1573812724819}{4694215680} - \frac{7985}{192} \pi^2
\right. \right. \nonumber \\ && \left. \left.
- \frac{1712}{21} \gamma - \frac{856}{21}  \ln{(16 x)} \right] x^3
+ \frac{357185}{7938} \pi x^{7/2} \right\}.
\label{eq:TaylorPhase}
\end{eqnarray}

\subsubsection{TaylorT3}

The TaylorT3 approximant is closely related to TaylorT2.
It is obtained by introducing the dimensionless
time variable
\begin{equation}
\tau \equiv \frac{\nu c^3}{5 G m}(t_0 - t),
\end{equation}
where $\nu=m_1m_2/m^2$ and $\tau^{-1/4} = O(\epsilon)$. The TaylorT2 expression 
$t(x)$ is inverted to obtain $x(\tau)$, and truncated at the desired PN order.
Then $x(\tau)$ is integrated to
obtain 
\begin{equation}
\Phi(\tau) = \Phi_0 - \int_{\tau_0}^{\tau} d\tau \frac{5 x^{3/2}}{\nu}. 
\end{equation}
This procedure yields for an equal-mass binary \cite{Blanchet2006}:
\begin{eqnarray}
x &=& \frac{1}{4} \tau^{-1/4} \left\{ 
1 + \frac{487}{2016} \tau^{-1/4} -\frac{1}{5} \pi \tau ^{-3/8}
\right. \nonumber \\* &+& \left.
 \frac{1875101}{16257024} \tau ^{-1/2}
- \frac{1391}{6720} \pi \tau ^{-5/8}
\right. \nonumber \\* &+& \left.
 \left[ - \frac{999777207379}{1502149017600} + \frac{1597}{12288} \pi^2
+ \frac{107}{420} \gamma 
\right. \right. \nonumber \\ &-& \left. \left.
 \frac{107}{3360} \ln{\left(\frac{\tau }{256}\right)}
\right] \tau ^{-3/4} - \frac{88451}{282240} \pi \tau ^{-7/8}
\right\} \\* 
\Phi &=& \Phi_0 - 4 \tau^{5/8} \left\{ 
1 + \frac{2435}{4032} \tau^{-1/4} - \frac{3}{4} \pi \tau ^{-3/8} 
\right. \nonumber \\* &+& \left.
 \frac{1760225}{1806336} \tau ^{-1/2} 
- \frac{1165}{5376} \pi \tau ^{-5/8} \ln{\tau} 
\right. \nonumber \\* &+& \left.
 \left[ \frac{24523613019127}{3605157642240} - \frac{42997}{40960} \pi^2
- \frac{107}{56} \gamma  \right. \right. \nonumber \\ &+& \left. \left.
 \frac{107}{448} \ln{ \left(\frac{\tau}{256}\right)} \right] \tau ^{-3/4}
+ \frac{28325105}{21676032} \pi \tau ^{-7/8} \right\}
\end{eqnarray}

This is the post-Newtonian approximant used in visual comparisons
by~\cite{Buonanno-Cook-Pretorius:2007} and in the NR-PN comparisons in
~\cite{Hannam2007} at 3PN order in phase.

\subsubsection{TaylorT4}
In addition to simply numerically integrating the flux-energy
equation~(\ref{eq:fluxenergy}), as is
done for
TaylorT1, one may instead re-expand the right side
of~(\ref{eq:fluxenergy}) as a
single series and truncate at the appropriate PN order before
doing the integration.
The phase evolution $\Phi(t)$ can
thus be obtained by numerically integrating the ODEs
\begin{eqnarray}\label{eq:xdot}
\frac{dx}{dt} &=& \frac{16 c^3}{5 G m} x^5 \left\{
1 - \frac{487}{168} x + 4 \pi x^{3/2}
+ \frac{274229}{72576} x^2 
\right. \nonumber \\* &-& \left.  \frac{254}{21} \pi  x^{5/2}
+ \left[ \frac{178384023737}{3353011200} + \frac{1475}{192} \pi^2 
- \frac{1712}{105} \gamma 
\right. \right. \nonumber \\ &-& \left. \left.
 \frac{856}{105} \ln{(16x)} \right] x^3
+ \frac{3310}{189} \pi x^{7/2} \right\}
\\*
\frac{d\Phi}{dt} &=& \frac{x^{3/2} c^3}{G m}.
\end{eqnarray}
This approximant was not considered in~\cite{Damour2001}, however for
consistency with their notation, we call it TaylorT4. TaylorT4 is the
primary approximant used in the NR-PN comparisons
in~\cite{Baker2006d,Baker2006e}, and one of the several approximants
considered in the NR-PN comparisons in~\cite{Pan2007}.
Ref.\cite{Buonanno-Cook-Pretorius:2007} pointed out that TaylorT4 at
3.5PN order in phase is close to TaylorT3 at 3PN order in phase, and
therefore should give similar agreement with numerical results.  

\section{PN-NR Comparison Procedure}
\label{sec:PNComparison}

\subsection{What to compare?}

There are many ways to compare numerical relativity and post-Newtonian
results.  For example, the post-Newtonian orbital phase $\Phi(t)$
could be compared with the coordinate phase of the black hole
trajectories.  However, this and many other comparisons are difficult
to make in a coordinate-independent manner without expending
significant effort to understand the relationship between the gauge
choices used in post-Newtonian theory and in the NR simulations.  Therefore, in
order to obtain the most meaningful comparison possible, we attempt to
minimize gauge effects by comparing gravitational waveforms as seen by
an observer at infinity.  The waveform quantity most easily obtained
from the numerical relativity code is the Newman-Penrose quantity
$\Psi_4$, and we will compare its $(2,2)$ component
[cf. Eq.~(\ref{eq:Psi4Ylm})], split into phase $\phi$ and amplitude
$A$ according to Eq.~(\ref{eq:A-phi-definition}) and extrapolated to
infinite extraction radius.

The post-Newtonian formulae in Section~\ref{sec:PN} yield the metric
perturbation components $h_+$ and $h_\times$, which---for a gravitational
wave at infinity---are related to $\Psi_4$ by
\begin{equation}
\label{eq:Psi4vsh}
  \Psi_4(t) = \frac{\partial^2}{\partial t^2}\left(h_+(t) - i h_\times(t)\right).
\end{equation}
We numerically differentiate the post-Newtonian expressions for $h_+(t)$
and $h_\times(t)$ twice before computing amplitude and phase using
Eq.~(\ref{eq:A-phi-definition}).  Note that $\phi(t)$ will differ
slightly from the phase computed from the metric perturbation
directly, as $\tan^{-1}(h_\times/h_+)$, because both the amplitude and
phase of the metric perturbation are time dependent.  For the same
reason, $\phi(t)$ is not precisely equal to twice the orbital phase.

As in Ref.~\cite{Hannam2007}, we compare $\Psi_4$ rather than
$h_{+,\times}$ to avoid difficulties arising with fixing the
integration constants when integrating the numerically obtained
$\Psi_4$ (see~\cite{Pfeiffer-Brown-etal:2007} for more details).  Both $\Psi_4$ and $h_{+,\times}$ contain the same
information, so differences between both procedures should be minimal.

\subsection{Matching procedure}
\label{sec:matching-procedure}

Each of the post-Newtonian waveforms has an arbitrary time offset $t_0$ and an
arbitrary phase offset $\phi_0$.  These constants can be thought of as
representing the absolute time of merger and the orientation of the
binary at merger, and we are free to adjust them in order to match NR
and PN waveforms.  Following~\cite{Baker2006d,Hannam2007}, we choose these
constants by demanding that the PN and NR gravitational wave phase and
gravitational wave frequency agree at some fiducial frequency
$\omega_m$.  Specifically, we proceed as follows:
We start with a NR waveform $\Psi_4^{\rm NR}(t)$ and an unshifted
PN waveform $\Psi_4^{\rm PN'}(t)$ that has an arbitrary time and phase
shift.  After selecting the matching frequency $\omega_m$, we can find
(to essentially unlimited accuracy) the time $t_c$ such that
the derivative of the PN phase satisfies $\dot\phi_{\rm PN'}(t_c) =\omega_m$,
where $\phi_{\rm PN'}(t)$ is the phase associated with $\Psi_4^{\rm PN'}(t)$.
Similarly, we find the time $t_m$ such that $\dot\phi_{\rm NR}(t_m)=\omega_m$.
The time $t_m$ cannot be found to unlimited accuracy, and the uncertainty in
$t_m$ is due mainly to residual eccentricity of the NR waveform, as discussed
in Section~\ref{sec:Errors-Eccentricity}.  Once we have $t_m$ and $t_c$, we
leave the NR waveform untouched, but we construct a new, shifted, PN waveform
\begin{equation}
\label{eq:DefShiftedPNWaveform}
\Psi_4^{\rm PN}(t) = \Psi_4^{\rm PN'}(t+t_c-t_m) 
e^{i\left(\phi_{\rm PN'}(t_c)-\phi_{\rm NR}(t_m)\right)}.
\end{equation}
The phase of this new PN waveform is therefore
\begin{equation}
\label{eq:DefShiftedPNPhase}
\phi_{\rm PN}(t) = \phi_{\rm PN'}(t+t_c-t_m)-\phi_{\rm PN'}(t_c)+\phi_{\rm NR}(t_m),
\end{equation}
which satisfies $\phi_{\rm PN}(t_m)=\phi_{\rm NR}(t_m)$ and
$\dot\phi_{\rm PN}(t_m)=\omega_m$ as desired.  All our comparisons are then
made using the new shifted waveform $\Psi_4^{\rm PN}(t)$ rather
than the unshifted waveform $\Psi_4^{\rm PN'}(t)$.

\subsection{Choice of Masses}

The post-Newtonian expressions as written in Section~\ref{sec:PN} involve the
total mass $m$, which corresponds to the the sum of the bare masses of
the point particles in post-Newtonian theory.  When comparing PN to
NR, the question then arises as to which of the many definitions of
the mass of a numerically-generated binary black hole solution should
correspond to the post-Newtonian parameter $m$.  For non-spinning
black holes at very large separation, $m$ reduces to the sum of the
irreducible masses of the two holes.  Neglecting tidal heating, the
irreducible masses should be conserved during the inspiral, so that we
identify $m$ with the sum of the irreducible masses of the initial
data 30c.  
As discussed in
Sec.~\ref{sec:SummaryErrors} the black hole spins are sufficiently
small so that there is no discernible difference between irreducible
mass of the black holes and the Christodoulou mass,
Eq.~(\ref{eq:Christoudoulou-mass}).  Of course, the latter would be 
more appropriate for spinning black holes.

\section{Estimation of uncertainties}
\label{sec:SummaryErrors}

To make precise statements about agreement or disagreement between
numerical and post-Newtonian waveforms, it is essential to
know the size of the uncertainties in this comparison.  When
discussing these uncertainties, we must strive to include all effects that may
cause our numerical waveform to differ from the post-Newtonian
waveforms we compare to.  For instance, in addition to considering
effects such as numerical truncation error, we also
account for the fact that NR and PN waveforms correspond to
slightly different physical scenarios: The PN waveforms have
identically zero spin and eccentricity, whereas the numerical
simulations have some small residual spin and eccentricity.
Table~\ref{tab:Errors} lists all effects we have considered; we discuss
these in detail below starting in
Sec.~\ref{sec:Errors-Numerical}.  All uncertainties are quoted in
terms of phase and amplitude differences, and apply to waveform
comparisons via matching at a fixed $\omega_m$ according to the
procedure in Sec.~\ref{sec:matching-procedure}.

Most of the effects responsible for our uncertainties are time dependent, so
that it is difficult to arrive at a single number describing each
effect.  For simplicity, the error bounds in Table~\ref{tab:Errors}
ignore the junk-radiation noise that occurs in the numerical waveform
for $t-r^\ast\lesssim 1000m$.  The extent to which this noise affects
the PN-NR comparisons presented below in
Sections~\ref{sec:ComparisonWithPN}
and~\ref{sec:ComparisonDifferentPN} will be evident from the noise in
the graphs in these sections. Note that all four matching frequencies
$\omega_m$ occur after the noise disappears at $t-r^\ast\sim 1000m$.
Furthermore, the post-Newtonian waveforms end at different times
depending on the PN order and on which particular post-Newtonian
approximant is used.  Therefore, in order to produce a single number
for each effect listed in Table~\ref{tab:Errors}, we consider only the
part of the waveform prior to some cutoff time, which we choose to be
the time at which the numerical waveform reaches gravitational wave
frequency $m\omega=0.1$.  

\begin{table}
\caption{
\label{tab:Errors}Summary of uncertainties in the comparison
between numerical relativity and post-Newtonian expansions.  
Quoted
error estimates ignore the junk-radiation noise
at $t\lesssim 1000m$ and apply to times
before the numerical waveform reaches
gravitational wave frequency $m\omega=0.1$.
Uncertainties apply to
waveform comparisons via matching at a fixed $\omega_m$ according to the
procedure in Sec.~\ref{sec:matching-procedure}, and
represent the maximum values for all four different matching frequencies
$\omega_m$ that we consider, unless noted otherwise.
}
\begin{tabular}{lll}
Effect & $\delta\phi$ (radians) & $\delta A/A$        \\\hline
Numerical truncation error &  \quad0.003 &0.001       \\
Finite outer boundary      & \quad0.005  & 0.002      \\
Extrapolation $r\to\infty$   & \quad0.005& 0.002      \\
Wave extraction at $r_{\rm areal}$=const? & \quad0.002 & $10^{-4}$ \\
Drift of mass $m$ & \quad0.002 & $10^{-4}$\\
Coordinate time = proper time? & \quad0.002 &$10^{-4}$\\
Lapse spherically symmetric? & \quad0.01 &$4\times 10^{-4}$ \\
residual eccentricity & \quad0.02\footnote{For the case of
matching at $m\omega_m=0.04$, the phase uncertainty due to residual
eccentricity increases to 0.05 radians, thus increasing the
root-mean-square sum to 0.06 radians.} &0.004\\
residual spins & \quad0.03 &$0.001$ \\
\hline
{\bf root-mean-square sum} & \quad{\bf 0.04}\footnotemark[1] 
& {\bf 0.005}
\end{tabular}
\end{table}

\subsection{Errors in numerical approximations}
\label{sec:Errors-Numerical}

The first three error sources listed in Table~\ref{tab:Errors}
have already been discussed in detail in
Section~\ref{sec:NRGeneration}.  We estimate numerical truncation
error using the difference between the two highest resolution runs
after the waveforms have been shifted to agree at some matching
frequency $\omega_m$.  For $m\omega_m=0.1$ this difference is shown as
the curves labeled '30c-1/N5' in the lower panels of
Figs.~\ref{fig:PhaseConvergence} and~\ref{fig:AmplitudeConvergence},
and corresponds to a phase difference of 0.003 radians and a relative
amplitude difference of 0.001.  For other values of $\omega_m$ the
differences are similar.  The effect of the outer boundary is
estimated by the difference between the runs 30c-1/N6 and 30c-2/N6,
which for $m\omega_m=0.1$ is shown as the curves labeled '30c-2/N6' in
the lower panels of Figs.~\ref{fig:PhaseConvergence}
and~\ref{fig:AmplitudeConvergence}, and amount to phase
differences of 0.005 radians and relative
amplitude differences of 0.002.  Errors associated with
extrapolation to infinity have been discussed in detail in
Figs.~\ref{fig:ExtrapolationConvergence_Phi} and
\ref{fig:ExtrapolationRadii_Phi}.  Specifically,
Fig.~\ref{fig:ExtrapolationConvergence_Phi} shows that increasing the
extrapolation order between 3 and 4
changes the extrapolated phase by less than
$0.005$ radians, and Fig.~\ref{fig:ExtrapolationRadii_Phi} confirms
that the extrapolated result is robust under changes of extraction
radii.

\subsection{Constancy of extraction radii}
\label{sec:Errors-Extraction}

If the physical locations of the coordinate-stationary extraction
radii happen to change during the evolution, then the extracted
gravitational waves will accrue a timing error equal to the
light-travel time between the original location and the final
location.  From Fig.~\ref{fig:ArealRadius}, we see that the drift in
areal radius is less than $0.02m$, resulting in a time uncertainty of
$\delta t=0.02m$.  This time uncertainty translates into a phase
uncertainty via 
\begin{equation}
\label{eq:PhaseErrFromTimeError}
\delta\phi=m\omega \times (\delta t/m)
\end{equation}
which yields $ \delta\phi \approx 0.002$, when $m\omega=0.1$ (the
value at the end of the PN comparison) was used.

To estimate the effect of this time uncertainty on the amplitude, we
first note that to lowest order in the post-Newtonian parameter $x$
(defined in Eq.~(\ref{eq:xDefinition})), the wave amplitude of
$\Psi_4$ scales like $x^4$. Also, from
Eq.~(\ref{eq:xdot}), we have $dx/dt=16/(5m)x^5$.  Therefore,
\begin{equation}
\label{eq:AmpErrorFromTimeError}
\frac{\delta A}{A} \sim \frac{d\ln A}{dx} \frac{dx}{dt} \delta t
\sim \frac{64}{5}(m\omega/2)^{8/3}\frac{\delta t}{m},
\end{equation}
where we have used the fact that the gravitational wave frequency $\omega$ is
approximately twice the orbital frequency.
For a time uncertainty $\delta t=0.02m$, Eq.~(\ref{eq:AmpErrorFromTimeError})
gives $\delta A/A \approx 10^{-4}$ for $m\omega=0.1$.

\subsection{Constancy of mass}
\label{sec:Errors-Mass}

Our comparisons with post-Newtonian formulae assume a constant
post-Newtonian mass parameter $m$, which we set equal to the total
irreducible mass of the black holes in the numerical simulation.  If
the total mass of the numerical simulation is not constant, this will
lead to errors in the comparison. For example, changes in $t/m$ caused
by a changing mass will lead to phase differences.
Figure~\ref{fig:AhMass} demonstrates that the irreducible mass is
conserved to a fractional accuracy of about $\delta m/m\approx 5\times
10^{-6}$.

This change in irreducible mass could be caused by numerical errors, or
by a physical increase of the mass of each black hole through tidal
heating.  For our simulations, $m(t)$ {\em
  decreases} during the run (this is not apparent from
Fig.~\ref{fig:AhMass} which plots absolute values), thus contradicting
the second law of black hole thermodynamics.  Moreover, the increase
in $m(t)$ through tidal heating is much  smaller
than the observed variations in $m(t)$ (see, e.g. ~\cite{Poisson2004}).  
Therefore, the variations in
$m(t)$ are numerical errors, and we need to bound the influence of
these errors on the comparison to post-Newtonian expansions.

Over an evolution time of $t/m=4000$, the observed mass uncertainty of
$\delta m/m\approx 5\times 10^{-6}$ results in an 
uncertainty in the overall time interval of $\delta (t/m)=(t/m) \times
(\delta m/m)\approx 0.02$.  This time uncertainty translates into a
phase uncertainty of $\delta\phi \approx 0.002$, using
Eq.~(\ref{eq:PhaseErrFromTimeError}) for $m\omega=0.1$.  Note that the
effect of the black-hole spins on the mass is negligible relative to
the numerical drift of $5\times 10^{-6}$. This is because the spins of
the holes are bounded by $S/\Mirr<2\times 10^{-4}$ and the spin enters
quadratically into the Christodoulou
formula~(\ref{eq:Christoudoulou-mass}).  The error in the
gravitational wave amplitude caused by time uncertainties due to
varying mass is $\delta A/A \approx10^{-4}$ using
Eq.~(\ref{eq:AmpErrorFromTimeError}) for $m\omega=0.1$.  An error in
the mass will affect the amplitude not only via a time offset, but
also because the amplitude is proportional to $(\omega m/2)^{8/3}$ (to
lowest PN order).  However, this additional error is very small,
$\delta A/A\approx (8/3)\delta m/m\approx 10^{-5}$.

\subsection{Time coordinate ambiguity}
\label{sec:Errors-Lapse}

\begin{figure}
\includegraphics[scale=0.49]{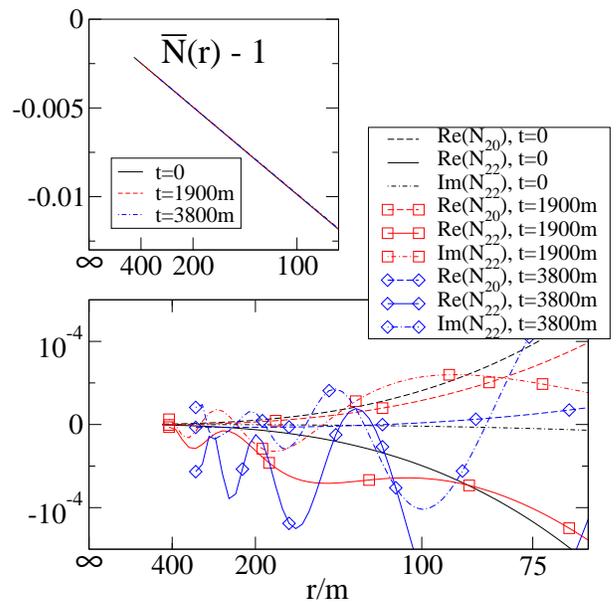}
\caption{\label{fig:LapseConvergence}Asymptotic behavior of the lapse
  at large radii for times $t/m\!=\!0,1900,3800$.  The top figure displays
  the angular average of the lapse as a function of radius at $t=0,
  1900m, 7800m$.  The bottom figure shows the dominant higher
  multipole moments of the lapse.  Both horizontal axes are spaced in
  $1/r$.  }
\end{figure}

We now turn to two possible sources of error that have not yet been
discussed, both of which are related to ambiguity in the time
coordinate.  The basic issue is that the time variable $t$ in
post-Newtonian expansions corresponds to proper time in the
asymptotically flat region, but the time $t$ in numerical simulations
is coordinate time.  These two quantities agree only if the lapse
function $N$ approaches unity at large distances.  To verify this, we
decompose $N$ in spherical harmonics centered on the center of mass of
the system,
\begin{equation}
N(r,\theta,\varphi)
=\sum_{l=0}^{\infty}\sum_{m=-l}^{l}N_{lm}(r)Y_{lm}(\theta,\varphi).
\end{equation}
The angular average of the lapse function, $\bar N(r)\equiv
\sqrt{4\pi}N_{00}(r)$ should then approach unity for $r\to\infty$, and
all other modes $N_{lm}(r)$ should decay to zero.  The top panel of
Fig.~\ref{fig:LapseConvergence} plots $\bar N(r)-1$ vs. $m/r$ for
three different evolution times.  Fitting $\bar N(r)-1$ for $r>100m$
to a polynomial in $m/r$ gives a y-intercept of $<5\times 10^{-6}$ for
all three times, and for polynomial orders of two through five.
Therefore, the coordinate time of the evolution agrees with proper
time at infinity to better than $\delta t/m=t/m \times 5\times
10^{-6}\approx 0.02$, which induces a phase error of at most
$\delta\phi\approx 0.002$ and an amplitude error of $\delta A/A\approx
10^{-4}$ [cf. Eqs.~(\ref{eq:PhaseErrFromTimeError}) 
and~(\ref{eq:AmpErrorFromTimeError})].

The second source of error related to the lapse is shown in
the lower panel of Fig.~\ref{fig:LapseConvergence}, which presents the
three dominant higher order moments $N_{lm}(r)$.  All these modes decay to
zero as $r\to\infty$, except, perhaps, the real part of the $N_{22}$
mode at $t/m=3800$.  This mode seems to approach a value of about
$5\times 10^{-5}$.  At $t=1900m$, this mode still decays nicely to
zero, hence the maximum time uncertainty introduced by this effect at
late times is $\delta t=1900m \times 5\times 10^{-5}\approx 0.1m$,
resulting in a potential phase uncertainty of $\delta \phi\approx0.01$ and a
potential amplitude uncertainty of $\delta A/A\approx 4\times 10^{-4}$.

\subsection{Eccentricity}
\label{sec:Errors-Eccentricity}

We estimated the eccentricity during the numerical simulation with
several of the methods described in
~\cite{Buonanno-Cook-Pretorius:2007,Pfeiffer-Brown-etal:2007,
  Husa-Hannam-etal:2007}, and have found consistently $e \lesssim6\times
10^{-5}$.  This eccentricity can affect our comparison to a
post-Newtonian waveform of a quasi-circular (i.e. zero eccentricity)
inspiral in three ways. 

\subsubsection{Change in rate of inspiral}

The first effect arises because an eccentric binary has a different
inspiral rate than a non-eccentric binary; physically, this can be
understood by noting that the gravitational flux and orbital energy
depend upon the eccentricity, and therefore modify the rate at which
the orbital frequency evolves assuming energy balance.
Reference~\cite{Krolak95} has derived the first-order correction in
the phase of the gravitational wave due to this effect.  Converting
their result to our notation and restricting to the equal mass case
yields
\begin{equation}
\frac{1}{(dx/dt)} = \frac{5 G m}{16 c^3 x^5} \left( 1 
- \frac{157}{24} e_i^2 \left( \frac{x_i}{x} \right)^{19/6} \right),
\end{equation}
where $e_i$ is the initial eccentricity and $x_i$ is the initial value
of the orbital frequency parameter.  Substituting this into
Eq.~(\ref{eq:T2phaseIntegral}) yields
\begin{equation}
\Phi = \Phi_0 - \frac{1}{8} x^{-5/2} + \frac{785}{2176} e_i^2 x_i^{19/6} x^{-17/3}.
\end{equation}
Using $e_i = 6 \times 10^{-5}$ and integrating over the frequency
range from the start of our simulation to the matching frequency of
$m\omega = 0.1$ yields a phase shift of $\sim -2 \times 10^{-6}$,
which is dwarfed by many other error sources, such as the
uncertainty in the numerical mass $m$,
cf. Sec.~\ref{sec:Errors-Mass}.

\subsubsection{Uncertainty in matching time}
\label{sec:EccentricityMatchingTime}

The second way in which eccentricity affects our comparison
is by introducing errors in our procedure for matching the PN and NR
waveforms.  Recall that our matching procedure involves determining
the time $t_m$ at which the gravitational wave frequency
$\omega$ takes a certain value $m\omega_m$; eccentricity
modulates the instantaneous gravitational wave frequency $\omega(t)$ via
\begin{equation}\label{eq:omega-variation}
\omega(t)=\bar\omega(t)\big[1+2e\cos(\Omega_r t)\big],
\end{equation}
where $\bar\omega(t)$ represents the averaged ``non-eccentric''
evolution of the gravitational wave frequency, and $\Omega_r$ is the
frequency of radial oscillations, which is approximately equal to 
the orbital frequency.
We see that $\omega$ can differ from $\bar\omega$ by as much as
$2e\bar\omega\approx 2e\omega$.  This could induce
an error in the determination of $t_m$ by as much as
\begin{equation}\label{eq:deltat-eccentricity}
|\delta t_m|=\frac{|\delta \omega|}{\dot{\omega}}
       \approx \frac{2 e \omega}{\dot{\omega}}
\end{equation}
We can simplify this expression by using Eq.~(\ref{eq:xdot}) to lowest
order, and
by noting that the gravitational wave frequency is approximately twice
the orbital frequency.  We find
\begin{equation}
|\delta t_m|\le e \frac{5m}{12}\left(\frac{m\omega}{2}\right)^{-8/3}.
\end{equation}
This uncertainty is largest at {\em small} frequencies, because the
frequency changes much more slowly.  For $m\omega=0.04$, we find
$|\delta t_m|\lesssim 0.9m$, and for $m\omega=0.1$, we find $|\delta
t_m|\lesssim 0.1m$.  

To determine how uncertainties in $t_m$ translate into phase differences, 
recall
that in the matching procedure described in
Section~\ref{sec:matching-procedure}, $t_m$ enters into the phase of the
shifted PN waveform according to Eq.~(\ref{eq:DefShiftedPNPhase}). Therefore
the phase difference that we compute between the PN and NR waveforms is
\begin{align}
\Delta\phi(t) &=\phi_{\rm PN}(t)-\phi_{\rm NR}(t) \nonumber \\
&=\phi_{\rm PN'}(t\!+\!t_c\!-\!t_m) - \phi_{\rm NR}(t)
\label{eq:DeltaPhiofT}
+ \phi_{\rm NR}(t_m) - \phi_{\rm PN'}(t_c).
\end{align}
Then the error in $\Delta\phi$ is
found by Taylor expanding Eq.~(\ref{eq:DeltaPhiofT}):
\begin{align}
\delta\phi\equiv\delta(\Delta\phi(t))&=\left(\dot\phi_{\rm PN'}(t+t_c-t_m)
-\dot\phi_{\rm NR}(t_m)\right)\delta t_m \nonumber \\
&=\left(\dot\phi_{\rm PN}(t)-\omega_m\right)\delta t_m.
\end{align}

Our simulations (and therefore the comparisons to post-Newtonian
theory) start at $m\omega\approx 0.033$, so that the maximal error
$\delta\phi$ within our comparison at times {\em before}
the matching frequency will be 
\begin{equation}\label{eq:deltaphi-before}
|\delta\phi_{\rm before}|\le |0.033-\omega_m|\;|\delta t_m|
\end{equation}
Combining Eqs.~(\ref{eq:deltaphi-before}) and
(\ref{eq:deltat-eccentricity}), and using $e\approx 6\times 10^{-5}$,
we find that $\delta\phi_{\rm before}<0.01$ radians for all four of
our matching frequencies $m\omega_m=0.04, 0.05, 0.063, 0.1$.
The maximum error $\delta\phi$ within our comparison at times {\em after}
the matching frequency is 
\begin{equation}\label{eq:deltaphi-after}
|\delta\phi_{\rm after}|\le |0.1-\omega_m|\;|\delta t_m|,
\end{equation}
because we end our comparisons to post-Newtonian theory at $m\omega=0.1$.
Eq.~(\ref{eq:deltaphi-after})
evaluates to $0.05$ radians for $m\omega_m=0.04$, and is
less than about $0.02$ radians for the three higher matching
frequencies.

The error in the gravitational wave amplitude caused by an error in
$t_m$ can be estimated by Eq.~(\ref{eq:AmpErrorFromTimeError}).  A
conservative estimate using $\delta t=0.9m$ still gives a small error,
$\delta A/A\approx 0.004$.  

Note that the bounds on $\delta\phi_{\rm before}$ and $\delta\phi_{\rm
  after}$ are proportional to the eccentricity of the numerical
simulation.  Even with eccentricity as low as $6\times 10^{-5}$, this
effect is one of our largest sources or error for the PN-NR
comparison.  (cf. Table~\ref{tab:Errors}).  This is the reason why the
simpler eccentricity removal procedure of Husa {\em et
  al.}~\cite{Husa-Hannam-etal:2007} (resulting in $e=0.0016$) is not
adequate for our purposes.

\subsubsection{Periodic modulation of phase and amplitude}

The third effect of orbital eccentricity is a periodic modulation of
the gravitational wave phase and amplitude.  If we assume that
$\bar\omega(t)$ varies on much longer time scales than 
$1/\Omega_r$ (which is true at large separation) then time integration of
Eq.~(\ref{eq:omega-variation}) yields
\begin{equation}
\label{eq:PhiVariation}
\phi(t) = \bar\phi(t) + 2e \frac{\bar\omega}{\Omega_r} \sin(\Omega_r t).
\end{equation}
Because $\Omega_r\approx \Omega\approx \bar\omega/2$, we therefore find
that the gravitational wave phase consists of the sum of the desired
``circular'' phase, $\bar\phi(t)$, plus an oscillatory component with
amplitude $4e\approx 2\times 10^{-4}$.  This oscillatory component,
however, is much smaller than other uncertainties of the comparison,
for instance the uncertainty in determination of $t_m$.

Residual eccentricity will also cause a modulation of the gravitational
wave amplitude in a manner similar to that of the phase.
This is because eccentricity explicitly
enters the post-Newtonian amplitude formula at 0PN
order~\cite{Wahlquist87}.  This term is proportional to $e$,
and since $e\lesssim 6\times 10^{-5}$ its contribution to the
amplitude error is small compared to the effect due to uncertainty in $t_m$.

While oscillations in phase and amplitude due to eccentricity are tiny
and dwarfed by other uncertainties in the PN-NR comparison, their
characteristic oscillatory behavior makes them nevertheless visible on
some of the graphs we present below, for instance, both panels of 
Fig.~\ref{fig:NR-TaylorT4}.

\subsection{Spin}
\label{sec:Errors-Spin}

We now turn our attention to effects of the small residual spins of
the black holes.  References~
\cite{Faye-Blanchet-Buonanno:2006,Blanchet-Buonanno-Faye:2006} compute
spin-orbit coupling up to 2.5 post-Newtonian order, and find that the
orbital phase, Eq.~(\ref{eq:TaylorPhase}), acquires the following spin
contributions
\begin{align}
\nonumber
\Phi_S(x)=&-\frac{1}{32\nu}\sum_{i=1,2}\chi_i
\bigg\{\left(\frac{565}{24}\frac{m_i^2}{m^2}+\frac{125\nu}{8}\right)x^{-1}\\
\nonumber
&-\Big[\left(\frac{681145}{4032}+\frac{965\nu}{28}\right)\frac{m^2_i}{m^2}\\
&\qquad  +\frac{37265\nu}{448}+\frac{1735\nu^2}{56}\Big]\ln x
\bigg\},
\end{align}
where $\chi_i=\mathbf{S}_i\cdot\hat{\mathbf{L}}/m_i^2$ is the
projection of the dimensionless spin of the $i$-th hole onto the
orbital angular momentum.  For equal-mass binaries with spins
$\chi_1=\chi_2\equiv \chi$, this reduces to
\begin{align}
\label{eq:SpinOrbitPhaseContribution}
\Phi_S(x)=&-\chi
\left(\frac{235}{96}x^{-1}
-\frac{270625}{16128}\ln x\right).
\end{align}
Our comparisons to post-Newtonian theory are performed over the
orbital frequency range of $0.0167\leq m\Omega\leq 0.05$,
corresponding to $0.065\leq x\leq 0.136$.  The gravitational wave phase
is approximately twice the orbital phase, so that the spin-orbit
coupling contributes
\begin{equation}
\delta\phi_S=2\big[\Phi_S(0.065)-\Phi_S(0.136)\Big]\approx - 64\, \chi
\end{equation}
to the gravitational wave phase.  In Sec.~\ref{sec:InspiralEvolution}
we estimated $|\mathbf{S}|/\Mirr^2<5\times 10^{-4}$, where $\Mirr$ is
the irreducible mass of either black hole.  Because $\chi\leq
|\mathbf{S}|/\Mirr^2\approx 5\times 10^{-4}$, the residual black hole
spins contribute less than $0.03$ radians to the overall gravitational
wave phase. 

We now turn to errors in the amplitude comparison caused by residual spin.
From Eq.~(\ref{eq:SpinOrbitPhaseContribution}) we can compute the error
in orbital frequency as
\begin{align}
\delta \Omega = \dot{\Phi}_s &= \nonumber
\chi\frac{\dot{x}}{x}\left(\frac{235}{96}x^{-1}+\frac{270625}{16128}\right)\\
&=
\chi x^4\frac{16}{5m}\left(\frac{235}{96}x^{-1}+\frac{270625}{16128}\right),
\end{align}
where we have used Eq.~(\ref{eq:xdot}).
Because the amplitude of $\Psi_4$ scales like $\Omega^{8/3}$, we arrive at
\begin{equation}\label{eq:SpinDeltaAoverA}
\frac{\delta A}{A} =\frac{8}{3}\frac{\delta\Omega}{\Omega}
=\chi x^{5/2}\frac{128}{15}\left(\frac{235}{96}x^{-1}
+\frac{270625}{16128}\right),
\end{equation}
which for $m\omega_m = 0.1$ (i.e. $x=0.136$) gives $\delta A/A = 2.0
\chi\ \sim 1.0 \times 10^{-3}$.

Spin-orbit terms also contribute directly to the
amplitude~\cite{kidder95,Will96}.  The leading order contribution (for
an equal-mass binary with equal spins) contributes a term $\delta A/A
\sim (4/3) \chi x^{3/2}$, which is the same order of magnitude as the
previous error, $10^{-3}$.

\section{Results}
\label{sec:Results}

\subsection{Comparison with individual post-Newtonian approximants}
\label{sec:ComparisonWithPN}

\begin{figure*}
\includegraphics[scale=0.47]{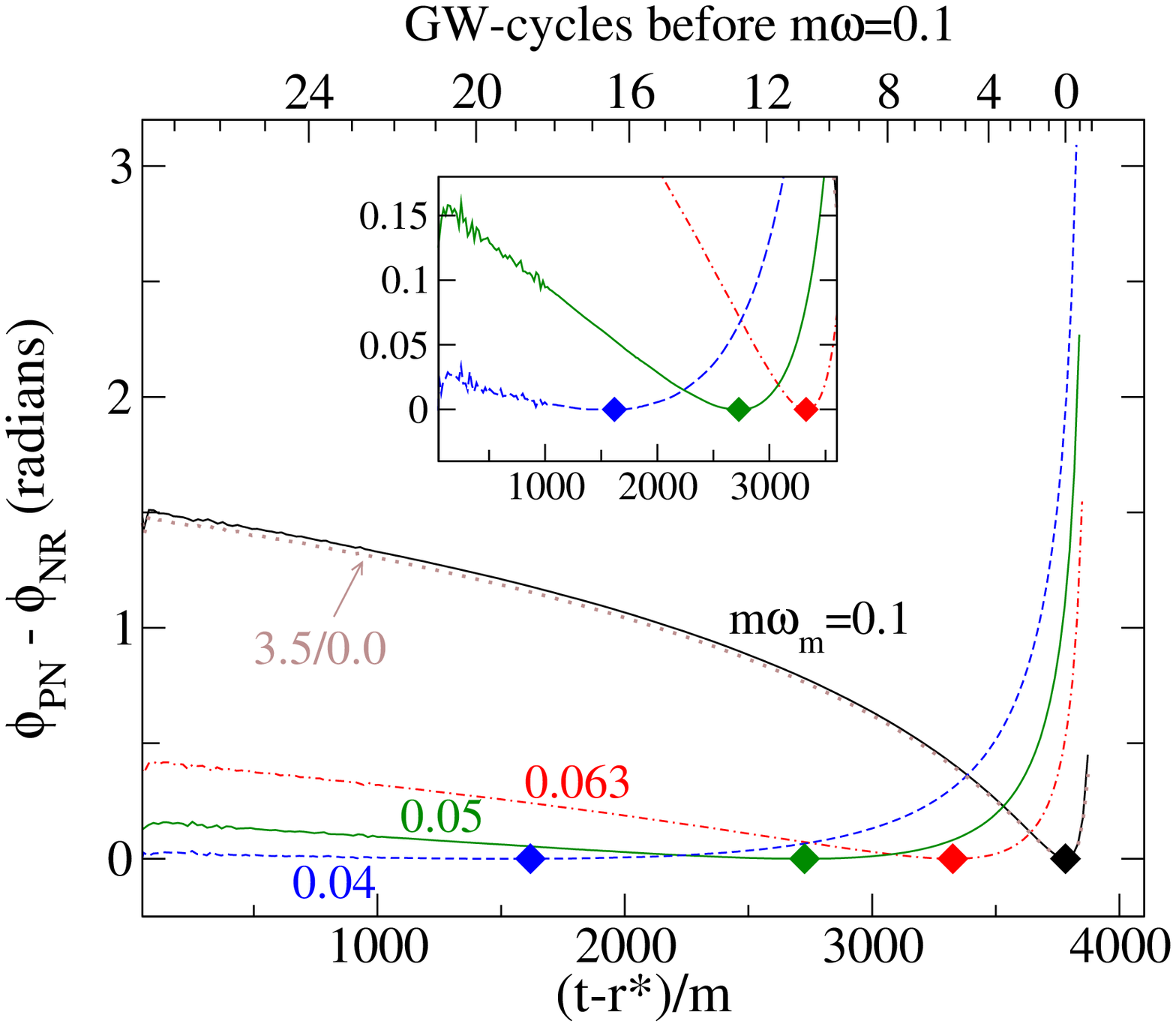}
\qquad\includegraphics[scale=0.47]{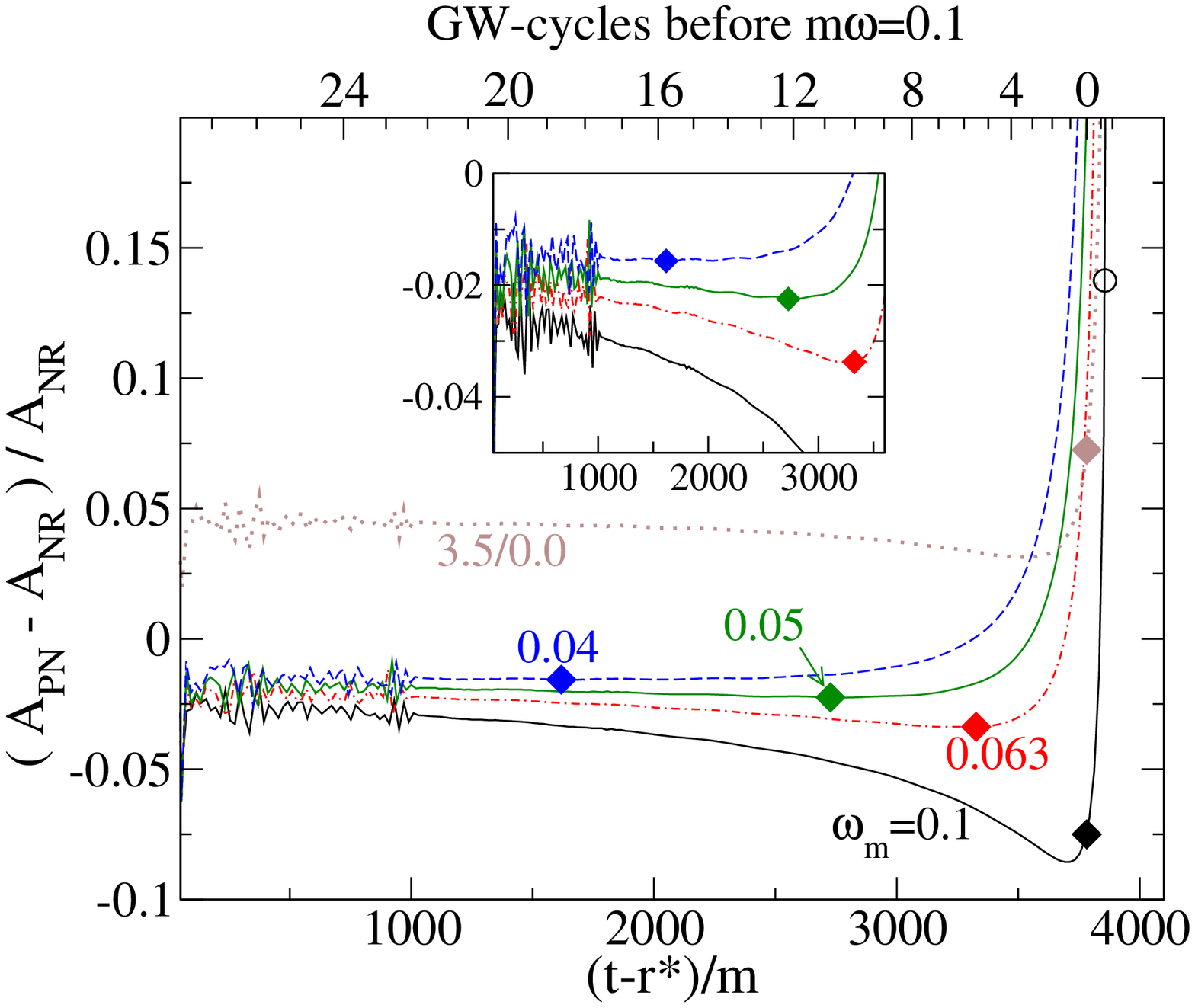}
\caption{\label{fig:NR-TaylorT1} Comparison of numerical simulation
  with {\bf TaylorT1 3.5/2.5} waveforms.  Left:
  Difference in gravitational wave phase.  Right: Relative amplitude
  difference.  Plotted are comparisons for four values of $\omega_m$. 
  The filled diamond on each curve shows the point at
  which $\dot\phi=\omega_m$.  The insets show enlargements for small
  differences and early times.  Also shown is the difference between the
  numerical and restricted (i.e. 3.5PN phase, 0PN amplitude) Taylor
  T1 for $m\omega_m=0.1$.  }
\end{figure*}

We compare our simulations with four different post-Newtonian
approximants: the TaylorT1, TaylorT2, TaylorT3, and TaylorT4 waveforms.  These
four waveforms agree with each other up to their respective
post-Newtonian expansion orders, but they differ in the way that the
uncontrolled higher order terms enter.  We start with the comparison
to TaylorT1.

\subsubsection{TaylorT1 (3.5PN phase, 2.5PN amplitude)}
\label{sec:ComparisonWithPN-TaylorT1}

Figure~\ref{fig:NR-TaylorT1} compares the numerical simulation to
TaylorT1 3.5/2.5 waveforms (i.e. expansion order 3.5PN in phase and
2.5PN in amplitude, the highest expansion orders currently available
for generic direction, cf.~\ref{sec:PN-PolarizationWaveforms}).  The left panel
shows the phase difference, where we find differences of more than a
radian for all four matching frequencies we consider: $\omega_m=0.04$,
$0.05$, $0.063$, and $0.01$.

For our largest matching frequency, $m\omega_m=0.1$, the
phase differences are small toward the end of the run by construction.
Nevertheless, a phase difference of more than 0.5 radians builds up in
the $\sim 1.5$ cycles after the matching point before the TaylorT1
template generation fails.  Recall that $m\omega_m=0.1$ occurs about
2.2 gravitational wave cycles before our simulations fail, which is
still about 1.5 cycles before merger.
However, the largest phase disagreement for $m\omega_m=0.1$ builds up
at early times, reaching 1.5 radians at the beginning of our simulation,
about $28$ cycles before the matching ($\sim 30$ cycles before the end
of the simulation), 
and still showing no sign of flattening even at
the start of our simulation.

\begin{figure}
\includegraphics[scale=0.49]{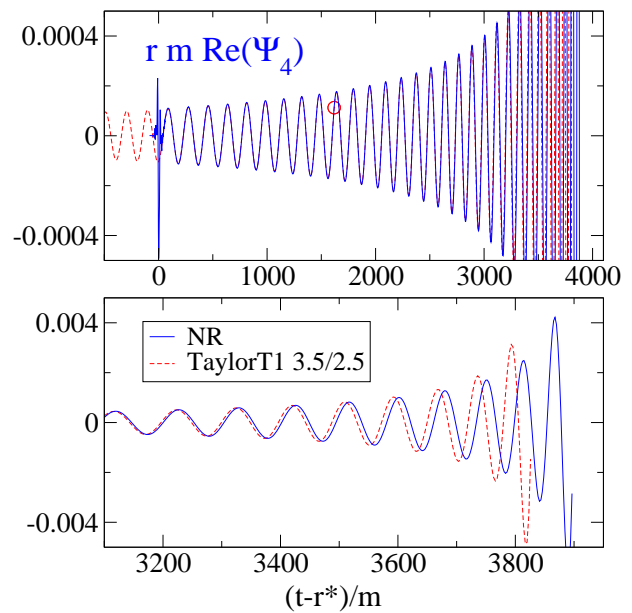}
\caption{\label{fig:NR-TaylorT1-Waveform} Numerical and {\bf TaylorT1 3.5/2.5}
 waveforms.  The PN waveform is matched
  to the numerical one at $m\omega_m=0.04$, indicated by the
  small circle.  The lower panel shows a detailed view of the last
  10 gravitational wave
  cycles.
}
\end{figure}
\begin{figure*}
\includegraphics[scale=0.47]{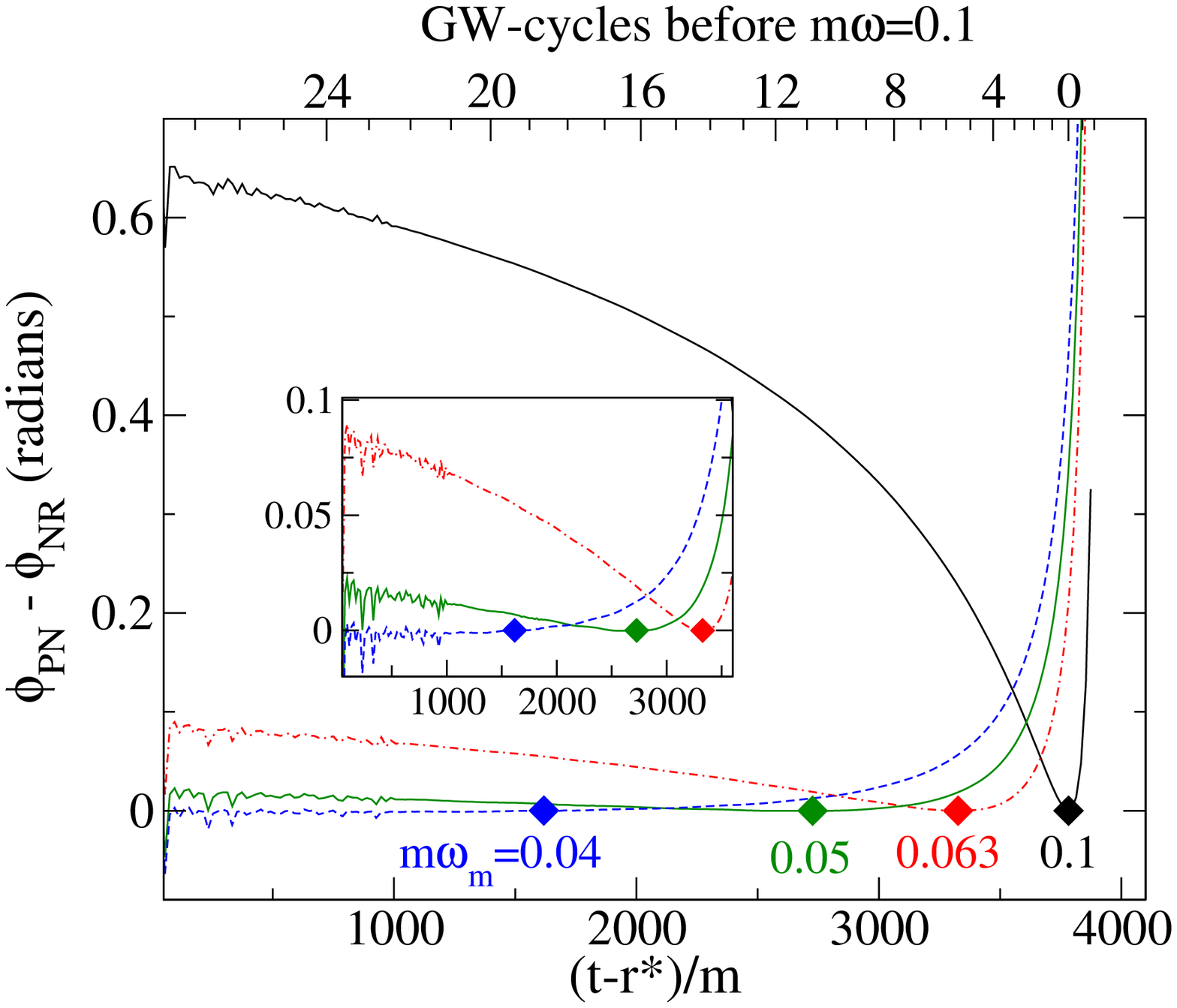}
\qquad\includegraphics[scale=0.47]{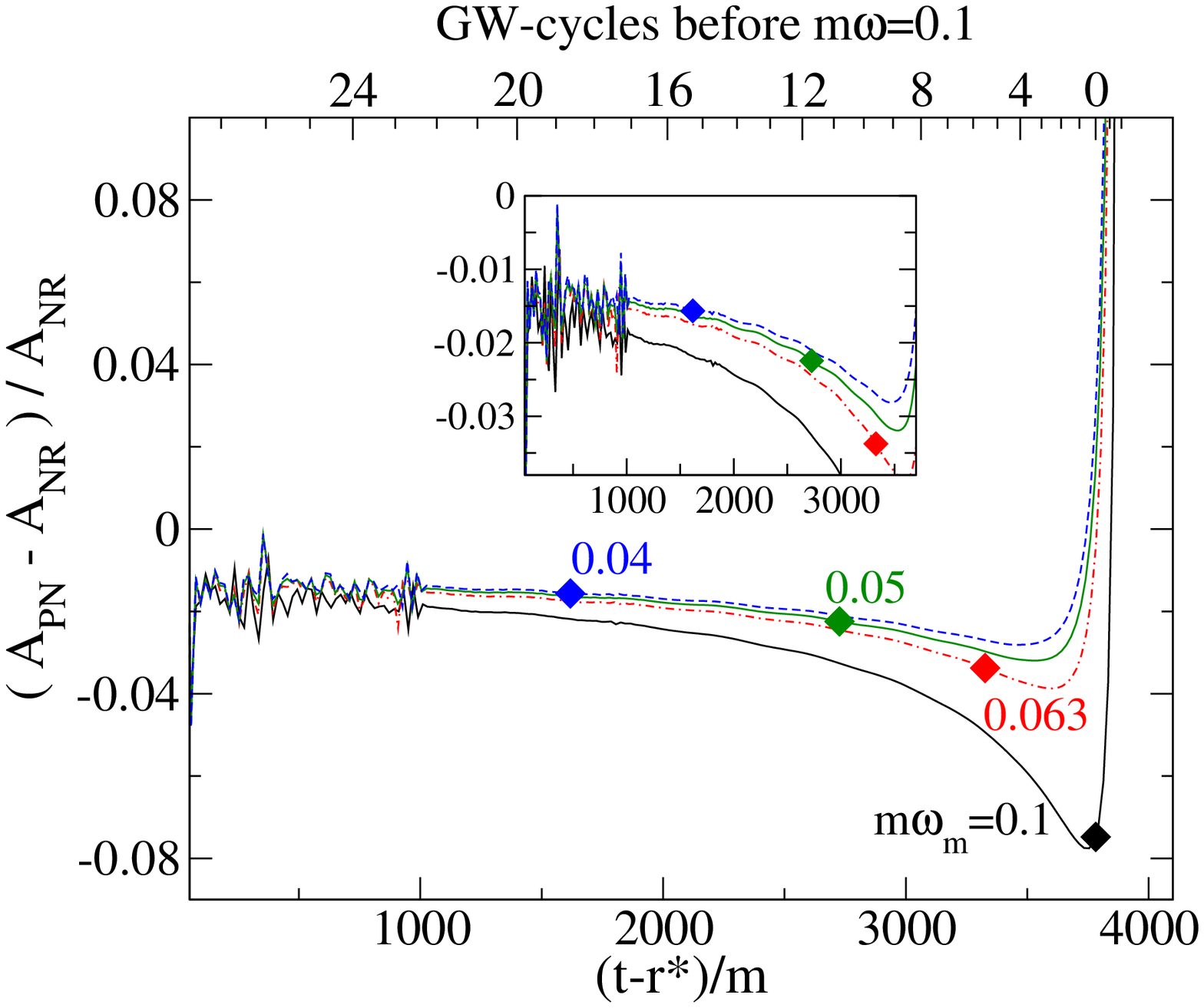}
\caption{\label{fig:NR-TaylorT2} Comparison of numerical simulation
  with {\bf TaylorT2 3.5/2.5} waveforms.  Left: Difference in
  gravitational wave phase.  Right: Relative amplitude difference.
  Plotted are comparisons for four values of $\omega_m$.  The filled
  diamond on each curve shows the point at which $\dot\phi=\omega_m$.
  The insets show enlargements for small differences and early times.
}
\end{figure*}

To achieve phase coherence with the early inspiral waveform, it is
therefore necessary to match earlier than $m\omega_m=0.1$.  The left
panel of Fig.~\ref{fig:NR-TaylorT1} clearly shows that phase
differences at earlier times become smaller when the matching point
itself is moved to earlier time.  For instance, $m\omega_m=0.063$
(about eight gravitational wave cycles before the end of our
simulation), results in phase differences less than 0.5 radians during
the 22 earlier cycles of our evolution.  However, the phase difference
$\phi_{\rm PN}-\phi_{\rm NR}$ does not level off at early times within
the length of our simulation, so it seems quite possible that the
phase difference may grow to a full radian or more if the numerical
simulations could cover many more cycles.  We thus estimate that for
TaylorT1, to achieve 1-radian phase coherence with the early inspiral
may require matching more than 10 cycles before merger.  To achieve
more stringent error bounds in phase coherence will require matching even
earlier: for instance it appears one needs to use $m\omega_m=0.04$
(about 20 cycles before the end of our simulation) for a phase error
of less than $\lesssim 0.1$ radians.

While matching at small $\omega_m$ yields good phase coherence early
in the run, it produces much larger phase differences late in the run.  For
example, matching at $m\omega_m=0.04$ results in a phase difference
of almost 2 radians at frequency $m\omega=0.1$.  This rather dramatic
disagreement is illustrated in Fig.~\ref{fig:NR-TaylorT1-Waveform},
which plots both the numerical and the TaylorT1 waveform, matched at
$m\omega_m=0.04$.

The left panel of Fig.~\ref{fig:NR-TaylorT1} also includes a
comparison to the so-called restricted TaylorT1 template, where only
the leading order amplitude terms are used ({\em i.e.} 0PN in
amplitude).  The reason that higher-order amplitude terms affect the
phase differences at all is because we are plotting gravitational-wave
phase, not orbital phase. However, we see that the effect of these
higher-order amplitude terms on the phase difference is small.

We now turn our attention to comparing the amplitudes of the
post-Newtonian and numerical waveforms.  The right panel of
Fig.~\ref{fig:NR-TaylorT1} shows relative amplitude differences
between TaylorT1 3.5/2.5 and the numerical
waveforms.  At early times, the amplitudes agree to within 2 or 3 per
cent, the agreement being somewhat better when the matching is
performed at early times.  At late times, the amplitudes disagree
dramatically; a large fraction of this disagreement lies probably in
the fact the post-Newtonian point of merger (i.e. the point at which the
amplitude diverges) occurs at a different time than the numerical point of
merger.  We also plot the amplitude of the restricted TaylorT1
template.  The disagreement between restricted TaylorT1 and the numerical
result is much larger, about 5 per cent.

Hannam et al.~\cite{Hannam2007} performed a similar comparison,
matching their waveforms with a restricted TaylorT1 waveform
(i.e. 3.5/0.0) generated using the LIGO Algorithm Library
(LAL)~\cite{LAL}. The phase difference they observe for waveforms
matched at $m\omega=0.1$ is consistent with our results within
numerical errors.  When matching TaylorT1 3.5/0.0 early in their
simulation (at $m\omega=0.0455$), however, Hannam et al. find a
cumulative phase difference of 0.6 radians at $m\omega=0.1$.  From
Fig.~\ref{fig:NR-TaylorT1} we find a quite different value of 1.5
radians for our simulation.  This disagreement might be caused by the
use of the finite extraction radius $R=90m$ for the gravitational wave
phase in Hannam et. al.:
Figure~\ref{fig:ExtrapolationConvergence_Phi} shows that extracting
at a finite radius leads to
a systematic phase error, which will induce a systematic error in
determination of the matching time of Hannam et al.  This error is
amplified by the increasing gravitational wave frequency toward
merger.

\begin{figure*}
\includegraphics[scale=0.47]{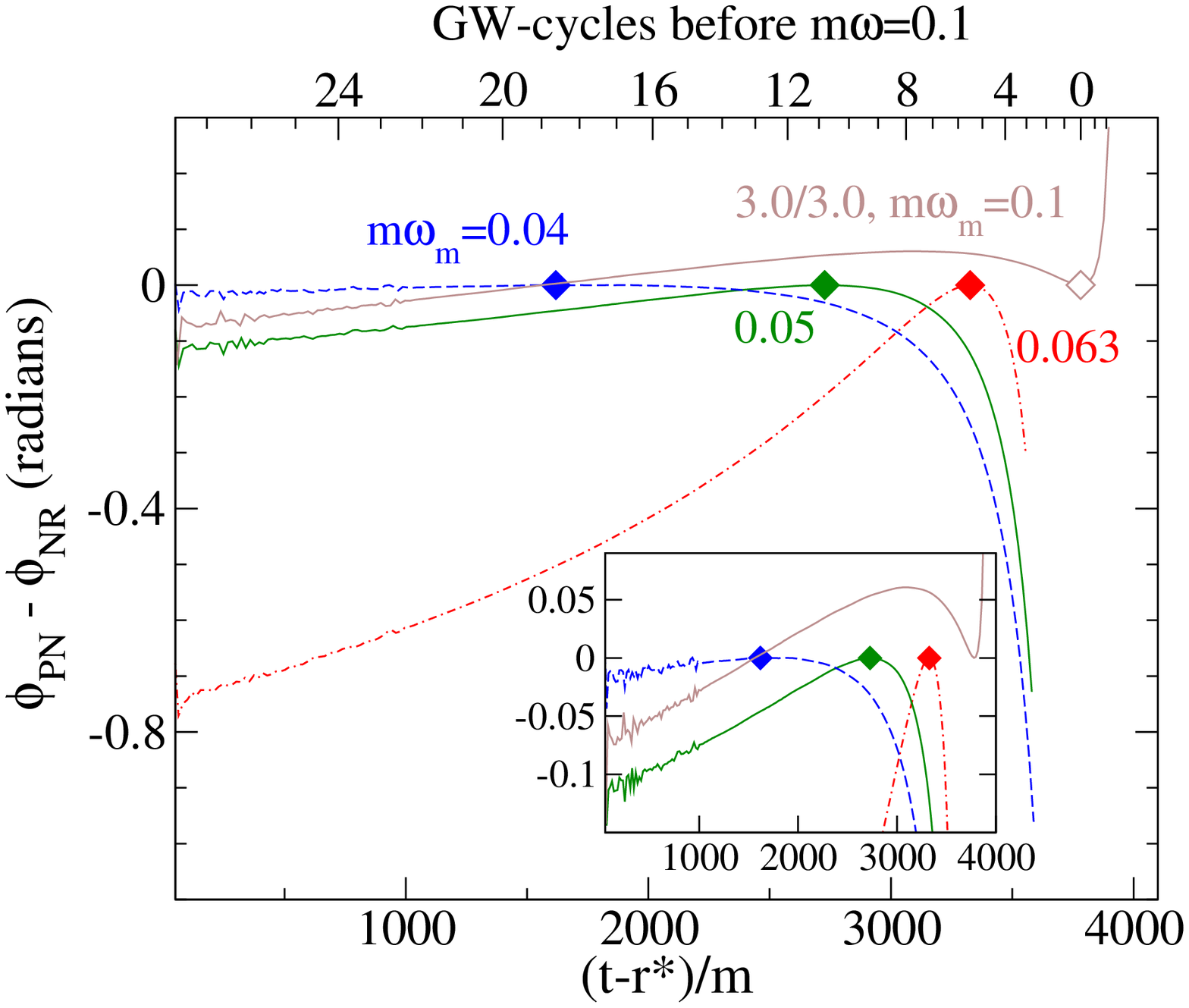}
\qquad
\includegraphics[scale=0.47]{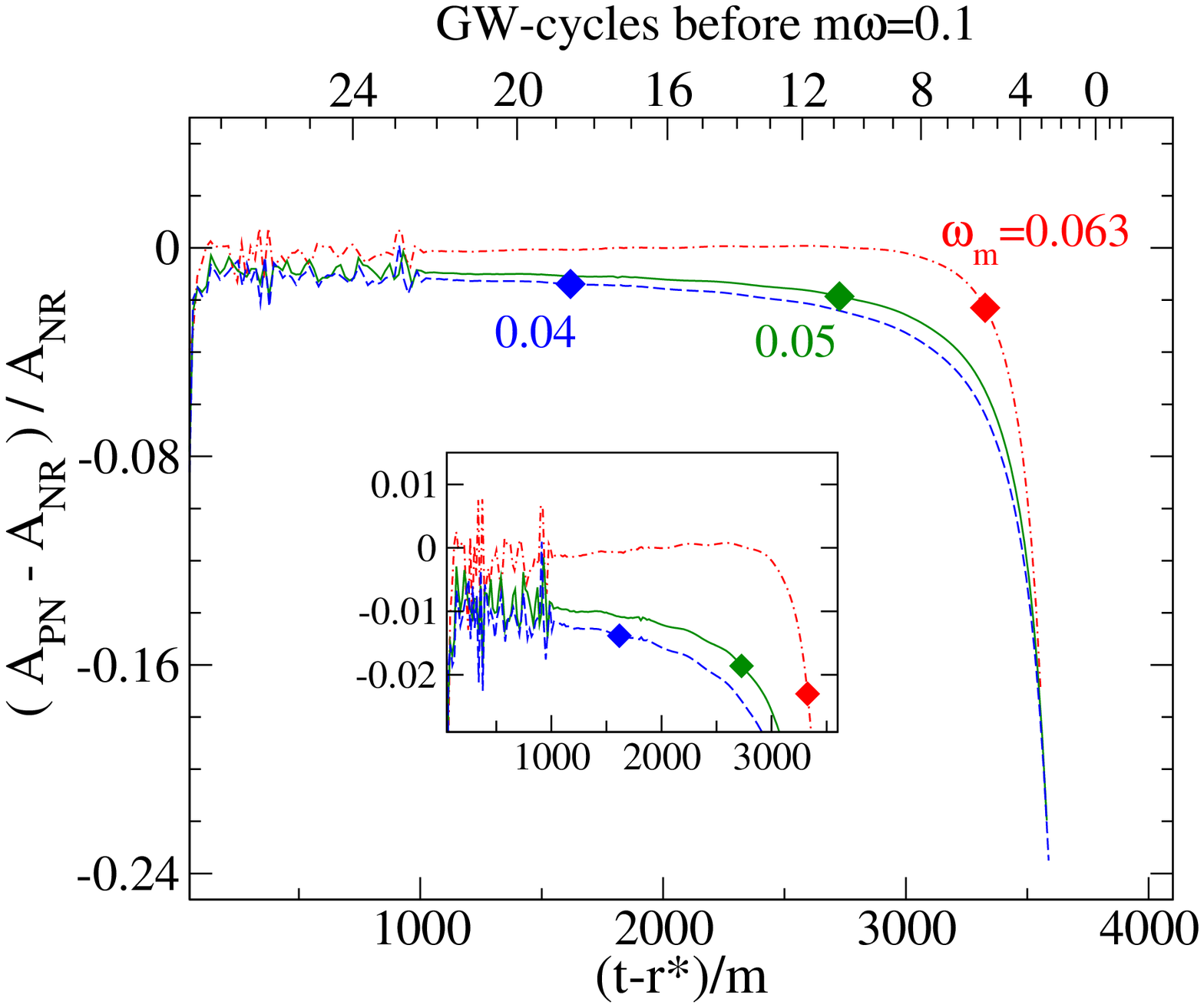}
\caption{\label{fig:NR-TaylorT3} Comparison of numerical simulation
  with {\bf TaylorT3 3.5/2.5} waveforms.  Left: Difference in
  gravitational wave phase.  Right: Relative amplitude difference.
  Plotted are comparisons for three values $\omega_m$.  The filled
  diamond on each curve shows the point at which $\dot\phi=\omega_m$.
  The lines end when the frequency of the TaylorT3 waveform reaches
  its maximum, which happens before $m\omega=0.1$, so that the
  matching frequency $m\omega_m=0.1$ is absent.  The left plot also
  contains TaylorT3 3.0/3.0, matched at $m\omega_m=0.1$.  The insets
  show enlargements for small differences.}
\end{figure*}


\subsubsection{TaylorT2 (3.5PN phase, 2.5PN amplitude)}
\label{sec:ComparisonWithPN-TaylorT2}

Figure~\ref{fig:NR-TaylorT2} presents the comparison between the
numerical waveform and the TaylorT2 approximant.  The overall trends
are very similar to the TaylorT1 comparison of
Fig.~\ref{fig:NR-TaylorT1}, however, the phase differences are smaller
by about a factor of 2 when matching at $m\omega_m=0.1$, and smaller
by a factor of 3 to 4 when matching earlier.  To our knowledge TaylorT2
has never been compared to a numerical simulation; we include it here
mainly for completeness.


\subsubsection{TaylorT3 (3.5PN and 3.0PN phase, 2.5PN amplitude)}

Figure~\ref{fig:NR-TaylorT3} is the same as Fig.\ref{fig:NR-TaylorT1}
except it compares numerical simulations to the TaylorT3 family of
waveforms.  Two differences between TaylorT1 and TaylorT3 are readily
apparent from comparing these two figures. The first is that we do not
match TaylorT3 3.5/2.5 waveforms at $m\omega_m=0.1$. This is because
the frequency of TaylorT3 3.5/2.5 waveforms reaches a maximum shortly
before the formal coalescence time of the post-Newtonian template, and
then {\em decreases}.  The maximal frequency is less than $0.1$, so
that matching at $m\omega_m=0.1$ is not possible.  For this reason, we
have also shown in Fig.~\ref{fig:NR-TaylorT3} a comparison with a
TaylorT3 3.0/3.0 waveform matched at $m\omega_m=0.1$.  The other major
difference between the TaylorT3 3.5/2.5 and TaylorT1 3.5/2.5
comparison is that the phase difference, $\phi_{\rm PN}-\phi_{\rm
NR}$, has a different sign.  While TaylorT1 3.5/2.5 spirals in {\em
more rapidly} than the numerical simulation, TaylorT3 3.5/2.5 {\em
lags behind}.  Interestingly, the phase differences from the numerical
simulation for both TaylorT1 3.5/2.5 and TaylorT3 3.5/2.5 are of about
equal magnitude (but opposite sign).  The TaylorT3 3.0/3.0 comparison
matched at $m\omega_m=0.1$ has smaller phase differences than does the
TaylorT3 3.5/2.5 comparison, but the slope of the 3.0/3.0 curve in
Fig.~\ref{fig:NR-TaylorT3} is nonzero at early times, so it appears
that Taylor T3 3.0/3.0 will accumulate significant phase differences
at even earlier times, prior to the start of our simulation.  In
Fig.~\ref{fig:PNvsNR-2.0-2.5-3.0-3.5} it can be seen that matching
TaylorT3 3.0/3.0 at $m\omega_m=0.04$ leads to a good match early, but
leads to a phase difference of $0.6$ radians by $m\omega=0.1$.

Hannam et al.~\cite{Hannam2007} match a TaylorT3 3.0/0.0 waveform at
$m\omega_m=0.1$ and $m\omega_m=0.0455$.  Matching at $m\omega_m=0.1$ again
gives phase differences consistent with our results within numerical errors.
Matching at  $m\omega_m=0.0455$, Hannam et al. find a phase difference of
0.9 radians, while we find a smaller value of 0.5 radians.  Again, this
difference could be due to the finite extraction radius used by Hannam et al.

\subsubsection{TaylorT4 (3.5PN phase, 2.5PN amplitude)}
\begin{figure*}
\includegraphics[scale=0.47]{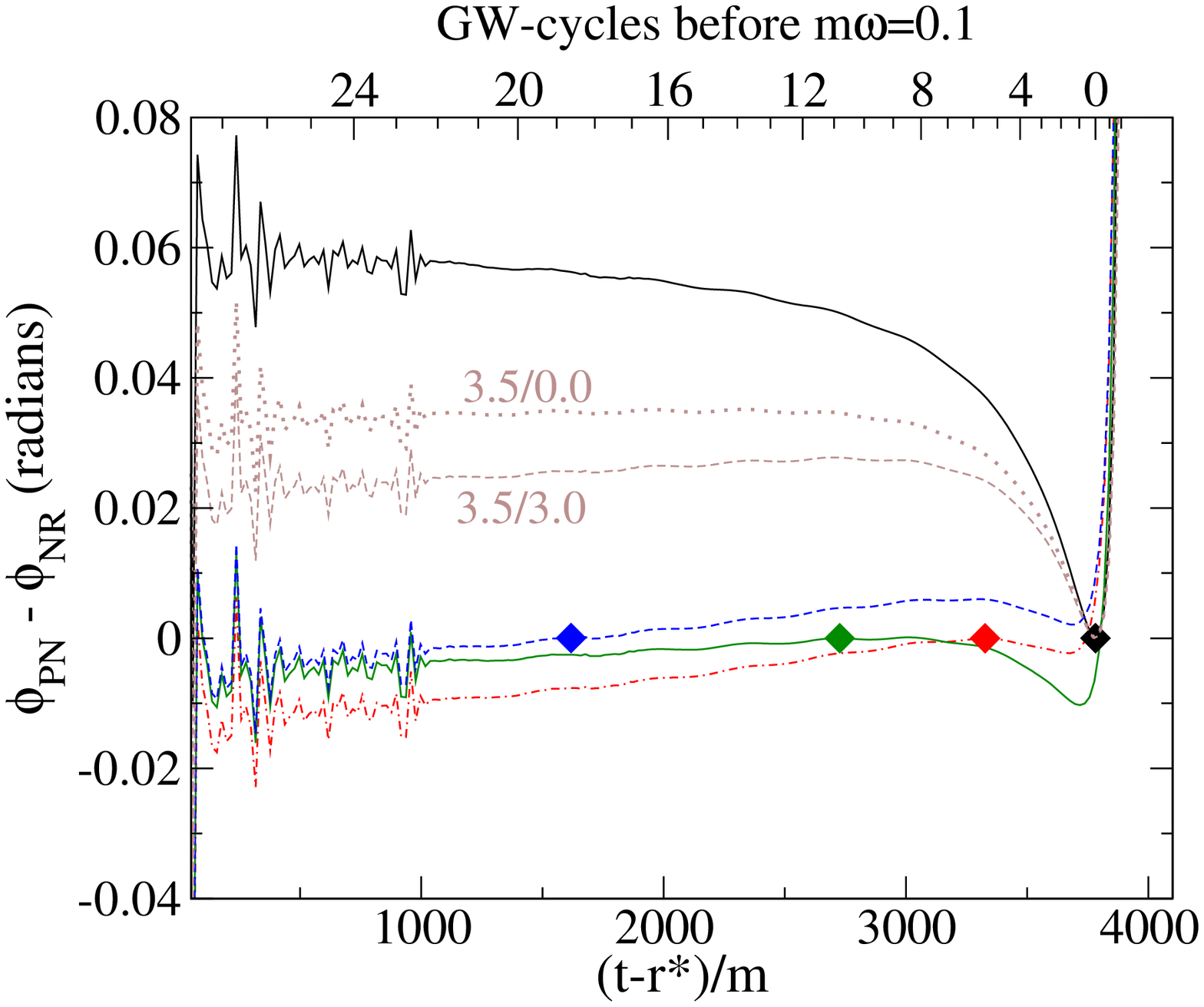}
\qquad
\includegraphics[scale=0.47]{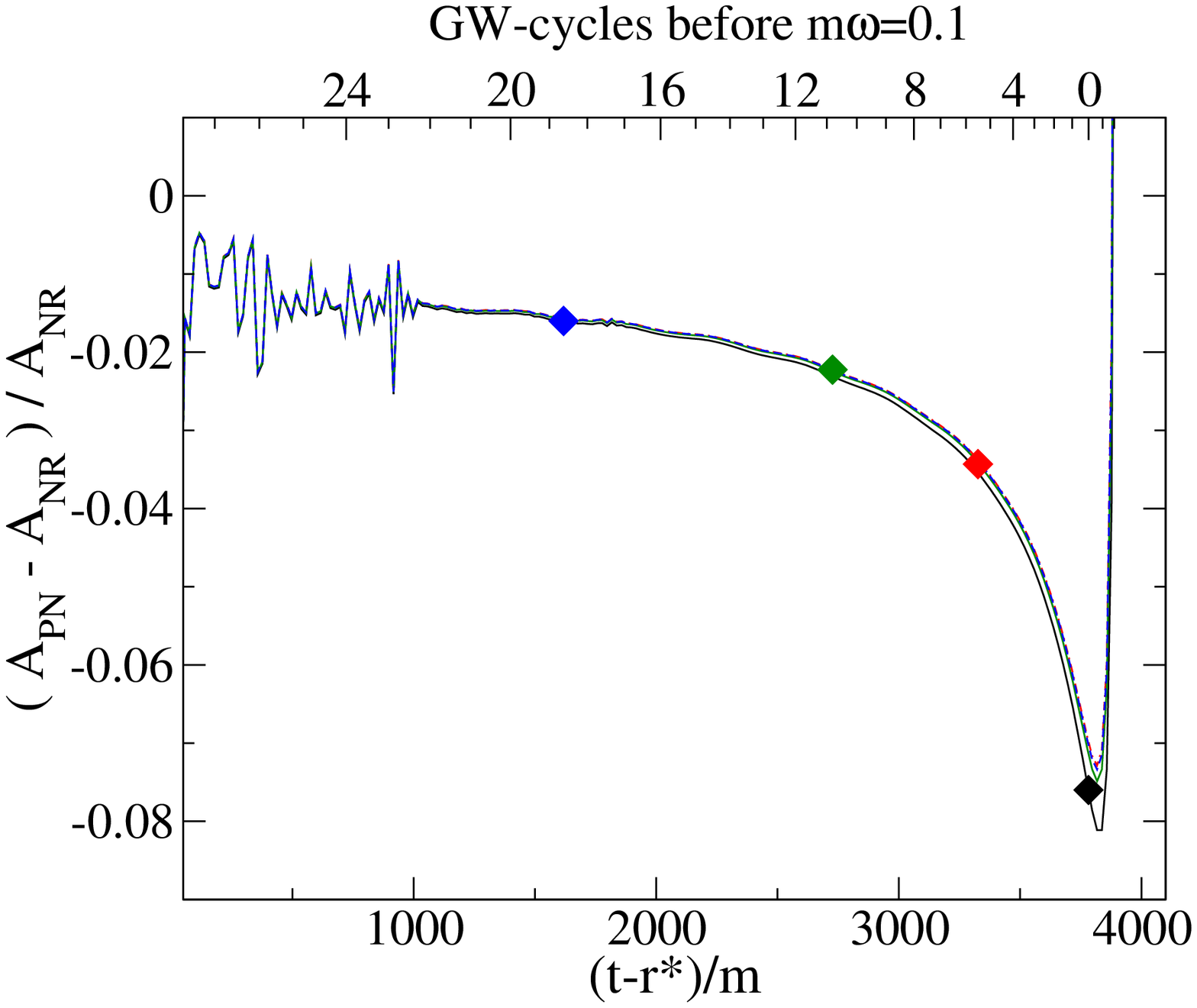}
\caption{\label{fig:NR-TaylorT4} Comparison of numerical simulation
  with {\bf TaylorT4 3.5/2.5} waveforms.  Left:
  Difference in gravitational wave phase.  Right: Relative amplitude
  difference.  Plotted are comparisons for four values of $\omega_m$.
  The filled diamond on each curve shows the point at which
  $\dot\phi=\omega_m$.  The left plot also includes two phase
  comparisons with expansions of different PN order in amplitude, as
  labeled, for $m\omega_m=0.1$.  }
\end{figure*}

\begin{figure}
\includegraphics[scale=0.49]{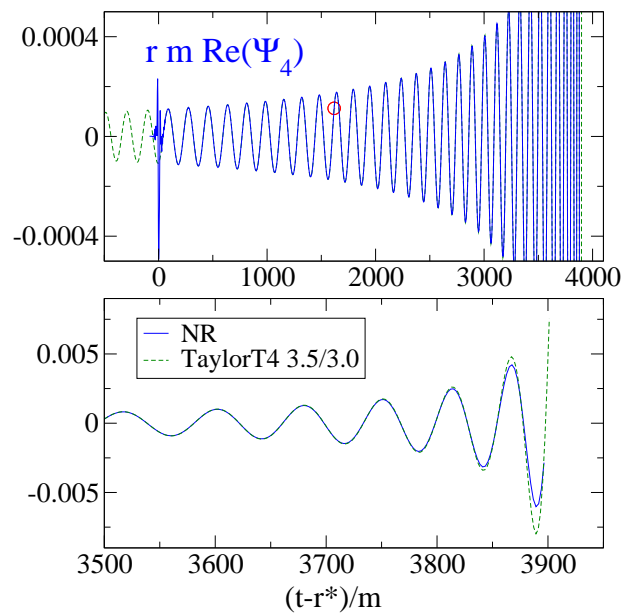}
\caption{\label{fig:NR-TaylorT4-Waveform} Numerical and {\bf TaylorT4
  3.5/3.0} waveforms.  The PN waveform is matched to the numerical one
  at $m\omega_m=0.04$, indicated by the small circle.  The lower panel
  shows a detailed view of the end of the waveform.  }
\end{figure}

Figure~\ref{fig:NR-TaylorT4} is the same as
Figs.~\ref{fig:NR-TaylorT1} and~\ref{fig:NR-TaylorT3} except it
compares numerical simulations to the TaylorT4 PN waveforms.  The
agreement between TaylorT4 waveforms and the numerical results is
astonishingly good, far better than the agreement between NR and
either TaylorT1 or TaylorT3. The gravitational wave phase difference
lies within our error bounds for the entire comparison region $m\omega\le 0.1$,
agreeing to $0.05$ radians or better over 29 of 30 gravitational wave cycles.  
Ref.~\cite{Baker2006d} found
agreement between TaylorT4 and their numerical simulation to the level
of their numerical accuracy ($\sim 2$ radians), agreeing to roughly
$0.5$ radians in the gravitational frequency range of $0.054 \leq
m\omega \leq 0.1$.  Ref.~\cite{Pan2007} found that NR agrees better
with TaylorT4 than with TaylorT1, but the larger systematic and
numerical errors of the numerical waveforms used in these studies 
did not allow them to see the surprising degree to
 which NR and TaylorT4 agree.  The gravitational wave amplitude of
 TaylorT4 agrees with the NR waveform to about 1--2 percent at early
 times, and 8 percent at late times. In
 Fig.~\ref{fig:NR-TaylorT4-Waveform} we plot the NR and TaylorT4
 waveforms; the two waveforms are visually indistinguishable on the
 plot, except for small amplitude differences in the final cycles.

On the left panel of Fig.~\ref{fig:NR-TaylorT4} we also show phase
comparisons using PN waveforms computed to 3.5PN order in phase but to
0PN and 3.0PN orders in amplitude, for the case $m\omega_m=0.1$.  The
PN order of the amplitude expansion affects the phase comparison
because we are plotting differences in gravitational-wave phase and
not orbital phase.  The differences between using 0PN, 2.5PN, and
3.0PN amplitude expansions are evident on the scale of the graph, but
because these differences are smaller than our estimated uncertainties
(see Table~\ref{tab:Errors}), we cannot reliably conclude which of these
most closely agrees with the true waveform.

\begin{figure}
\includegraphics[scale=0.49]{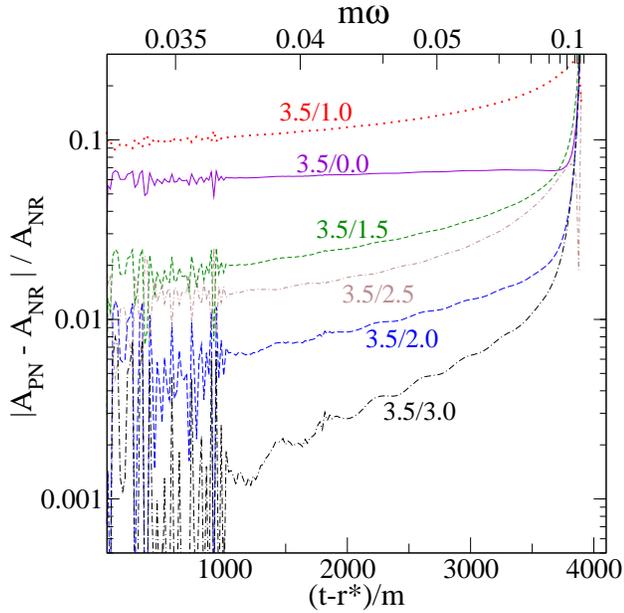}
\caption{\label{fig:NR-TaylorT4-AmpVsPN} {\bf TaylorT4} amplitude
  comparison for different PN orders. Shown is the relative difference
  in gravitational wave amplitude between TaylorT4 and numerical
  $Y_{22}$ waveforms as a function of time.  Matching is performed at
  $m\omega_m=0.04$.  All curves
  use 3.5PN order in phase but different PN orders (as labeled) in the
  amplitude expansion.  }
\end{figure}

Figure~\ref{fig:NR-TaylorT4-AmpVsPN} presents amplitude differences
between NR and TaylorT4 as the post-Newtonian order of the amplitude
expansion is varied, but the phase expansion remains at 3.5PN.  The
2.5PN amplitude curve was already included in the right panel of
Fig.~\ref{fig:NR-TaylorT4}.  We see clearly that higher order
amplitude corrections generally result in smaller differences.  The 3PN
amplitude correction to the $(2,2)$ mode recently derived by
Kidder~\cite{Kidder07a} improves agreement dramatically over the
widely known 2.5PN amplitude formulae.  Unfortunately, the 3PN
amplitude correction to the entire waveform, including all $Y_{lm}$
modes, is not known.\footnote{To get the complete waveform to 3PN
  order, only the $(2,2)$ mode must be known to 3PN order; 
  other modes must be known
  to smaller PN orders.  For an equal mass, non-spinning binary, all
  modes except the $(3,2)$ mode
  are currently known to sufficient order to get a complete
  3PN waveform~\cite{Kidder07a}. 
  }


\subsection{Comparing different post-Newtonian approximants}
\label{sec:ComparisonDifferentPN}

The previous section presented detailed comparisons of our numerical
waveforms with four different post-Newtonian approximants.  We now turn our
attention to some comparisons between these approximants.  In this
section we also explore further how the post-Newtonian order influences
agreement between numerical and post-Newtonian waveforms.

\begin{figure}
\includegraphics[scale=0.49]{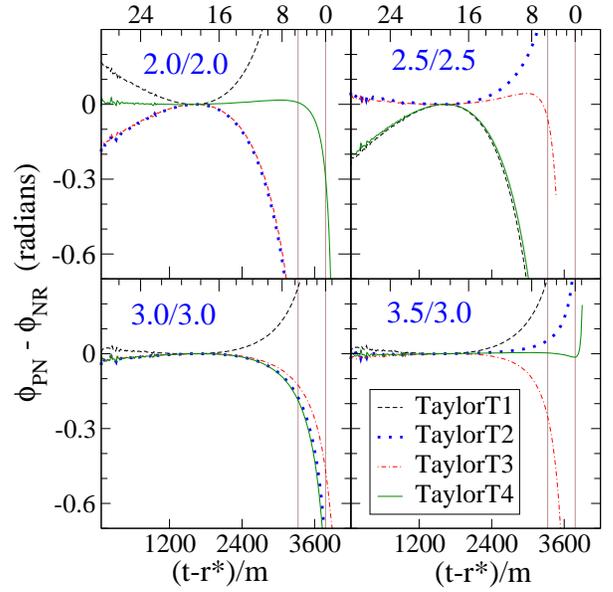}
\caption{\label{fig:PNvsNR-2.0-2.5-3.0-3.5} Phase comparison for
  different PN approximants at different PN orders, matched at
  $m\omega_m=0.04$.  Shown is the difference in gravitational wave
  phase between each post-Newtonian approximant and the numerical
  $Y_{22}$ waveforms as a function of time.  
The two vertical brown
  lines indicate when the numerical simulation reaches $m\omega=0.063$
  and $0.1$, respectively; the labels along the top horizontal axes
  give the number of gravitational-wave cycles before $m\omega=0.1$.  }
\end{figure}

\begin{figure}
\includegraphics[scale=0.49]{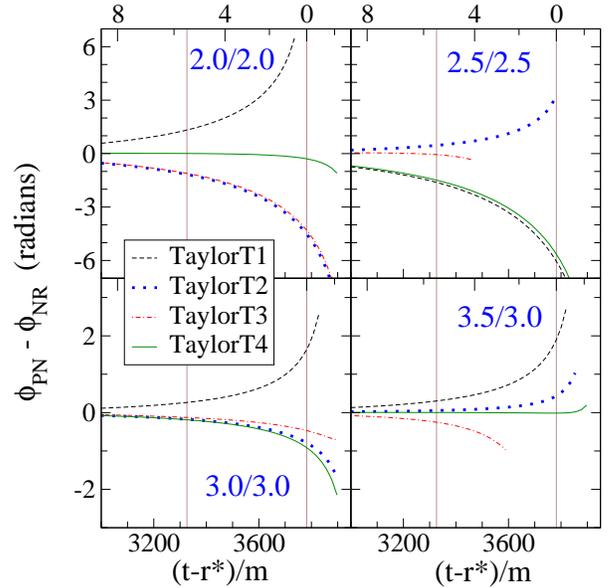}
\caption{\label{fig:PNvsNR-2.0-2.5-3.0-3.5-LateTime} Same as
  Fig.~\ref{fig:PNvsNR-2.0-2.5-3.0-3.5}, but showing only the last
  stage of the inspiral.  The horizontal axis ends at the estimated
  time of merger, $(t-r^\ast)_{\rm CAH}=3950m$, cf. Sec~\ref{sec:EstimatedTimeOfMerger}.   The top and bottom panels
use different vertical scales.}
\end{figure}

Figure~\ref{fig:PNvsNR-2.0-2.5-3.0-3.5} presents phase differences as
a function of time for all four PN approximants we consider here
and for different PN orders.  
The post-Newtonian and numerical waveforms are matched at
$m\omega_m=0.04$, about 9 cycles after the beginning of the numerical
waveform, and about 21 cycles before its end.  We find that some PN
approximants at some particular orders agree exceedingly well with the
numerical results.  The best match is easily TaylorT4 at 3.5PN order,
and the next best match is TaylorT4 at 2.0PN order.  Some approximants
behave significantly worse, such as the TaylorT1 and TaylorT4
waveforms at 2.5PN order.  The 2.5PN and 3PN TaylorT3 waveforms agree
very well with the numerical waveform at early times, but at late
times they accumulate a large phase difference; the 2.5PN TaylorT3
waveform ends even before the
numerical waveform reaches $m\omega=0.1$ (the rightmost vertical brown
line in Fig.~\ref{fig:PNvsNR-2.0-2.5-3.0-3.5}).

We also find that all four PN approximants, when computed to 3PN
order or higher, match the numerical waveform (and each other) quite
closely at early times, when all PN approximants are expected to be
accurate.  However, at late times, $t-r^\ast>2500m$, the four PN
approximants begin to diverge, indicating that PN is beginning to
break down.

Figure~\ref{fig:PNvsNR-2.0-2.5-3.0-3.5-LateTime} is an
enlargement of Fig.~\ref{fig:PNvsNR-2.0-2.5-3.0-3.5} for the last 10
gravitational wave cycles before merger.  This figure shows in more
detail how the different PN approximants behave near merger.

\begin{figure}
\includegraphics[scale=0.47]{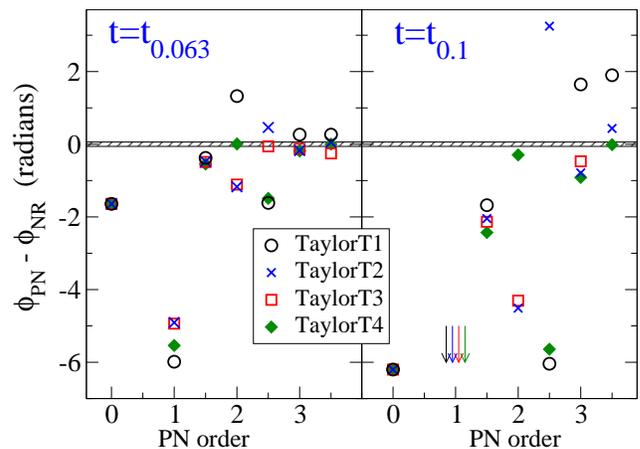}
\caption{\label{fig:PNvsNR_byOrder}Phase differences between numerical
  and post-Newtonian waveforms at two selected times close to merger.
  Waveforms are matched at $m\omega_m=0.04$, and phase differences are
  computed at the time when the numerical simulation reaches
  $m\omega=0.063$ (left panel) and $m\omega=0.1$ (right panel).
  Differences are plotted versus PN order (equal order in phase and
  amplitude, except the '3.5 PN' points are 3.5/3.0).  On the right
  plot, the 1PN data points are off scale, clustering at $-15$
  radians.  The thin black bands indicate upper bounds on the
  uncertainty of the comparison as discussed in
  Sec.~\ref{sec:Errors-Numerical}.  }
\end{figure}

Figure~\ref{fig:PNvsNR_byOrder} presents similar results in a
different format.  We compute the phase differences between the
numerical waveform and the various post-Newtonian approximants at the
times when the numerical waveform reaches gravitational wave
frequencies $m\omega=0.063$ and $m\omega=0.1$ (the times corresponding
to these frequencies are also indicated by brown lines in
Fig.~\ref{fig:PNvsNR-2.0-2.5-3.0-3.5}). We then plot these phase
differences as a function of the post-Newtonian order (using equal
order in phase and amplitude, except for 3.5PN order, where we use
3.0PN in amplitude).  Three PN approximants end before $t_{0.1}$:
TaylorT1 2.0/2.0, TaylorT3 2.5/2.5, TaylorT3
3.5/3.0. These data points therefore cannot be included in the right
panel of Fig.~\ref{fig:PNvsNR_byOrder}.

The general trend seen in Fig.~\ref{fig:PNvsNR_byOrder} is that the
phase difference decreases with increasing PN order.  However, this
convergence is not monotonic, and the scatter in
Fig.~\ref{fig:PNvsNR_byOrder} can be larger than the phase differences
themselves.  For example, the 0PN waveforms are about as good as the
2.5PN waveforms for TaylorT1 and TaylorT4, and the 2PN TaylorT4
waveform agrees with the numerical results much better than do either the
2.5PN or 3PN TaylorT4 waveforms.  Considering Fig.~\ref{fig:PNvsNR_byOrder}, it
seems difficult to make statements about the convergence with PN order
for any particular PN approximant, or statements about which PN orders
are generally ``good''.  Given that at fixed PN order the different
approximants differ merely by the treatment of uncontrolled
higher-order terms, the scatter in Fig.~\ref{fig:PNvsNR_byOrder} in
some sense represents the truncation error at each PN order.  While
some PN approximants at certain orders may show better agreement with
the numerical simulation, we are not aware of any means to predict
this besides direct comparisons to numerical simulations (as is done
here).  In particular, Fig.~\ref{fig:PNvsNR_byOrder} suggests that the
remarkable agreement between our numerical results and the 3.5PN
TaylorT4 approximant may be simply due to luck; clearly, more PN-NR
comparisons are needed, with different mass ratios and spins, to see
if this is the case.


\section{Conclusions}
\label{sec:conclusions}

We have described numerical simulations of an equal mass,
non-spinning binary black hole spacetime covering 15 orbits of
inspiral just prior to the merger of the two black holes.  Using a
multi-domain pseudospectral method we are able to extract the
gravitational wave content measured by a distant observer with a phase
accuracy of better than 0.02 radians over the roughly 30 cycles of
gravitational radiation observed.  We demonstrate that in order to
achieve this accuracy it is necessary to accurately extrapolate the
waveform from data obtained at extraction surfaces sufficiently far
from the center of mass of the system.  When comparing to zero-spin,
zero-eccentricity PN formulae, our phase uncertainty increases to 0.05
radians because the numerical simulation has a small but nonzero
orbital eccentricity and small but nonzero spins on the holes.

Judging from the case in which we match at $m\omega_m=0.04$, our
numerical simulations are consistent (within our estimated phase
uncertainty) with all PN approximants (at the highest PN order) from
the beginning of our inspiral until about 15 gravitational wave cycles
prior to the merger of the binary.  This agreement provides an
important validation of our numerical simulation.  It also establishes
a regime in which the 3.5-th order post-Newtonian waveforms are
accurate to this level, at least for an equal mass, non-spinning black
hole binary.
After this point, the various PN
approximants begin to diverge, suggesting that the approximation is
beginning to break down.  Since there are many different PN
approximants (including Pad\'e~\cite{Damour2001} and
effective-one-body~\cite{Buonanno99,Damour01c,Damour03,Buonanno2007}
which were not discussed in this paper) it may be possible to find a
clever way to push the PN expansion beyond its breaking point.

Indeed, we find that one approximant, TaylorT4 at 3.5PN in phase,
works astonishingly well, agreeing with our numerical waveforms for
almost the entire 30-cycle length of our runs.  Given the wide scatter
plot of predictions by various PN approximants, it is likely that
TaylorT4 3.5/3.0 simply got lucky for the equal mass non-spinning
black hole binary.  In fact, the assumption of adiabaticity
({\em i.e.,} circular orbits) is known to lead to much larger
phase differences relative to a non-adiabatic inspiral (see Fig. 4 of
\cite{Buonanno00} and~\cite{Miller2004}) than the phase differences
between NR and TaylorT4 we find in Fig.~\ref{fig:NR-TaylorT4}.  Thus
it seems that the uncontrolled higher order terms of TaylorT4 3.5/3.0
balance the error introduced by the adiabaticity assumption to a
remarkable degree.  It remains to be determined whether similar
cancellations occur when the black hole masses are unequal or when
the holes have nonzero spin.

Regardless of the robustness of TaylorT4, it seems evident that
numerical simulations are needed in order to know which, if any, PN
approximant yields the correct waveform after the various approximants
begin to diverge.  For there is no {\it a priori} reason why TaylorT4
should be a better choice than TaylorT1 as they differ only in whether
the ratio of gravitational wave flux to the derivative of the orbital
energy with respect to frequency is left as a ratio of post-Newtonian
expansions or re-expanded as a single post-Newtonian expansion.

The surprising accuracy of TaylorT4 3.5/3.0 in the gravitational frequency
range from $m\omega=0.035$ through $m\omega=0.15$, for the equal mass,
non-spinning inspiral of two black holes, in principle could form a
basis for evaluating the errors of numerical simulations.  Instead of
worrying about errors due to different formulations, initial data,
boundary conditions, extraction methods, etc., perhaps a long inspiral
simulation could be compared with TaylorT4 3.5/3.0 in order to get a
direct estimate of the phase error.  
Similarly, because of its good agreement, TaylorT4 3.5/3.0 could also
be used to address questions that require much longer waveforms than
currently available, for instance the question of when lower order
post-Newtonian waveforms become unreliable. 

We find that the 3PN contributions to the amplitude of the
$(2,2)$ modes improve their accuracy with respect to the numerical
waveforms.  This suggests that for accurate parameter estimation, it
may be desirable to compute the full 3PN amplitude for the
polarization waveforms. Despite the formidable nature of the calculation
required, it would also be interesting to see how the inclusion of 4PN order
corrections to the phasing would affect our comparisons.

Much work still needs to be done to improve the comparison between NR
and PN.  Our primary goal is to push our simulations through merger
and ringdown so that we may compare various resummed PN approximants
and the effective-one-body approximants during the last cycle of
inspiral and merger, as well as test TaylorT4 3.5/3.0 closer to merger.  
We also intend to do long inspirals with arbitrary masses
and spins in order to test the robustness of PN over a range of these
parameters.

Furthermore we wish to improve our initial data.  There is a large
amount of 'junk radiation' present in the initial data that limits how
early we can match PN and NR waveforms.  Reduction of this junk
radiation~\cite{Thesis:Lovelace} would improve the accuracy of our
simulations as well.

Finally, we have done just a simple comparison between NR and PN,
without including any treatment of effects that are important for real
gravitational wave detectors such as limited bandwidth and detector
noise.  In order to more directly address the suitability of PN
formulae for analyzing data from gravitational wave detectors, it will
be necessary to fold in the properties of the detector, to consider
specific values for the total mass of the binary, and to fit for the
mass from the waveforms rather than assuming that the PN and NR
waveforms correspond to the same mass.  We leave this for future work.

\begin{acknowledgments}

It is a pleasure to acknowledge useful discussions with Stuart
Anderson, Alessandra Buonanno, Mark Hannam, Ian Hinder, Luis Lehner,
Lee Lindblom, Geoffrey Lovelace, Sean McWilliams, Robert Owen, Yi Pan, Oliver Rinne and Kip
Thorne.  In particular, we would like to thank Alessandra Buonanno for
a careful reading of this manuscript, Rob Owen for estimating the BH
spin, Geoffrey Lovelace for his help constructing initial data, Oliver
Rinne for providing improved boundary conditions, and Lee Lindblom for
his guidance and input throughout this project. This work was
supported in part by grants from the Sherman Fairchild Foundation to
Caltech and Cornell, and from the Brinson Foundation to Caltech; by
NSF grants PHY-0601459, PHY-0652995, DMS-0553302 and NASA grant
NNG05GG52G at Caltech; by NSF grants PHY-0652952, DMS-0553677,
PHY-0652929, and NASA grant NNG05GG51G at Cornell; and by the
Z.\ Smith Reynolds Foundation and NSF grant PHY-0555617 at Wake
Forest.  We thank NASA/JPL for providing computing facilities that
contributed to this work. Some of the simulations discussed here were
produced with LIGO Laboratory computing facilities. LIGO was
constructed by the California Institute of Technology and
Massachusetts Institute of Technology with funding from the National
Science Foundation and operates under cooperative agreement
PHY-0107417. This paper has been assigned LIGO document number
LIGO-P070101-00-Z.

\end{acknowledgments}



\begin{thebibliography}{144}
\expandafter\ifx\csname natexlab\endcsname\relax\def\natexlab#1{#1}\fi
\expandafter\ifx\csname bibnamefont\endcsname\relax
  \def\bibnamefont#1{#1}\fi
\expandafter\ifx\csname bibfnamefont\endcsname\relax
  \def\bibfnamefont#1{#1}\fi
\expandafter\ifx\csname citenamefont\endcsname\relax
  \def\citenamefont#1{#1}\fi
\expandafter\ifx\csname url\endcsname\relax
  \def\url#1{\texttt{#1}}\fi
\expandafter\ifx\csname urlprefix\endcsname\relax\def\urlprefix{URL }\fi
\providecommand{\bibinfo}[2]{#2}
\providecommand{\eprint}[2][]{\url{#2}}

\bibitem[{\citenamefont{Pretorius}(2005{\natexlab{a}})}]{Pretorius2005a}
\bibinfo{author}{\bibfnamefont{F.}~\bibnamefont{Pretorius}},
  \bibinfo{journal}{Phys.\ Rev.\ Lett.} \textbf{\bibinfo{volume}{95}},
  \bibinfo{pages}{121101} (\bibinfo{year}{2005}{\natexlab{a}}).

\bibitem[{\citenamefont{Pretorius}(2006)}]{Pretorius2006}
\bibinfo{author}{\bibfnamefont{F.}~\bibnamefont{Pretorius}},
  \bibinfo{journal}{Class. Quant. Grav.} \textbf{\bibinfo{volume}{23}},
  \bibinfo{pages}{S529} (\bibinfo{year}{2006}).

\bibitem[{\citenamefont{Campanelli
  et~al.}(2006{\natexlab{a}})\citenamefont{Campanelli, Lousto, Marronetti, and
  Zlochower}}]{Campanelli2006a}
\bibinfo{author}{\bibfnamefont{M.}~\bibnamefont{Campanelli}},
  \bibinfo{author}{\bibfnamefont{C.~O.} \bibnamefont{Lousto}},
  \bibinfo{author}{\bibfnamefont{P.}~\bibnamefont{Marronetti}},
  \bibnamefont{and}
  \bibinfo{author}{\bibfnamefont{Y.}~\bibnamefont{Zlochower}},
  \bibinfo{journal}{Phys.\ Rev.\ Lett.} \textbf{\bibinfo{volume}{96}},
  \bibinfo{pages}{111101} (\bibinfo{year}{2006}{\natexlab{a}}).

\bibitem[{\citenamefont{Baker et~al.}(2006{\natexlab{a}})\citenamefont{Baker,
  Centrella, Choi, Koppitz, and van Meter}}]{Baker2006a}
\bibinfo{author}{\bibfnamefont{J.~G.} \bibnamefont{Baker}},
  \bibinfo{author}{\bibfnamefont{J.}~\bibnamefont{Centrella}},
  \bibinfo{author}{\bibfnamefont{D.-I.} \bibnamefont{Choi}},
  \bibinfo{author}{\bibfnamefont{M.}~\bibnamefont{Koppitz}}, \bibnamefont{and}
  \bibinfo{author}{\bibfnamefont{J.}~\bibnamefont{van Meter}},
  \bibinfo{journal}{Phys.\ Rev.\ Lett.} \textbf{\bibinfo{volume}{96}},
  \bibinfo{pages}{111102} (\bibinfo{year}{2006}{\natexlab{a}}).

\bibitem[{\citenamefont{Campanelli
  et~al.}(2006{\natexlab{b}})\citenamefont{Campanelli, Lousto, and
  Zlochower}}]{Campanelli-Lousto-Zlochower:2006}
\bibinfo{author}{\bibfnamefont{M.}~\bibnamefont{Campanelli}},
  \bibinfo{author}{\bibfnamefont{C.~O.} \bibnamefont{Lousto}},
  \bibnamefont{and}
  \bibinfo{author}{\bibfnamefont{Y.}~\bibnamefont{Zlochower}},
  \bibinfo{journal}{Phys.\ Rev.\ D} \textbf{\bibinfo{volume}{73}},
  \bibinfo{pages}{061501(R)} (\bibinfo{year}{2006}{\natexlab{b}}),
  \eprint{gr-qc/0601091}.

\bibitem[{\citenamefont{Diener et~al.}(2006)\citenamefont{Diener, Herrmann,
  Pollney, Schnetter, Seidel, Takahashi, Thornburg, and
  Ventrella}}]{Diener2006}
\bibinfo{author}{\bibfnamefont{P.}~\bibnamefont{Diener}},
  \bibinfo{author}{\bibfnamefont{F.}~\bibnamefont{Herrmann}},
  \bibinfo{author}{\bibfnamefont{D.}~\bibnamefont{Pollney}},
  \bibinfo{author}{\bibfnamefont{E.}~\bibnamefont{Schnetter}},
  \bibinfo{author}{\bibfnamefont{E.}~\bibnamefont{Seidel}},
  \bibinfo{author}{\bibfnamefont{R.}~\bibnamefont{Takahashi}},
  \bibinfo{author}{\bibfnamefont{J.}~\bibnamefont{Thornburg}},
  \bibnamefont{and}
  \bibinfo{author}{\bibfnamefont{J.}~\bibnamefont{Ventrella}},
  \bibinfo{journal}{Phys.\ Rev.\ Lett.} \textbf{\bibinfo{volume}{96}},
  \bibinfo{pages}{121101} (\bibinfo{year}{2006}), \eprint{gr-qc/0512108}.

\bibitem[{\citenamefont{Scheel et~al.}(2006)\citenamefont{Scheel, Pfeiffer,
  Lindblom, Kidder, Rinne, and Teukolsky}}]{Scheel2006}
\bibinfo{author}{\bibfnamefont{M.~A.} \bibnamefont{Scheel}},
  \bibinfo{author}{\bibfnamefont{H.~P.} \bibnamefont{Pfeiffer}},
  \bibinfo{author}{\bibfnamefont{L.}~\bibnamefont{Lindblom}},
  \bibinfo{author}{\bibfnamefont{L.~E.} \bibnamefont{Kidder}},
  \bibinfo{author}{\bibfnamefont{O.}~\bibnamefont{Rinne}}, \bibnamefont{and}
  \bibinfo{author}{\bibfnamefont{S.~A.} \bibnamefont{Teukolsky}},
  \bibinfo{journal}{Phys.\ Rev.\ D} \textbf{\bibinfo{volume}{74}},
  \bibinfo{pages}{104006} (\bibinfo{year}{2006}), \eprint{gr-qc/0607056}.

\bibitem[{\citenamefont{Br{\"u}gmann et~al.}(2008)\citenamefont{Br{\"u}gmann,
  Gonzalez, Hannam, Husa, Sperhake, and Tichy}}]{Bruegmann2006}
\bibinfo{author}{\bibfnamefont{B.}~\bibnamefont{Br{\"u}gmann}},
  \bibinfo{author}{\bibfnamefont{J.~A.} \bibnamefont{Gonzalez}},
  \bibinfo{author}{\bibfnamefont{M.}~\bibnamefont{Hannam}},
  \bibinfo{author}{\bibfnamefont{S.}~\bibnamefont{Husa}},
  \bibinfo{author}{\bibfnamefont{U.}~\bibnamefont{Sperhake}}, \bibnamefont{and}
  \bibinfo{author}{\bibfnamefont{W.}~\bibnamefont{Tichy}},
  \bibinfo{journal}{Phys.\ Rev.\ D} \textbf{\bibinfo{volume}{77}},
  \bibinfo{eid}{024027} (pages~\bibinfo{numpages}{25}) (\bibinfo{year}{2008}),
  \bibinfo{note}{gr-qc/0610128},
  \urlprefix\url{http://link.aps.org/abstract/PRD/v77/e024027}.

\bibitem[{\citenamefont{Marronetti et~al.}(2007)\citenamefont{Marronetti,
  Tichy, Br{\"u}gmann, Gonzalez, Hannam, Husa, and Sperhake}}]{Marronetti2007}
\bibinfo{author}{\bibfnamefont{P.}~\bibnamefont{Marronetti}},
  \bibinfo{author}{\bibfnamefont{W.}~\bibnamefont{Tichy}},
  \bibinfo{author}{\bibfnamefont{B.}~\bibnamefont{Br{\"u}gmann}},
  \bibinfo{author}{\bibfnamefont{J.}~\bibnamefont{Gonzalez}},
  \bibinfo{author}{\bibfnamefont{M.}~\bibnamefont{Hannam}},
  \bibinfo{author}{\bibfnamefont{S.}~\bibnamefont{Husa}}, \bibnamefont{and}
  \bibinfo{author}{\bibfnamefont{U.}~\bibnamefont{Sperhake}},
  \bibinfo{journal}{Class.\ Quantum Grav.} \textbf{\bibinfo{volume}{24}},
  \bibinfo{pages}{S43} (\bibinfo{year}{2007}), \eprint{gr-qc/0701123}.

\bibitem[{\citenamefont{Szil\'agyi et~al.}(2007)\citenamefont{Szil\'agyi,
  Pollney, Rezzolla, Thornburg, and Winicour}}]{Szilagyi2007}
\bibinfo{author}{\bibfnamefont{B.}~\bibnamefont{Szil\'agyi}},
  \bibinfo{author}{\bibfnamefont{D.}~\bibnamefont{Pollney}},
  \bibinfo{author}{\bibfnamefont{L.}~\bibnamefont{Rezzolla}},
  \bibinfo{author}{\bibfnamefont{J.}~\bibnamefont{Thornburg}},
  \bibnamefont{and} \bibinfo{author}{\bibfnamefont{J.}~\bibnamefont{Winicour}},
  \bibinfo{journal}{Class.\ Quantum Grav.} \textbf{\bibinfo{volume}{24}},
  \bibinfo{pages}{S275} (\bibinfo{year}{2007}), \eprint{gr-qc/0612150}.

\bibitem[{\citenamefont{Herrmann
  et~al.}(2007{\natexlab{a}})\citenamefont{Herrmann, Hinder, Shoemaker, and
  Laguna}}]{Herrmann2007b}
\bibinfo{author}{\bibfnamefont{F.}~\bibnamefont{Herrmann}},
  \bibinfo{author}{\bibfnamefont{I.}~\bibnamefont{Hinder}},
  \bibinfo{author}{\bibfnamefont{D.}~\bibnamefont{Shoemaker}},
  \bibnamefont{and} \bibinfo{author}{\bibfnamefont{P.}~\bibnamefont{Laguna}},
  \bibinfo{journal}{Class.\ Quantum Grav.} \textbf{\bibinfo{volume}{24}},
  \bibinfo{pages}{S33} (\bibinfo{year}{2007}{\natexlab{a}}),
  \eprint{gr-qc/0601026}.

\bibitem[{\citenamefont{Sperhake}(2007)}]{Sperhake2006}
\bibinfo{author}{\bibfnamefont{U.}~\bibnamefont{Sperhake}},
  \bibinfo{journal}{Phys.\ Rev.\ D} p. \bibinfo{pages}{104015}
  (\bibinfo{year}{2007}), \eprint{gr-qc/0606079}.

\bibitem[{\citenamefont{Etienne et~al.}(2007)\citenamefont{Etienne, Faber, Liu,
  Shapiro, and Baumgarte}}]{Etienne2007}
\bibinfo{author}{\bibfnamefont{Z.~B.} \bibnamefont{Etienne}},
  \bibinfo{author}{\bibfnamefont{J.~A.} \bibnamefont{Faber}},
  \bibinfo{author}{\bibfnamefont{Y.~T.} \bibnamefont{Liu}},
  \bibinfo{author}{\bibfnamefont{S.~L.} \bibnamefont{Shapiro}},
  \bibnamefont{and} \bibinfo{author}{\bibfnamefont{T.~W.}
  \bibnamefont{Baumgarte}}, \bibinfo{journal}{Phys.\ Rev.\ D}
  \textbf{\bibinfo{volume}{76}}, \bibinfo{pages}{101503}
  (\bibinfo{year}{2007}), \eprint{arXiv:0707.2083 [gr-qc]}.

\bibitem[{\citenamefont{Baker et~al.}(2006{\natexlab{b}})\citenamefont{Baker,
  Centrella, Choi, Koppitz, and van Meter}}]{Baker2006b}
\bibinfo{author}{\bibfnamefont{J.~G.} \bibnamefont{Baker}},
  \bibinfo{author}{\bibfnamefont{J.}~\bibnamefont{Centrella}},
  \bibinfo{author}{\bibfnamefont{D.-I.} \bibnamefont{Choi}},
  \bibinfo{author}{\bibfnamefont{M.}~\bibnamefont{Koppitz}}, \bibnamefont{and}
  \bibinfo{author}{\bibfnamefont{J.}~\bibnamefont{van Meter}},
  \bibinfo{journal}{Phys.\ Rev.\ D} \textbf{\bibinfo{volume}{73}},
  \bibinfo{pages}{104002} (\bibinfo{year}{2006}{\natexlab{b}}).

\bibitem[{\citenamefont{Baker et~al.}(2007{\natexlab{a}})\citenamefont{Baker,
  Campanelli, Pretorius, and Zlochower}}]{Baker-Campanelli-etal:2007}
\bibinfo{author}{\bibfnamefont{J.~G.} \bibnamefont{Baker}},
  \bibinfo{author}{\bibfnamefont{M.}~\bibnamefont{Campanelli}},
  \bibinfo{author}{\bibfnamefont{F.}~\bibnamefont{Pretorius}},
  \bibnamefont{and}
  \bibinfo{author}{\bibfnamefont{Y.}~\bibnamefont{Zlochower}},
  \bibinfo{journal}{Class.\ Quantum Grav.} \textbf{\bibinfo{volume}{24}},
  \bibinfo{pages}{S25} (\bibinfo{year}{2007}{\natexlab{a}}),
  \eprint{gr-qc/0701016}.

\bibitem[{\citenamefont{Baker et~al.}(2006{\natexlab{c}})\citenamefont{Baker,
  Centrella, Choi, Koppitz, van Meter, and Miller}}]{Baker2006c}
\bibinfo{author}{\bibfnamefont{J.~G.} \bibnamefont{Baker}},
  \bibinfo{author}{\bibfnamefont{J.}~\bibnamefont{Centrella}},
  \bibinfo{author}{\bibfnamefont{D.-I.} \bibnamefont{Choi}},
  \bibinfo{author}{\bibfnamefont{M.}~\bibnamefont{Koppitz}},
  \bibinfo{author}{\bibfnamefont{J.~R.} \bibnamefont{van Meter}},
  \bibnamefont{and} \bibinfo{author}{\bibfnamefont{M.~C.}
  \bibnamefont{Miller}}, \bibinfo{journal}{Astrophys.\ J.}
  \textbf{\bibinfo{volume}{653}}, \bibinfo{pages}{L93}
  (\bibinfo{year}{2006}{\natexlab{c}}), \eprint{astro-ph/0603204}.

\bibitem[{\citenamefont{Gonzalez
  et~al.}(2007{\natexlab{a}})\citenamefont{Gonzalez, Sperhake, Br{\"u}gmann,
  Hannam, and Husa}}]{Gonzalez2007}
\bibinfo{author}{\bibfnamefont{J.~A.} \bibnamefont{Gonzalez}},
  \bibinfo{author}{\bibfnamefont{U.}~\bibnamefont{Sperhake}},
  \bibinfo{author}{\bibfnamefont{B.}~\bibnamefont{Br{\"u}gmann}},
  \bibinfo{author}{\bibfnamefont{M.}~\bibnamefont{Hannam}}, \bibnamefont{and}
  \bibinfo{author}{\bibfnamefont{S.}~\bibnamefont{Husa}},
  \bibinfo{journal}{Phys.\ Rev.\ Lett.} \textbf{\bibinfo{volume}{98}},
  \bibinfo{pages}{091101} (\bibinfo{year}{2007}{\natexlab{a}}),
  \eprint{gr-qc/0702052}.

\bibitem[{\citenamefont{Koppitz et~al.}(2007)\citenamefont{Koppitz, Pollney,
  Reisswig, Rezzolla, Thornburg, Diener, and Schnetter}}]{Koppitz2007}
\bibinfo{author}{\bibfnamefont{M.}~\bibnamefont{Koppitz}},
  \bibinfo{author}{\bibfnamefont{D.}~\bibnamefont{Pollney}},
  \bibinfo{author}{\bibfnamefont{C.}~\bibnamefont{Reisswig}},
  \bibinfo{author}{\bibfnamefont{L.}~\bibnamefont{Rezzolla}},
  \bibinfo{author}{\bibfnamefont{J.}~\bibnamefont{Thornburg}},
  \bibinfo{author}{\bibfnamefont{P.}~\bibnamefont{Diener}}, \bibnamefont{and}
  \bibinfo{author}{\bibfnamefont{E.}~\bibnamefont{Schnetter}},
  \bibinfo{journal}{Phys.\ Rev.\ Lett.} \textbf{\bibinfo{volume}{99}},
  \bibinfo{pages}{041102} (\bibinfo{year}{2007}), \eprint{gr-qc/0701163}.

\bibitem[{\citenamefont{Campanelli
  et~al.}(2007{\natexlab{a}})\citenamefont{Campanelli, Lousto, Zlochower, and
  Merritt}}]{Campanelli2007}
\bibinfo{author}{\bibfnamefont{M.}~\bibnamefont{Campanelli}},
  \bibinfo{author}{\bibfnamefont{C.~O.} \bibnamefont{Lousto}},
  \bibinfo{author}{\bibfnamefont{Y.}~\bibnamefont{Zlochower}},
  \bibnamefont{and} \bibinfo{author}{\bibfnamefont{D.}~\bibnamefont{Merritt}},
  \bibinfo{journal}{Phys.\ Rev.\ Lett.} \textbf{\bibinfo{volume}{98}},
  \bibinfo{pages}{231102} (\bibinfo{year}{2007}{\natexlab{a}}),
  \eprint{gr-qc/0702133}.

\bibitem[{\citenamefont{Gonzalez
  et~al.}(2007{\natexlab{b}})\citenamefont{Gonzalez, Hannam, Sperhake,
  Br{\"u}gmann, and Husa}}]{Gonzalez2007b}
\bibinfo{author}{\bibfnamefont{J.~A.} \bibnamefont{Gonzalez}},
  \bibinfo{author}{\bibfnamefont{M.~D.} \bibnamefont{Hannam}},
  \bibinfo{author}{\bibfnamefont{U.}~\bibnamefont{Sperhake}},
  \bibinfo{author}{\bibfnamefont{B.}~\bibnamefont{Br{\"u}gmann}},
  \bibnamefont{and} \bibinfo{author}{\bibfnamefont{S.}~\bibnamefont{Husa}},
  \bibinfo{journal}{Phys.\ Rev.\ Lett.} \textbf{\bibinfo{volume}{98}},
  \bibinfo{pages}{231101} (\bibinfo{year}{2007}{\natexlab{b}}),
  \eprint{gr-qc/0702052}.

\bibitem[{\citenamefont{Herrmann
  et~al.}(2007{\natexlab{b}})\citenamefont{Herrmann, Hinder, Shoemaker, Laguna,
  and Matzner}}]{Herrmann2007}
\bibinfo{author}{\bibfnamefont{F.}~\bibnamefont{Herrmann}},
  \bibinfo{author}{\bibfnamefont{I.}~\bibnamefont{Hinder}},
  \bibinfo{author}{\bibfnamefont{D.}~\bibnamefont{Shoemaker}},
  \bibinfo{author}{\bibfnamefont{P.}~\bibnamefont{Laguna}}, \bibnamefont{and}
  \bibinfo{author}{\bibfnamefont{R.~A.} \bibnamefont{Matzner}},
  \bibinfo{journal}{Astrophys.\ J.} \textbf{\bibinfo{volume}{661}},
  \bibinfo{pages}{430} (\bibinfo{year}{2007}{\natexlab{b}}),
  \eprint{gr-qc/0701143}.

\bibitem[{\citenamefont{Sopuerta et~al.}(2007)\citenamefont{Sopuerta, Yunes,
  and Laguna}}]{Sopuerta-Yunes-Laguna:2007}
\bibinfo{author}{\bibfnamefont{C.~F.} \bibnamefont{Sopuerta}},
  \bibinfo{author}{\bibfnamefont{N.}~\bibnamefont{Yunes}}, \bibnamefont{and}
  \bibinfo{author}{\bibfnamefont{P.}~\bibnamefont{Laguna}},
  \bibinfo{journal}{Astrophys.\ J.} \textbf{\bibinfo{volume}{656}},
  \bibinfo{pages}{L9} (\bibinfo{year}{2007}), \eprint{astro-ph/0611110}.

\bibitem[{\citenamefont{Choi et~al.}(2007)\citenamefont{Choi, Kelly, Boggs,
  Baker, Centrella, and van Meter}}]{Choi-Kelly-Boggs-etal:2007}
\bibinfo{author}{\bibfnamefont{D.-I.} \bibnamefont{Choi}},
  \bibinfo{author}{\bibfnamefont{B.~J.} \bibnamefont{Kelly}},
  \bibinfo{author}{\bibfnamefont{W.~D.} \bibnamefont{Boggs}},
  \bibinfo{author}{\bibfnamefont{J.~G.} \bibnamefont{Baker}},
  \bibinfo{author}{\bibfnamefont{J.}~\bibnamefont{Centrella}},
  \bibnamefont{and} \bibinfo{author}{\bibfnamefont{J.}~\bibnamefont{van
  Meter}}, \bibinfo{journal}{Phys.\ Rev.\ D} \textbf{\bibinfo{volume}{76}},
  \bibinfo{pages}{104026} (\bibinfo{year}{2007}), \eprint{gr-qc/0702016}.

\bibitem[{\citenamefont{Campanelli
  et~al.}(2007{\natexlab{b}})\citenamefont{Campanelli, Lousto, Zlochower, and
  Merritt}}]{Campanelli2007a}
\bibinfo{author}{\bibfnamefont{M.}~\bibnamefont{Campanelli}},
  \bibinfo{author}{\bibfnamefont{C.~O.} \bibnamefont{Lousto}},
  \bibinfo{author}{\bibfnamefont{Y.}~\bibnamefont{Zlochower}},
  \bibnamefont{and} \bibinfo{author}{\bibfnamefont{D.}~\bibnamefont{Merritt}},
  \bibinfo{journal}{Astrophys.\ J.} \textbf{\bibinfo{volume}{659}},
  \bibinfo{pages}{L5} (\bibinfo{year}{2007}{\natexlab{b}}).

\bibitem[{\citenamefont{Br{\"u}gmann et~al.}(2007)\citenamefont{Br{\"u}gmann,
  Gonzalez, Hannam, Husa, and Sperhake}}]{Bruegmann-Gonzalez-Hannam-etal:2007}
\bibinfo{author}{\bibfnamefont{B.}~\bibnamefont{Br{\"u}gmann}},
  \bibinfo{author}{\bibfnamefont{J.}~\bibnamefont{Gonzalez}},
  \bibinfo{author}{\bibfnamefont{M.}~\bibnamefont{Hannam}},
  \bibinfo{author}{\bibfnamefont{S.}~\bibnamefont{Husa}}, \bibnamefont{and}
  \bibinfo{author}{\bibfnamefont{U.}~\bibnamefont{Sperhake}}
  (\bibinfo{year}{2007}), \eprint{arXiv:0707.0135}.

\bibitem[{\citenamefont{Baker et~al.}(2007{\natexlab{b}})\citenamefont{Baker,
  Boggs, Centrella, Kelly, McWilliams, Miller, and van Meter}}]{Baker2007}
\bibinfo{author}{\bibfnamefont{J.~G.} \bibnamefont{Baker}},
  \bibinfo{author}{\bibfnamefont{W.~D.} \bibnamefont{Boggs}},
  \bibinfo{author}{\bibfnamefont{J.}~\bibnamefont{Centrella}},
  \bibinfo{author}{\bibfnamefont{B.~J.} \bibnamefont{Kelly}},
  \bibinfo{author}{\bibfnamefont{S.~T.} \bibnamefont{McWilliams}},
  \bibinfo{author}{\bibfnamefont{M.~C.} \bibnamefont{Miller}},
  \bibnamefont{and} \bibinfo{author}{\bibfnamefont{J.~R.} \bibnamefont{van
  Meter}}, \bibinfo{journal}{Astrophys.\ J.} \textbf{\bibinfo{volume}{668}},
  \bibinfo{pages}{1140} (\bibinfo{year}{2007}{\natexlab{b}}),
  \eprint{astro-ph/0702390}.

\bibitem[{\citenamefont{Herrmann
  et~al.}(2007{\natexlab{c}})\citenamefont{Herrmann, Hinder, Shoemaker, Laguna,
  and Matzner}}]{Herrmann2007c}
\bibinfo{author}{\bibfnamefont{F.}~\bibnamefont{Herrmann}},
  \bibinfo{author}{\bibfnamefont{I.}~\bibnamefont{Hinder}},
  \bibinfo{author}{\bibfnamefont{D.~M.} \bibnamefont{Shoemaker}},
  \bibinfo{author}{\bibfnamefont{P.}~\bibnamefont{Laguna}}, \bibnamefont{and}
  \bibinfo{author}{\bibfnamefont{R.~A.} \bibnamefont{Matzner}}
  (\bibinfo{year}{2007}{\natexlab{c}}), \eprint{arXiv:0706.2541v1 [gr-qc]}.

\bibitem[{\citenamefont{{Schnittman} et~al.}(2007)\citenamefont{{Schnittman},
  {Buonanno}, {van Meter}, {Baker}, {Boggs}, {Centrella}, {Kelly}, and
  {McWilliams}}}]{Schnittman2007}
\bibinfo{author}{\bibfnamefont{J.~D.} \bibnamefont{{Schnittman}}},
  \bibinfo{author}{\bibfnamefont{A.}~\bibnamefont{{Buonanno}}},
  \bibinfo{author}{\bibfnamefont{J.~R.} \bibnamefont{{van Meter}}},
  \bibinfo{author}{\bibfnamefont{J.~G.} \bibnamefont{{Baker}}},
  \bibinfo{author}{\bibfnamefont{W.~D.} \bibnamefont{{Boggs}}},
  \bibinfo{author}{\bibfnamefont{J.}~\bibnamefont{{Centrella}}},
  \bibinfo{author}{\bibfnamefont{B.~J.} \bibnamefont{{Kelly}}},
  \bibnamefont{and} \bibinfo{author}{\bibfnamefont{S.~T.}
  \bibnamefont{{McWilliams}}} (\bibinfo{year}{2007}), \eprint{arXiv:0707.0301v1
  [gr-qc]}.

\bibitem[{\citenamefont{Campanelli
  et~al.}(2007{\natexlab{c}})\citenamefont{Campanelli, Lousto, Zlochower, and
  Merritt}}]{Campanelli2007b}
\bibinfo{author}{\bibfnamefont{M.}~\bibnamefont{Campanelli}},
  \bibinfo{author}{\bibfnamefont{C.~O.} \bibnamefont{Lousto}},
  \bibinfo{author}{\bibfnamefont{Y.}~\bibnamefont{Zlochower}},
  \bibnamefont{and} \bibinfo{author}{\bibfnamefont{D.}~\bibnamefont{Merritt}},
  \bibinfo{journal}{Phys.\ Rev.\ D} \textbf{\bibinfo{volume}{75}},
  \bibinfo{pages}{064030} (\bibinfo{year}{2007}{\natexlab{c}}),
  \eprint{gr-qc/0612076}.

\bibitem[{\citenamefont{Campanelli
  et~al.}(2006{\natexlab{c}})\citenamefont{Campanelli, Lousto, and
  Zlochower}}]{Campanelli2006d}
\bibinfo{author}{\bibfnamefont{M.}~\bibnamefont{Campanelli}},
  \bibinfo{author}{\bibfnamefont{C.~O.} \bibnamefont{Lousto}},
  \bibnamefont{and}
  \bibinfo{author}{\bibfnamefont{Y.}~\bibnamefont{Zlochower}},
  \bibinfo{journal}{Phys.\ Rev.\ D} \textbf{\bibinfo{volume}{74}},
  \bibinfo{pages}{084023} (\bibinfo{year}{2006}{\natexlab{c}}),
  \eprint{gr-qc/0608275}.

\bibitem[{\citenamefont{Campanelli
  et~al.}(2006{\natexlab{d}})\citenamefont{Campanelli, Lousto, and
  Zlochower}}]{Campanelli2006c}
\bibinfo{author}{\bibfnamefont{M.}~\bibnamefont{Campanelli}},
  \bibinfo{author}{\bibfnamefont{C.~O.} \bibnamefont{Lousto}},
  \bibnamefont{and}
  \bibinfo{author}{\bibfnamefont{Y.}~\bibnamefont{Zlochower}},
  \bibinfo{journal}{Phys.\ Rev.\ D} \textbf{\bibinfo{volume}{74}},
  \bibinfo{pages}{041501} (\bibinfo{year}{2006}{\natexlab{d}}),
  \eprint{gr-qc/0604012}.

\bibitem[{\citenamefont{Buonanno
  et~al.}(2007{\natexlab{a}})\citenamefont{Buonanno, Cook, and
  Pretorius}}]{Buonanno-Cook-Pretorius:2007}
\bibinfo{author}{\bibfnamefont{A.}~\bibnamefont{Buonanno}},
  \bibinfo{author}{\bibfnamefont{G.~B.} \bibnamefont{Cook}}, \bibnamefont{and}
  \bibinfo{author}{\bibfnamefont{F.}~\bibnamefont{Pretorius}},
  \bibinfo{journal}{Phys.\ Rev.\ D} \textbf{\bibinfo{volume}{75}},
  \bibinfo{pages}{124018} (\bibinfo{year}{2007}{\natexlab{a}}).

\bibitem[{\citenamefont{Ajith et~al.}(2007)\citenamefont{Ajith, Babak, Chen,
  Hewitson, Krishnan, Whelan, Br{\"u}gmann, Diener, Gonzalez, Hannam
  et~al.}}]{Ajith-Babak-Chen-etal:2007}
\bibinfo{author}{\bibfnamefont{P.}~\bibnamefont{Ajith}},
  \bibinfo{author}{\bibfnamefont{S.}~\bibnamefont{Babak}},
  \bibinfo{author}{\bibfnamefont{Y.}~\bibnamefont{Chen}},
  \bibinfo{author}{\bibfnamefont{M.}~\bibnamefont{Hewitson}},
  \bibinfo{author}{\bibfnamefont{B.}~\bibnamefont{Krishnan}},
  \bibinfo{author}{\bibfnamefont{J.~T.} \bibnamefont{Whelan}},
  \bibinfo{author}{\bibfnamefont{B.}~\bibnamefont{Br{\"u}gmann}},
  \bibinfo{author}{\bibfnamefont{P.}~\bibnamefont{Diener}},
  \bibinfo{author}{\bibfnamefont{J.}~\bibnamefont{Gonzalez}},
  \bibinfo{author}{\bibfnamefont{M.}~\bibnamefont{Hannam}},
  \bibnamefont{et~al.}, \emph{\bibinfo{title}{Phenomenological template family
  for black-hole coalescence waveforms}} (\bibinfo{year}{2007}),
  \bibinfo{note}{arXiv:0704.3764}, \eprint{arXiv:0704.3764}.

\bibitem[{\citenamefont{Berti et~al.}(2007)\citenamefont{Berti, Cardoso,
  Gonzalez, Sperhake, Hannam, Husa, and
  Br{\"u}gmann}}]{Berti-Cardoso-etal:2007}
\bibinfo{author}{\bibfnamefont{E.}~\bibnamefont{Berti}},
  \bibinfo{author}{\bibfnamefont{V.}~\bibnamefont{Cardoso}},
  \bibinfo{author}{\bibfnamefont{J.~A.} \bibnamefont{Gonzalez}},
  \bibinfo{author}{\bibfnamefont{U.}~\bibnamefont{Sperhake}},
  \bibinfo{author}{\bibfnamefont{M.}~\bibnamefont{Hannam}},
  \bibinfo{author}{\bibfnamefont{S.}~\bibnamefont{Husa}}, \bibnamefont{and}
  \bibinfo{author}{\bibfnamefont{B.}~\bibnamefont{Br{\"u}gmann}},
  \bibinfo{journal}{Phys.\ Rev.\ D} \textbf{\bibinfo{volume}{76}},
  \bibinfo{pages}{064034} (\bibinfo{year}{2007}).

\bibitem[{\citenamefont{Pan et~al.}(2007)\citenamefont{Pan, Buonanno, Baker,
  Centrella, Kelly, McWilliams, Pretorius, and van Meter}}]{Pan2007}
\bibinfo{author}{\bibfnamefont{Y.}~\bibnamefont{Pan}},
  \bibinfo{author}{\bibfnamefont{A.}~\bibnamefont{Buonanno}},
  \bibinfo{author}{\bibfnamefont{J.~G.} \bibnamefont{Baker}},
  \bibinfo{author}{\bibfnamefont{J.}~\bibnamefont{Centrella}},
  \bibinfo{author}{\bibfnamefont{B.~J.} \bibnamefont{Kelly}},
  \bibinfo{author}{\bibfnamefont{S.~T.} \bibnamefont{McWilliams}},
  \bibinfo{author}{\bibfnamefont{F.}~\bibnamefont{Pretorius}},
  \bibnamefont{and} \bibinfo{author}{\bibfnamefont{J.~R.} \bibnamefont{van
  Meter}}, \bibinfo{journal}{arXiv:0704.1964v1 [gr-qc]}
  (\bibinfo{year}{2007}).

\bibitem[{\citenamefont{Buonanno
  et~al.}(2007{\natexlab{b}})\citenamefont{Buonanno, Pan, Baker, Centrella,
  Kelly, McWilliams, and van Meter}}]{Buonanno2007}
\bibinfo{author}{\bibfnamefont{A.}~\bibnamefont{Buonanno}},
  \bibinfo{author}{\bibfnamefont{Y.}~\bibnamefont{Pan}},
  \bibinfo{author}{\bibfnamefont{J.~G.} \bibnamefont{Baker}},
  \bibinfo{author}{\bibfnamefont{J.}~\bibnamefont{Centrella}},
  \bibinfo{author}{\bibfnamefont{B.~J.} \bibnamefont{Kelly}},
  \bibinfo{author}{\bibfnamefont{S.~T.} \bibnamefont{McWilliams}},
  \bibnamefont{and} \bibinfo{author}{\bibfnamefont{J.~R.} \bibnamefont{van
  Meter}}, \bibinfo{journal}{Phys.\ Rev.\ D} \textbf{\bibinfo{volume}{76}},
  \bibinfo{pages}{104049} (\bibinfo{year}{2007}{\natexlab{b}}),
  \eprint{arXiv:0706.3732v2 [gr-qc]}.

\bibitem[{\citenamefont{{Baumgarte} et~al.}(2006)\citenamefont{{Baumgarte},
  {Brady}, {Creighton}, {Lehner}, {Pretorius}, and {DeVoe}}}]{Baumgarte:2006en}
\bibinfo{author}{\bibfnamefont{T.}~\bibnamefont{{Baumgarte}}},
  \bibinfo{author}{\bibfnamefont{P.}~\bibnamefont{{Brady}}},
  \bibinfo{author}{\bibfnamefont{J.~D.} \bibnamefont{{Creighton}}},
  \bibinfo{author}{\bibfnamefont{L.}~\bibnamefont{{Lehner}}},
  \bibinfo{author}{\bibfnamefont{F.}~\bibnamefont{{Pretorius}}},
  \bibnamefont{and} \bibinfo{author}{\bibfnamefont{R.}~\bibnamefont{{DeVoe}}}
  (\bibinfo{year}{2006}), \eprint{gr-qc/0612100}.

\bibitem[{\citenamefont{Baker et~al.}(2007{\natexlab{c}})\citenamefont{Baker,
  van Meter, McWilliams, Centrella, and Kelly}}]{Baker2006d}
\bibinfo{author}{\bibfnamefont{J.~G.} \bibnamefont{Baker}},
  \bibinfo{author}{\bibfnamefont{J.~R.} \bibnamefont{van Meter}},
  \bibinfo{author}{\bibfnamefont{S.~T.} \bibnamefont{McWilliams}},
  \bibinfo{author}{\bibfnamefont{J.}~\bibnamefont{Centrella}},
  \bibnamefont{and} \bibinfo{author}{\bibfnamefont{B.~J.} \bibnamefont{Kelly}},
  \bibinfo{journal}{Phys.\ Rev.\ Lett.} \textbf{\bibinfo{volume}{99}},
  \bibinfo{pages}{181101} (\bibinfo{year}{2007}{\natexlab{c}}).

\bibitem[{\citenamefont{Baker et~al.}(2007{\natexlab{d}})\citenamefont{Baker,
  McWilliams, van Meter, Centrella, Choi, Kelly, and Koppitz}}]{Baker2006e}
\bibinfo{author}{\bibfnamefont{J.~G.} \bibnamefont{Baker}},
  \bibinfo{author}{\bibfnamefont{S.~T.} \bibnamefont{McWilliams}},
  \bibinfo{author}{\bibfnamefont{J.~R.} \bibnamefont{van Meter}},
  \bibinfo{author}{\bibfnamefont{J.}~\bibnamefont{Centrella}},
  \bibinfo{author}{\bibfnamefont{D.-I.} \bibnamefont{Choi}},
  \bibinfo{author}{\bibfnamefont{B.~J.} \bibnamefont{Kelly}}, \bibnamefont{and}
  \bibinfo{author}{\bibfnamefont{M.}~\bibnamefont{Koppitz}},
  \bibinfo{journal}{Phys.\ Rev.\ D} \textbf{\bibinfo{volume}{75}},
  \bibinfo{pages}{124024} (\bibinfo{year}{2007}{\natexlab{d}}),
  \eprint{gr-qc/0612117}.

\bibitem[{\citenamefont{Pfeiffer et~al.}(2007)\citenamefont{Pfeiffer, Brown,
  Kidder, Lindblom, Lovelace, and Scheel}}]{Pfeiffer-Brown-etal:2007}
\bibinfo{author}{\bibfnamefont{H.~P.} \bibnamefont{Pfeiffer}},
  \bibinfo{author}{\bibfnamefont{D.~A.} \bibnamefont{Brown}},
  \bibinfo{author}{\bibfnamefont{L.~E.} \bibnamefont{Kidder}},
  \bibinfo{author}{\bibfnamefont{L.}~\bibnamefont{Lindblom}},
  \bibinfo{author}{\bibfnamefont{G.}~\bibnamefont{Lovelace}}, \bibnamefont{and}
  \bibinfo{author}{\bibfnamefont{M.~A.} \bibnamefont{Scheel}},
  \bibinfo{journal}{Class.\ Quantum Grav.} \textbf{\bibinfo{volume}{24}},
  \bibinfo{pages}{S59} (\bibinfo{year}{2007}).

\bibitem[{\citenamefont{Hannam et~al.}(2007)\citenamefont{Hannam, Husa,
  Sperhake, Br{\"u}gmann, and Gonzalez}}]{Hannam2007}
\bibinfo{author}{\bibfnamefont{M.}~\bibnamefont{Hannam}},
  \bibinfo{author}{\bibfnamefont{S.}~\bibnamefont{Husa}},
  \bibinfo{author}{\bibfnamefont{U.}~\bibnamefont{Sperhake}},
  \bibinfo{author}{\bibfnamefont{B.}~\bibnamefont{Br{\"u}gmann}},
  \bibnamefont{and} \bibinfo{author}{\bibfnamefont{J.~A.}
  \bibnamefont{Gonzalez}}, \emph{\bibinfo{title}{Where post-{N}ewtonian and
  numerical-relativity waveforms meet}} (\bibinfo{year}{2007}),
  \bibinfo{note}{arXiv:0706.1305v2 [gr-qc]}, \eprint{arXiv:0706.1305v2
  [gr-qc]}.

\bibitem[{\citenamefont{Husa et~al.}(2007{\natexlab{a}})\citenamefont{Husa,
  Gonz{\'a}lez, Hannam, Br{\"u}gmann, and Sperhake}}]{Husa2007}
\bibinfo{author}{\bibfnamefont{S.}~\bibnamefont{Husa}},
  \bibinfo{author}{\bibfnamefont{J.~A.} \bibnamefont{Gonz{\'a}lez}},
  \bibinfo{author}{\bibfnamefont{M.}~\bibnamefont{Hannam}},
  \bibinfo{author}{\bibfnamefont{B.}~\bibnamefont{Br{\"u}gmann}},
  \bibnamefont{and} \bibinfo{author}{\bibfnamefont{U.}~\bibnamefont{Sperhake}},
  \emph{\bibinfo{title}{Reducing phase error in long numerical binary black
  hole evolutions with sixth order finite differencing}}
  (\bibinfo{year}{2007}{\natexlab{a}}), \bibinfo{note}{arXiv:0706.0740v1
  [gr-qc]}, \eprint{arXiv:0706.0740v1 [gr-qc]}.

\bibitem[{\citenamefont{Barish and Weiss}(1999)}]{Barish:1999}
\bibinfo{author}{\bibfnamefont{B.~C.} \bibnamefont{Barish}} \bibnamefont{and}
  \bibinfo{author}{\bibfnamefont{R.}~\bibnamefont{Weiss}},
  \bibinfo{journal}{Phys.\ Today} \textbf{\bibinfo{volume}{52 (Oct)}},
  \bibinfo{pages}{44} (\bibinfo{year}{1999}).

\bibitem[{\citenamefont{Waldman}(2006)}]{Waldman:2006}
\bibinfo{author}{\bibfnamefont{S.~J.} \bibnamefont{Waldman}},
  \bibinfo{journal}{Class.\ Quantum Grav.} \textbf{\bibinfo{volume}{23}},
  \bibinfo{pages}{S653} (\bibinfo{year}{2006}).

\bibitem[{\citenamefont{Hild}(2006)}]{Hild:2006}
\bibinfo{author}{\bibfnamefont{S.}~\bibnamefont{Hild}},
  \bibinfo{journal}{Class.\ Quantum Grav.} \textbf{\bibinfo{volume}{23}},
  \bibinfo{pages}{S643} (\bibinfo{year}{2006}).

\bibitem[{\citenamefont{Acernese et~al.}(2002)}]{Acernese:2002}
\bibinfo{author}{\bibfnamefont{F.}~\bibnamefont{Acernese}}
  \bibnamefont{et~al.}, \bibinfo{journal}{Class.\ Quantum Grav.}
  \textbf{\bibinfo{volume}{19}}, \bibinfo{pages}{1421} (\bibinfo{year}{2002}).

\bibitem[{\citenamefont{Acernese et~al.}(2006)\citenamefont{Acernese, Amico,
  Alshourbagy, Antonucci, Aoudia, Avino, Babusci, Ballardin, Barone, Barsotti
  et~al.}}]{Acernese-etal:2006}
\bibinfo{author}{\bibfnamefont{F.}~\bibnamefont{Acernese}},
  \bibinfo{author}{\bibfnamefont{P.}~\bibnamefont{Amico}},
  \bibinfo{author}{\bibfnamefont{M.}~\bibnamefont{Alshourbagy}},
  \bibinfo{author}{\bibfnamefont{F.}~\bibnamefont{Antonucci}},
  \bibinfo{author}{\bibfnamefont{S.}~\bibnamefont{Aoudia}},
  \bibinfo{author}{\bibfnamefont{S.}~\bibnamefont{Avino}},
  \bibinfo{author}{\bibfnamefont{D.}~\bibnamefont{Babusci}},
  \bibinfo{author}{\bibfnamefont{G.}~\bibnamefont{Ballardin}},
  \bibinfo{author}{\bibfnamefont{F.}~\bibnamefont{Barone}},
  \bibinfo{author}{\bibfnamefont{L.}~\bibnamefont{Barsotti}},
  \bibnamefont{et~al.}, \bibinfo{journal}{Class.\ Quantum Grav.}
  \textbf{\bibinfo{volume}{23}}, \bibinfo{pages}{S635} (\bibinfo{year}{2006}).

\bibitem[{\citenamefont{Fritschel}(2003)}]{Fritschel2003}
\bibinfo{author}{\bibfnamefont{P.}~\bibnamefont{Fritschel}}, in
  \emph{\bibinfo{booktitle}{Proc. SPIE}}, edited by
  \bibinfo{editor}{\bibfnamefont{M.}~\bibnamefont{Cruise}} \bibnamefont{and}
  \bibinfo{editor}{\bibfnamefont{P.}~\bibnamefont{Saulson}}
  (\bibinfo{year}{2003}), vol. \bibinfo{volume}{4856}, pp.
  \bibinfo{pages}{282--291}, \eprint{arXiv:gr-qc/0308090v1}.

\bibitem[{\citenamefont{Flanagan and Hughes}(1998)}]{Flanagan1998a}
\bibinfo{author}{\bibfnamefont{E.~E.} \bibnamefont{Flanagan}} \bibnamefont{and}
  \bibinfo{author}{\bibfnamefont{S.~A.} \bibnamefont{Hughes}},
  \bibinfo{journal}{Phys. Rev. D} \textbf{\bibinfo{volume}{57}},
  \bibinfo{pages}{4535} (\bibinfo{year}{1998}), \eprint{gr-qc/9701039}.

\bibitem[{\citenamefont{Peters}(1964)}]{Peters1964}
\bibinfo{author}{\bibfnamefont{P.~C.} \bibnamefont{Peters}},
  \bibinfo{journal}{Phys.\ Rev.} \textbf{\bibinfo{volume}{136}},
  \bibinfo{pages}{B1224} (\bibinfo{year}{1964}).

\bibitem[{\citenamefont{{Damour} et~al.}(1998)\citenamefont{{Damour}, {Iyer},
  and {Sathyaprakash}}}]{Damour98}
\bibinfo{author}{\bibfnamefont{T.}~\bibnamefont{{Damour}}},
  \bibinfo{author}{\bibfnamefont{B.~R.} \bibnamefont{{Iyer}}},
  \bibnamefont{and} \bibinfo{author}{\bibfnamefont{B.~S.}
  \bibnamefont{{Sathyaprakash}}}, \bibinfo{journal}{Phys.\ Rev.\ D}
  \textbf{\bibinfo{volume}{57}}, \bibinfo{pages}{885} (\bibinfo{year}{1998}).

\bibitem[{\citenamefont{{Buonanno} and {Damour}}(1999)}]{Buonanno99}
\bibinfo{author}{\bibfnamefont{A.}~\bibnamefont{{Buonanno}}} \bibnamefont{and}
  \bibinfo{author}{\bibfnamefont{T.}~\bibnamefont{{Damour}}},
  \bibinfo{journal}{Phys.\ Rev.\ D} \textbf{\bibinfo{volume}{59}},
  \bibinfo{pages}{084006} (\bibinfo{year}{1999}).

\bibitem[{\citenamefont{Buonanno et~al.}(2003)\citenamefont{Buonanno, Chen, and
  Vallisneri}}]{Buonanno:2002ft}
\bibinfo{author}{\bibfnamefont{A.}~\bibnamefont{Buonanno}},
  \bibinfo{author}{\bibfnamefont{Y.-B.} \bibnamefont{Chen}}, \bibnamefont{and}
  \bibinfo{author}{\bibfnamefont{M.}~\bibnamefont{Vallisneri}},
  \bibinfo{journal}{Phys. Rev.} \textbf{\bibinfo{volume}{D67}},
  \bibinfo{pages}{024016} (\bibinfo{year}{2003}), \bibinfo{note}{{\bf 74},
  029903(E) (2006)}, \eprint{gr-qc/0205122}.

\bibitem[{\citenamefont{{Damour}
  et~al.}(2001{\natexlab{a}})\citenamefont{{Damour}, {Iyer}, and
  {Sathyaprakash}}}]{Damour2001}
\bibinfo{author}{\bibfnamefont{T.}~\bibnamefont{{Damour}}},
  \bibinfo{author}{\bibfnamefont{B.~R.} \bibnamefont{{Iyer}}},
  \bibnamefont{and} \bibinfo{author}{\bibfnamefont{B.~S.}
  \bibnamefont{{Sathyaprakash}}}, \bibinfo{journal}{Phys.\ Rev.\ D}
  \textbf{\bibinfo{volume}{63}}, \bibinfo{pages}{044023}
  (\bibinfo{year}{2001}{\natexlab{a}}), \bibinfo{note}{{\bf 72}, 029902(E)
  (2005)}.

\bibitem[{\citenamefont{{Damour} et~al.}(2002)\citenamefont{{Damour}, {Iyer},
  and {Sathyaprakash}}}]{Damour02}
\bibinfo{author}{\bibfnamefont{T.}~\bibnamefont{{Damour}}},
  \bibinfo{author}{\bibfnamefont{B.~R.} \bibnamefont{{Iyer}}},
  \bibnamefont{and} \bibinfo{author}{\bibfnamefont{B.~S.}
  \bibnamefont{{Sathyaprakash}}}, \bibinfo{journal}{Phys.\ Rev.\ D}
  \textbf{\bibinfo{volume}{66}}, \bibinfo{pages}{027502}
  (\bibinfo{year}{2002}), \bibinfo{note}{{\bf 72}, 029901(E) (2005)}.

\bibitem[{\citenamefont{{Buonanno} and {Damour}}(2000)}]{Buonanno00}
\bibinfo{author}{\bibfnamefont{A.}~\bibnamefont{{Buonanno}}} \bibnamefont{and}
  \bibinfo{author}{\bibfnamefont{T.}~\bibnamefont{{Damour}}},
  \bibinfo{journal}{Phys.\ Rev.\ D} \textbf{\bibinfo{volume}{62}},
  \bibinfo{pages}{064015} (\bibinfo{year}{2000}).

\bibitem[{\citenamefont{{Damour}
  et~al.}(2000{\natexlab{a}})\citenamefont{{Damour}, {Jaranowski}, and
  {Sch{\"a}fer}}}]{2000PhRvD..62h4011D}
\bibinfo{author}{\bibfnamefont{T.}~\bibnamefont{{Damour}}},
  \bibinfo{author}{\bibfnamefont{P.}~\bibnamefont{{Jaranowski}}},
  \bibnamefont{and}
  \bibinfo{author}{\bibfnamefont{G.}~\bibnamefont{{Sch{\"a}fer}}},
  \bibinfo{journal}{Phys.\ Rev.\ D} \textbf{\bibinfo{volume}{62}},
  \bibinfo{pages}{084011} (\bibinfo{year}{2000}{\natexlab{a}}),
  \eprint{arXiv:gr-qc/0005034}.

\bibitem[{\citenamefont{{Damour}}(2001)}]{Damour01c}
\bibinfo{author}{\bibfnamefont{T.}~\bibnamefont{{Damour}}},
  \bibinfo{journal}{Phys.\ Rev.\ D} \textbf{\bibinfo{volume}{64}},
  \bibinfo{pages}{124013} (\bibinfo{year}{2001}).

\bibitem[{\citenamefont{{Damour} et~al.}(2003)\citenamefont{{Damour}, {Iyer},
  {Jaranowski}, and {Sathyaprakash}}}]{Damour03}
\bibinfo{author}{\bibfnamefont{T.}~\bibnamefont{{Damour}}},
  \bibinfo{author}{\bibfnamefont{B.~R.} \bibnamefont{{Iyer}}},
  \bibinfo{author}{\bibfnamefont{P.}~\bibnamefont{{Jaranowski}}},
  \bibnamefont{and} \bibinfo{author}{\bibfnamefont{B.~S.}
  \bibnamefont{{Sathyaprakash}}}, \bibinfo{journal}{Phys.\ Rev.\ D}
  \textbf{\bibinfo{volume}{67}}, \bibinfo{pages}{064028}
  (\bibinfo{year}{2003}).

\bibitem[{\citenamefont{{Buonanno} et~al.}(2006)\citenamefont{{Buonanno},
  {Chen}, and {Damour}}}]{Buonanno06}
\bibinfo{author}{\bibfnamefont{A.}~\bibnamefont{{Buonanno}}},
  \bibinfo{author}{\bibfnamefont{Y.}~\bibnamefont{{Chen}}}, \bibnamefont{and}
  \bibinfo{author}{\bibfnamefont{T.}~\bibnamefont{{Damour}}},
  \bibinfo{journal}{Phys.\ Rev.\ D} \textbf{\bibinfo{volume}{74}},
  \bibinfo{pages}{104005} (\bibinfo{year}{2006}).

\bibitem[{\citenamefont{Husa et~al.}(2007{\natexlab{b}})\citenamefont{Husa,
  Hannam, Gonz{\'a}lez, Sperhake, and Br{\"u}gmann}}]{Husa-Hannam-etal:2007}
\bibinfo{author}{\bibfnamefont{S.}~\bibnamefont{Husa}},
  \bibinfo{author}{\bibfnamefont{M.}~\bibnamefont{Hannam}},
  \bibinfo{author}{\bibfnamefont{J.~A.} \bibnamefont{Gonz{\'a}lez}},
  \bibinfo{author}{\bibfnamefont{U.}~\bibnamefont{Sperhake}}, \bibnamefont{and}
  \bibinfo{author}{\bibfnamefont{B.}~\bibnamefont{Br{\"u}gmann}},
  \emph{\bibinfo{title}{Reducing eccentricity in black-hole binary evolutions
  with initial parameters from post-newtonian inspiral}}
  (\bibinfo{year}{2007}{\natexlab{b}}), \eprint{arXiv:0706.0904}.

\bibitem[{\citenamefont{{Kidder}}(2007)}]{Kidder07a}
\bibinfo{author}{\bibfnamefont{L.~E.} \bibnamefont{{Kidder}}}
  (\bibinfo{year}{2007}), \eprint{arXiv:0710.0614}.

\bibitem[{\citenamefont{{York, Jr.}}(1999)}]{York1999}
\bibinfo{author}{\bibfnamefont{J.~W.} \bibnamefont{{York, Jr.}}},
  \bibinfo{journal}{Phys. Rev. Lett.} \textbf{\bibinfo{volume}{82}},
  \bibinfo{pages}{1350} (\bibinfo{year}{1999}).

\bibitem[{\citenamefont{Pfeiffer and York}(2003)}]{Pfeiffer2003b}
\bibinfo{author}{\bibfnamefont{H.~P.} \bibnamefont{Pfeiffer}} \bibnamefont{and}
  \bibinfo{author}{\bibfnamefont{J.~W.} \bibnamefont{York}},
  \bibinfo{journal}{Phys.\ Rev.\ D} \textbf{\bibinfo{volume}{67}},
  \bibinfo{pages}{044022} (\bibinfo{year}{2003}).

\bibitem[{\citenamefont{Pfeiffer et~al.}(2003)\citenamefont{Pfeiffer, Kidder,
  Scheel, and Teukolsky}}]{Pfeiffer2003}
\bibinfo{author}{\bibfnamefont{H.~P.} \bibnamefont{Pfeiffer}},
  \bibinfo{author}{\bibfnamefont{L.~E.} \bibnamefont{Kidder}},
  \bibinfo{author}{\bibfnamefont{M.~A.} \bibnamefont{Scheel}},
  \bibnamefont{and} \bibinfo{author}{\bibfnamefont{S.~A.}
  \bibnamefont{Teukolsky}}, \bibinfo{journal}{Comput.\ Phys.\ Commun.}
  \textbf{\bibinfo{volume}{152}}, \bibinfo{pages}{253} (\bibinfo{year}{2003}).

\bibitem[{\citenamefont{Cook}(2002)}]{Cook2002}
\bibinfo{author}{\bibfnamefont{G.~B.} \bibnamefont{Cook}},
  \bibinfo{journal}{Phys. Rev. D} \textbf{\bibinfo{volume}{65}},
  \bibinfo{pages}{084003} (\bibinfo{year}{2002}).

\bibitem[{\citenamefont{Cook and Pfeiffer}(2004)}]{Cook2004}
\bibinfo{author}{\bibfnamefont{G.~B.} \bibnamefont{Cook}} \bibnamefont{and}
  \bibinfo{author}{\bibfnamefont{H.~P.} \bibnamefont{Pfeiffer}},
  \bibinfo{journal}{Phys. Rev. D} \textbf{\bibinfo{volume}{70}},
  \bibinfo{pages}{104016} (\bibinfo{year}{2004}).

\bibitem[{\citenamefont{{Caudill} et~al.}(2006)\citenamefont{{Caudill}, {Cook},
  {Grigsby}, and {Pfeiffer}}}]{Caudill-etal:2006}
\bibinfo{author}{\bibfnamefont{M.}~\bibnamefont{{Caudill}}},
  \bibinfo{author}{\bibfnamefont{G.~B.} \bibnamefont{{Cook}}},
  \bibinfo{author}{\bibfnamefont{J.~D.} \bibnamefont{{Grigsby}}},
  \bibnamefont{and} \bibinfo{author}{\bibfnamefont{H.~P.}
  \bibnamefont{{Pfeiffer}}}, \bibinfo{journal}{Phys.\ Rev.\ D}
  \textbf{\bibinfo{volume}{74}}, \bibinfo{pages}{064011}
  (\bibinfo{year}{2006}), \eprint{gr-qc/0605053}.

\bibitem[{\citenamefont{Lindblom et~al.}(2006)\citenamefont{Lindblom, Scheel,
  Kidder, Owen, and Rinne}}]{Lindblom2006}
\bibinfo{author}{\bibfnamefont{L.}~\bibnamefont{Lindblom}},
  \bibinfo{author}{\bibfnamefont{M.~A.} \bibnamefont{Scheel}},
  \bibinfo{author}{\bibfnamefont{L.~E.} \bibnamefont{Kidder}},
  \bibinfo{author}{\bibfnamefont{R.}~\bibnamefont{Owen}}, \bibnamefont{and}
  \bibinfo{author}{\bibfnamefont{O.}~\bibnamefont{Rinne}},
  \bibinfo{journal}{Class.\ Quantum Grav.} \textbf{\bibinfo{volume}{23}},
  \bibinfo{pages}{S447} (\bibinfo{year}{2006}).

\bibitem[{\citenamefont{Friedrich}(1985)}]{Friedrich1985}
\bibinfo{author}{\bibfnamefont{H.}~\bibnamefont{Friedrich}},
  \bibinfo{journal}{Commun.\ Math.\ Phys.} \textbf{\bibinfo{volume}{100}},
  \bibinfo{pages}{525} (\bibinfo{year}{1985}).

\bibitem[{\citenamefont{Garfinkle}(2002)}]{Garfinkle2002}
\bibinfo{author}{\bibfnamefont{D.}~\bibnamefont{Garfinkle}},
  \bibinfo{journal}{Phys.\ Rev.\ D} \textbf{\bibinfo{volume}{65}},
  \bibinfo{pages}{044029} (\bibinfo{year}{2002}).

\bibitem[{\citenamefont{Pretorius}(2005{\natexlab{b}})}]{Pretorius2005c}
\bibinfo{author}{\bibfnamefont{F.}~\bibnamefont{Pretorius}},
  \bibinfo{journal}{Class.\ Quantum Grav.} \textbf{\bibinfo{volume}{22}},
  \bibinfo{pages}{425} (\bibinfo{year}{2005}{\natexlab{b}}).

\bibitem[{\citenamefont{Rinne}(2006)}]{Rinne2006}
\bibinfo{author}{\bibfnamefont{O.}~\bibnamefont{Rinne}},
  \bibinfo{journal}{Class.\ Quantum Grav.} \textbf{\bibinfo{volume}{23}},
  \bibinfo{pages}{6275} (\bibinfo{year}{2006}), \eprint{gr-qc/0606053}.

\bibitem[{\citenamefont{Rinne et~al.}(2007)\citenamefont{Rinne, Lindblom, and
  Scheel}}]{Rinne2007}
\bibinfo{author}{\bibfnamefont{O.}~\bibnamefont{Rinne}},
  \bibinfo{author}{\bibfnamefont{L.}~\bibnamefont{Lindblom}}, \bibnamefont{and}
  \bibinfo{author}{\bibfnamefont{M.~A.} \bibnamefont{Scheel}},
  \bibinfo{journal}{Class.\ Quantum Grav.} \textbf{\bibinfo{volume}{24}},
  \bibinfo{pages}{4053} (\bibinfo{year}{2007}).

\bibitem[{\citenamefont{Stewart}(1998)}]{Stewart1998}
\bibinfo{author}{\bibfnamefont{J.~M.} \bibnamefont{Stewart}},
  \bibinfo{journal}{Class.\ Quantum Grav.} \textbf{\bibinfo{volume}{15}},
  \bibinfo{pages}{2865} (\bibinfo{year}{1998}).

\bibitem[{\citenamefont{Friedrich and Nagy}(1999)}]{FriedrichNagy1999}
\bibinfo{author}{\bibfnamefont{H.}~\bibnamefont{Friedrich}} \bibnamefont{and}
  \bibinfo{author}{\bibfnamefont{G.}~\bibnamefont{Nagy}},
  \bibinfo{journal}{Comm. Math. Phys.} \textbf{\bibinfo{volume}{201}},
  \bibinfo{pages}{619} (\bibinfo{year}{1999}).

\bibitem[{\citenamefont{Bardeen and Buchman}(2002)}]{Bardeen2002}
\bibinfo{author}{\bibfnamefont{J.~M.} \bibnamefont{Bardeen}} \bibnamefont{and}
  \bibinfo{author}{\bibfnamefont{L.~T.} \bibnamefont{Buchman}},
  \bibinfo{journal}{Phys.\ Rev.\ D} \textbf{\bibinfo{volume}{65}},
  \bibinfo{pages}{064037} (\bibinfo{year}{2002}).

\bibitem[{\citenamefont{Szil\'agyi et~al.}(2002)\citenamefont{Szil\'agyi,
  Schmidt, and Winicour}}]{Szilagyi2002}
\bibinfo{author}{\bibfnamefont{B.}~\bibnamefont{Szil\'agyi}},
  \bibinfo{author}{\bibfnamefont{B.}~\bibnamefont{Schmidt}}, \bibnamefont{and}
  \bibinfo{author}{\bibfnamefont{J.}~\bibnamefont{Winicour}},
  \bibinfo{journal}{Phys.\ Rev.\ D} \textbf{\bibinfo{volume}{65}},
  \bibinfo{pages}{064015} (\bibinfo{year}{2002}).

\bibitem[{\citenamefont{Calabrese et~al.}(2003)\citenamefont{Calabrese, Pullin,
  Sarbach, Tiglio, and Reula}}]{Calabrese2003}
\bibinfo{author}{\bibfnamefont{G.}~\bibnamefont{Calabrese}},
  \bibinfo{author}{\bibfnamefont{J.}~\bibnamefont{Pullin}},
  \bibinfo{author}{\bibfnamefont{O.}~\bibnamefont{Sarbach}},
  \bibinfo{author}{\bibfnamefont{M.}~\bibnamefont{Tiglio}}, \bibnamefont{and}
  \bibinfo{author}{\bibfnamefont{O.}~\bibnamefont{Reula}},
  \bibinfo{journal}{Commun. Math. Phys.} \textbf{\bibinfo{volume}{240}},
  \bibinfo{pages}{377} (\bibinfo{year}{2003}).

\bibitem[{\citenamefont{Szil\'agyi and Winicour}(2003)}]{Szilagyi2003}
\bibinfo{author}{\bibfnamefont{B.}~\bibnamefont{Szil\'agyi}} \bibnamefont{and}
  \bibinfo{author}{\bibfnamefont{J.}~\bibnamefont{Winicour}},
  \bibinfo{journal}{Phys.\ Rev.\ D} \textbf{\bibinfo{volume}{68}},
  \bibinfo{pages}{041501(R)} (\bibinfo{year}{2003}).

\bibitem[{\citenamefont{Kidder et~al.}(2005)\citenamefont{Kidder, Lindblom,
  Scheel, Buchman, and Pfeiffer}}]{Kidder2005}
\bibinfo{author}{\bibfnamefont{L.~E.} \bibnamefont{Kidder}},
  \bibinfo{author}{\bibfnamefont{L.}~\bibnamefont{Lindblom}},
  \bibinfo{author}{\bibfnamefont{M.~A.} \bibnamefont{Scheel}},
  \bibinfo{author}{\bibfnamefont{L.~T.} \bibnamefont{Buchman}},
  \bibnamefont{and} \bibinfo{author}{\bibfnamefont{H.~P.}
  \bibnamefont{Pfeiffer}}, \bibinfo{journal}{Phys. Rev. D}
  \textbf{\bibinfo{volume}{71}}, \bibinfo{pages}{064020}
  (\bibinfo{year}{2005}).

\bibitem[{\citenamefont{Buchman and Sarbach}(2006)}]{Buchman2006}
\bibinfo{author}{\bibfnamefont{L.~T.} \bibnamefont{Buchman}} \bibnamefont{and}
  \bibinfo{author}{\bibfnamefont{O.~C.~A.} \bibnamefont{Sarbach}},
  \bibinfo{journal}{Class.\ Quantum Grav.} \textbf{\bibinfo{volume}{23}},
  \bibinfo{pages}{6709} (\bibinfo{year}{2006}).

\bibitem[{\citenamefont{Bj{\o}rhus}(1995)}]{Bjorhus1995}
\bibinfo{author}{\bibfnamefont{M.}~\bibnamefont{Bj{\o}rhus}},
  \bibinfo{journal}{SIAM J. Sci. Comput.} \textbf{\bibinfo{volume}{16}},
  \bibinfo{pages}{542} (\bibinfo{year}{1995}).

\bibitem[{\citenamefont{Gottlieb and Hesthaven}(2001)}]{Gottlieb2001}
\bibinfo{author}{\bibfnamefont{D.}~\bibnamefont{Gottlieb}} \bibnamefont{and}
  \bibinfo{author}{\bibfnamefont{J.~S.} \bibnamefont{Hesthaven}},
  \bibinfo{journal}{J. Comp. Appl. Math.} \textbf{\bibinfo{volume}{128}},
  \bibinfo{pages}{83} (\bibinfo{year}{2001}).

\bibitem[{\citenamefont{Hesthaven}(2000)}]{Hesthaven2000}
\bibinfo{author}{\bibfnamefont{J.~S.} \bibnamefont{Hesthaven}},
  \bibinfo{journal}{Appl. Numer. Math.} \textbf{\bibinfo{volume}{33}},
  \bibinfo{pages}{23} (\bibinfo{year}{2000}).

\bibitem[{\citenamefont{Gundlach}(1998)}]{Gundlach1998}
\bibinfo{author}{\bibfnamefont{C.}~\bibnamefont{Gundlach}},
  \bibinfo{journal}{Phys.\ Rev.\ D} \textbf{\bibinfo{volume}{57}},
  \bibinfo{pages}{863} (\bibinfo{year}{1998}).

\bibitem[{\citenamefont{Baumgarte et~al.}(1996)\citenamefont{Baumgarte, Cook,
  Scheel, Shapiro, and Teukolsky}}]{baumgarte_etal96}
\bibinfo{author}{\bibfnamefont{T.~W.} \bibnamefont{Baumgarte}},
  \bibinfo{author}{\bibfnamefont{G.~B.} \bibnamefont{Cook}},
  \bibinfo{author}{\bibfnamefont{M.~A.} \bibnamefont{Scheel}},
  \bibinfo{author}{\bibfnamefont{S.~L.} \bibnamefont{Shapiro}},
  \bibnamefont{and} \bibinfo{author}{\bibfnamefont{S.~A.}
  \bibnamefont{Teukolsky}}, \bibinfo{journal}{Phys.\ Rev.\ D}
  \textbf{\bibinfo{volume}{54}}, \bibinfo{pages}{4849} (\bibinfo{year}{1996}).

\bibitem[{\citenamefont{Pfeiffer et~al.}(2000)\citenamefont{Pfeiffer,
  Teukolsky, and Cook}}]{Pfeiffer2000}
\bibinfo{author}{\bibfnamefont{H.~P.} \bibnamefont{Pfeiffer}},
  \bibinfo{author}{\bibfnamefont{S.~A.} \bibnamefont{Teukolsky}},
  \bibnamefont{and} \bibinfo{author}{\bibfnamefont{G.~B.} \bibnamefont{Cook}},
  \bibinfo{journal}{Phys.\ Rev.\ D} \textbf{\bibinfo{volume}{63}},
  \bibinfo{pages}{104018} (\bibinfo{year}{2000}).

\bibitem[{\citenamefont{Brown and York}(1993)}]{BrownYork1993}
\bibinfo{author}{\bibfnamefont{J.~D.} \bibnamefont{Brown}} \bibnamefont{and}
  \bibinfo{author}{\bibfnamefont{J.~W.} \bibnamefont{York}},
  \bibinfo{journal}{Phys.\ Rev.\ D} \textbf{\bibinfo{volume}{47}},
  \bibinfo{pages}{1407} (\bibinfo{year}{1993}).

\bibitem[{\citenamefont{Ashtekar et~al.}(2001)\citenamefont{Ashtekar, Beetle,
  and Lewandowsky}}]{Ashtekar2001}
\bibinfo{author}{\bibfnamefont{A.}~\bibnamefont{Ashtekar}},
  \bibinfo{author}{\bibfnamefont{C.}~\bibnamefont{Beetle}}, \bibnamefont{and}
  \bibinfo{author}{\bibfnamefont{J.}~\bibnamefont{Lewandowsky}},
  \bibinfo{journal}{Phys.\ Rev.\ D} \textbf{\bibinfo{volume}{64}},
  \bibinfo{pages}{044016} (\bibinfo{year}{2001}).

\bibitem[{\citenamefont{Ashtekar and Krishnan}(2003)}]{Ashtekar2003}
\bibinfo{author}{\bibfnamefont{A.}~\bibnamefont{Ashtekar}} \bibnamefont{and}
  \bibinfo{author}{\bibfnamefont{B.}~\bibnamefont{Krishnan}},
  \bibinfo{journal}{Phys.\ Rev.\ D} \textbf{\bibinfo{volume}{68}},
  \bibinfo{pages}{104030} (\bibinfo{year}{2003}).

\bibitem[{\citenamefont{Dreyer et~al.}(2003)\citenamefont{Dreyer, Krishnan,
  Schnetter, and Shoemaker}}]{Dreyer2003}
\bibinfo{author}{\bibfnamefont{O.}~\bibnamefont{Dreyer}},
  \bibinfo{author}{\bibfnamefont{B.}~\bibnamefont{Krishnan}},
  \bibinfo{author}{\bibfnamefont{E.}~\bibnamefont{Schnetter}},
  \bibnamefont{and}
  \bibinfo{author}{\bibfnamefont{D.}~\bibnamefont{Shoemaker}},
  \bibinfo{journal}{Phys.\ Rev.\ D} \textbf{\bibinfo{volume}{67}},
  \bibinfo{pages}{024018} (\bibinfo{year}{2003}).

\bibitem[{\citenamefont{Cook and Whiting}(2007)}]{Cook2007}
\bibinfo{author}{\bibfnamefont{G.~B.} \bibnamefont{Cook}} \bibnamefont{and}
  \bibinfo{author}{\bibfnamefont{B.~F.} \bibnamefont{Whiting}},
  \bibinfo{journal}{Phys.\ Rev.\ D} \textbf{\bibinfo{volume}{76}},
  \bibinfo{pages}{041501(R)} (\bibinfo{year}{2007}).

\bibitem[{\citenamefont{Owen}(2007)}]{OwenThesis}
\bibinfo{author}{\bibfnamefont{R.}~\bibnamefont{Owen}}, Ph.D. thesis,
  \bibinfo{school}{California Institute of Technology} (\bibinfo{year}{2007}),
  \eprint{http://resolver.caltech.edu/CaltechETD:etd-05252007-143511}.

\bibitem[{\citenamefont{Fiske et~al.}(2005)\citenamefont{Fiske, Baker, van
  Meter, Choi, and Centrella}}]{Fiske2005}
\bibinfo{author}{\bibfnamefont{D.~R.} \bibnamefont{Fiske}},
  \bibinfo{author}{\bibfnamefont{J.~G.} \bibnamefont{Baker}},
  \bibinfo{author}{\bibfnamefont{J.~R.} \bibnamefont{van Meter}},
  \bibinfo{author}{\bibfnamefont{D.-I.} \bibnamefont{Choi}}, \bibnamefont{and}
  \bibinfo{author}{\bibfnamefont{J.~M.} \bibnamefont{Centrella}},
  \bibinfo{journal}{Phys.\ Rev.\ D} \textbf{\bibinfo{volume}{71}},
  \bibinfo{pages}{104036} (\bibinfo{year}{2005}).

\bibitem[{\citenamefont{Beetle et~al.}(2005)\citenamefont{Beetle, Bruni, Burko,
  and Nerozzi}}]{Beetle2005}
\bibinfo{author}{\bibfnamefont{C.}~\bibnamefont{Beetle}},
  \bibinfo{author}{\bibfnamefont{M.}~\bibnamefont{Bruni}},
  \bibinfo{author}{\bibfnamefont{L.~M.} \bibnamefont{Burko}}, \bibnamefont{and}
  \bibinfo{author}{\bibfnamefont{A.}~\bibnamefont{Nerozzi}},
  \bibinfo{journal}{Phys.\ Rev.\ D} \textbf{\bibinfo{volume}{72}},
  \bibinfo{pages}{024013} (\bibinfo{year}{2005}).

\bibitem[{\citenamefont{Nerozzi et~al.}(2005)\citenamefont{Nerozzi, Beetle,
  Bruni, Burko, and Pollney}}]{Nerozzi2005}
\bibinfo{author}{\bibfnamefont{A.}~\bibnamefont{Nerozzi}},
  \bibinfo{author}{\bibfnamefont{C.}~\bibnamefont{Beetle}},
  \bibinfo{author}{\bibfnamefont{M.}~\bibnamefont{Bruni}},
  \bibinfo{author}{\bibfnamefont{L.~M.} \bibnamefont{Burko}}, \bibnamefont{and}
  \bibinfo{author}{\bibfnamefont{D.}~\bibnamefont{Pollney}},
  \bibinfo{journal}{Phys.\ Rev.\ D} \textbf{\bibinfo{volume}{72}},
  \bibinfo{pages}{024014} (\bibinfo{year}{2005}).

\bibitem[{\citenamefont{Burko et~al.}(2006)\citenamefont{Burko, Baumgarte, and
  Beetle}}]{Burko2006}
\bibinfo{author}{\bibfnamefont{L.~M.} \bibnamefont{Burko}},
  \bibinfo{author}{\bibfnamefont{T.~W.} \bibnamefont{Baumgarte}},
  \bibnamefont{and} \bibinfo{author}{\bibfnamefont{C.}~\bibnamefont{Beetle}},
  \bibinfo{journal}{Phys.\ Rev.\ D} \textbf{\bibinfo{volume}{73}},
  \bibinfo{pages}{024002} (\bibinfo{year}{2006}).

\bibitem[{\citenamefont{Campanelli
  et~al.}(2006{\natexlab{e}})\citenamefont{Campanelli, Kelly, and
  Lousto}}]{Campanelli2006}
\bibinfo{author}{\bibfnamefont{M.}~\bibnamefont{Campanelli}},
  \bibinfo{author}{\bibfnamefont{B.~J.} \bibnamefont{Kelly}}, \bibnamefont{and}
  \bibinfo{author}{\bibfnamefont{C.~O.} \bibnamefont{Lousto}},
  \bibinfo{journal}{Phys.\ Rev.\ D} \textbf{\bibinfo{volume}{73}},
  \bibinfo{pages}{064005} (\bibinfo{year}{2006}{\natexlab{e}}).

\bibitem[{\citenamefont{Nerozzi et~al.}(2006)\citenamefont{Nerozzi, Bruni,
  Burko, and Re}}]{Nerozzi2006}
\bibinfo{author}{\bibfnamefont{A.}~\bibnamefont{Nerozzi}},
  \bibinfo{author}{\bibfnamefont{M.}~\bibnamefont{Bruni}},
  \bibinfo{author}{\bibfnamefont{L.~M.} \bibnamefont{Burko}}, \bibnamefont{and}
  \bibinfo{author}{\bibfnamefont{V.}~\bibnamefont{Re}},
  \bibinfo{journal}{gr-qc/0607066}  (\bibinfo{year}{2006}).

\bibitem[{\citenamefont{Pazos et~al.}(2006)\citenamefont{Pazos, Dorband, Nagar,
  Palenzuela, Schnetter, and Tiglio}}]{Pazos2006}
\bibinfo{author}{\bibfnamefont{E.}~\bibnamefont{Pazos}},
  \bibinfo{author}{\bibfnamefont{E.~N.} \bibnamefont{Dorband}},
  \bibinfo{author}{\bibfnamefont{A.}~\bibnamefont{Nagar}},
  \bibinfo{author}{\bibfnamefont{C.}~\bibnamefont{Palenzuela}},
  \bibinfo{author}{\bibfnamefont{E.}~\bibnamefont{Schnetter}},
  \bibnamefont{and} \bibinfo{author}{\bibfnamefont{M.}~\bibnamefont{Tiglio}}
  (\bibinfo{year}{2006}), \eprint{gr-qc/0612149}.

\bibitem[{\citenamefont{{Lehner} and {Moreschi}}(2007)}]{Lehner2007}
\bibinfo{author}{\bibfnamefont{L.}~\bibnamefont{{Lehner}}} \bibnamefont{and}
  \bibinfo{author}{\bibfnamefont{O.~M.} \bibnamefont{{Moreschi}}},
  \bibinfo{journal}{ArXiv e-prints}  (\bibinfo{year}{2007}),
  \eprint{arXiv:0706.1319v1 [gr-qc]}.

\bibitem[{\citenamefont{Kocsis and Loeb}(2007)}]{Kocsis2007}
\bibinfo{author}{\bibfnamefont{B.}~\bibnamefont{Kocsis}} \bibnamefont{and}
  \bibinfo{author}{\bibfnamefont{A.}~\bibnamefont{Loeb}},
  \bibinfo{journal}{Phys.\ Rev.\ D} \textbf{\bibinfo{volume}{76}},
  \bibinfo{eid}{084022} (pages~\bibinfo{numpages}{14}) (\bibinfo{year}{2007}),
  \eprint{arXiv:0704.1149v4 [astro-ph]},
  \urlprefix\url{http://link.aps.org/abstract/PRD/v76/e084022}.

\bibitem[{\citenamefont{Blanchet}(2006)}]{Blanchet2006}
\bibinfo{author}{\bibfnamefont{L.}~\bibnamefont{Blanchet}},
  \bibinfo{journal}{Living Rev. Relativity} \textbf{\bibinfo{volume}{9}},
  \bibinfo{pages}{4} (\bibinfo{year}{2006}).

\bibitem[{\citenamefont{Jaranowski and Sch{\"a}fer}(1998)}]{Jaranowski98a}
\bibinfo{author}{\bibfnamefont{P.}~\bibnamefont{Jaranowski}} \bibnamefont{and}
  \bibinfo{author}{\bibfnamefont{G.}~\bibnamefont{Sch{\"a}fer}},
  \bibinfo{journal}{Phys. Rev. D} \textbf{\bibinfo{volume}{57}},
  \bibinfo{pages}{7274} (\bibinfo{year}{1998}), \bibinfo{note}{{\bf 63},
  029902(E) (2000)}.

\bibitem[{\citenamefont{{Jaranowski} and {Sch{\"a}fer}}(1999)}]{Jaranowski99a}
\bibinfo{author}{\bibfnamefont{P.}~\bibnamefont{{Jaranowski}}}
  \bibnamefont{and}
  \bibinfo{author}{\bibfnamefont{G.}~\bibnamefont{{Sch{\"a}fer}}},
  \bibinfo{journal}{Phys.\ Rev.\ D} \textbf{\bibinfo{volume}{60}},
  \bibinfo{pages}{124003} (\bibinfo{year}{1999}).

\bibitem[{\citenamefont{{Damour}
  et~al.}(2000{\natexlab{b}})\citenamefont{{Damour}, {Jaranowski}, and
  {Sch{\"a}fer}}}]{Damour00a}
\bibinfo{author}{\bibfnamefont{T.}~\bibnamefont{{Damour}}},
  \bibinfo{author}{\bibfnamefont{P.}~\bibnamefont{{Jaranowski}}},
  \bibnamefont{and}
  \bibinfo{author}{\bibfnamefont{G.}~\bibnamefont{{Sch{\"a}fer}}},
  \bibinfo{journal}{Phys.\ Rev.\ D} \textbf{\bibinfo{volume}{62}},
  \bibinfo{pages}{021501(R)} (\bibinfo{year}{2000}{\natexlab{b}}),
  \bibinfo{note}{{\bf 63}, 029903(E) (2000)}.

\bibitem[{\citenamefont{{Damour}
  et~al.}(2001{\natexlab{b}})\citenamefont{{Damour}, {Jaranowski}, and
  {Sch{\"a}fer}}}]{Damour01a}
\bibinfo{author}{\bibfnamefont{T.}~\bibnamefont{{Damour}}},
  \bibinfo{author}{\bibfnamefont{P.}~\bibnamefont{{Jaranowski}}},
  \bibnamefont{and}
  \bibinfo{author}{\bibfnamefont{G.}~\bibnamefont{{Sch{\"a}fer}}},
  \bibinfo{journal}{Phys.\ Rev.\ D} \textbf{\bibinfo{volume}{63}},
  \bibinfo{pages}{044021} (\bibinfo{year}{2001}{\natexlab{b}}),
  \bibinfo{note}{{\bf 66}, 029901(E) (2002)}.

\bibitem[{\citenamefont{{Blanchet} and {Faye}}(2000)}]{Blanchet00a}
\bibinfo{author}{\bibfnamefont{L.}~\bibnamefont{{Blanchet}}} \bibnamefont{and}
  \bibinfo{author}{\bibfnamefont{G.}~\bibnamefont{{Faye}}},
  \bibinfo{journal}{Phys.\ Lett.\ A} \textbf{\bibinfo{volume}{271}},
  \bibinfo{pages}{58} (\bibinfo{year}{2000}).

\bibitem[{\citenamefont{{Blanchet} and {Faye}}(2001)}]{Blanchet01a}
\bibinfo{author}{\bibfnamefont{L.}~\bibnamefont{{Blanchet}}} \bibnamefont{and}
  \bibinfo{author}{\bibfnamefont{G.}~\bibnamefont{{Faye}}},
  \bibinfo{journal}{Phys.\ Rev.\ D} \textbf{\bibinfo{volume}{63}},
  \bibinfo{pages}{062005} (\bibinfo{year}{2001}).

\bibitem[{\citenamefont{{Damour}
  et~al.}(2001{\natexlab{c}})\citenamefont{{Damour}, {Jaranowski}, and
  {Sch{\"a}fer}}}]{Damour01b}
\bibinfo{author}{\bibfnamefont{T.}~\bibnamefont{{Damour}}},
  \bibinfo{author}{\bibfnamefont{P.}~\bibnamefont{{Jaranowski}}},
  \bibnamefont{and}
  \bibinfo{author}{\bibfnamefont{G.}~\bibnamefont{{Sch{\"a}fer}}},
  \bibinfo{journal}{Phys.\ Lett.\ B} \textbf{\bibinfo{volume}{513}},
  \bibinfo{pages}{147} (\bibinfo{year}{2001}{\natexlab{c}}).

\bibitem[{\citenamefont{{Blanchet}
  et~al.}(2004{\natexlab{a}})\citenamefont{{Blanchet}, {Damour}, and
  {Esposito-Far{\`e}se}}}]{Blanchet04}
\bibinfo{author}{\bibfnamefont{L.}~\bibnamefont{{Blanchet}}},
  \bibinfo{author}{\bibfnamefont{T.}~\bibnamefont{{Damour}}}, \bibnamefont{and}
  \bibinfo{author}{\bibfnamefont{G.}~\bibnamefont{{Esposito-Far{\`e}se}}},
  \bibinfo{journal}{Phys.\ Rev.\ D} \textbf{\bibinfo{volume}{69}},
  \bibinfo{pages}{124007} (\bibinfo{year}{2004}{\natexlab{a}}).

\bibitem[{\citenamefont{{Itoh} et~al.}(2001)\citenamefont{{Itoh}, {Futamase},
  and {Asada}}}]{Itoh01}
\bibinfo{author}{\bibfnamefont{Y.}~\bibnamefont{{Itoh}}},
  \bibinfo{author}{\bibfnamefont{T.}~\bibnamefont{{Futamase}}},
  \bibnamefont{and} \bibinfo{author}{\bibfnamefont{H.}~\bibnamefont{{Asada}}},
  \bibinfo{journal}{Phys.\ Rev.\ D} \textbf{\bibinfo{volume}{63}},
  \bibinfo{pages}{064038} (\bibinfo{year}{2001}).

\bibitem[{\citenamefont{{Itoh} and {Futamase}}(2003)}]{Itoh03}
\bibinfo{author}{\bibfnamefont{Y.}~\bibnamefont{{Itoh}}} \bibnamefont{and}
  \bibinfo{author}{\bibfnamefont{T.}~\bibnamefont{{Futamase}}},
  \bibinfo{journal}{Phys.\ Rev.\ D} \textbf{\bibinfo{volume}{68}},
  \bibinfo{pages}{121501(R)} (\bibinfo{year}{2003}).

\bibitem[{\citenamefont{{Itoh}}(2004)}]{Itoh04}
\bibinfo{author}{\bibfnamefont{Y.}~\bibnamefont{{Itoh}}},
  \bibinfo{journal}{Phys.\ Rev.\ D} \textbf{\bibinfo{volume}{69}}
  (\bibinfo{year}{2004}), \bibinfo{note}{064018}.

\bibitem[{\citenamefont{{Blanchet} and {Iyer}}(2003)}]{Blanchet03a}
\bibinfo{author}{\bibfnamefont{L.}~\bibnamefont{{Blanchet}}} \bibnamefont{and}
  \bibinfo{author}{\bibfnamefont{B.~R.} \bibnamefont{{Iyer}}},
  \bibinfo{journal}{Class.\ and Quantum Grav.} \textbf{\bibinfo{volume}{20}},
  \bibinfo{pages}{755} (\bibinfo{year}{2003}).

\bibitem[{\citenamefont{{de Andrade} et~al.}(2001)\citenamefont{{de Andrade},
  {Blanchet}, and {Faye}}}]{Andrade01}
\bibinfo{author}{\bibfnamefont{V.~C.} \bibnamefont{{de Andrade}}},
  \bibinfo{author}{\bibfnamefont{L.}~\bibnamefont{{Blanchet}}},
  \bibnamefont{and} \bibinfo{author}{\bibfnamefont{G.}~\bibnamefont{{Faye}}},
  \bibinfo{journal}{Class.\ and Quantum Grav.} \textbf{\bibinfo{volume}{18}},
  \bibinfo{pages}{753} (\bibinfo{year}{2001}).

\bibitem[{\citenamefont{Thorne}(1980)}]{thorne80}
\bibinfo{author}{\bibfnamefont{K.~S.} \bibnamefont{Thorne}},
  \bibinfo{journal}{Rev.\ Mod.\ Phys.} \textbf{\bibinfo{volume}{52}},
  \bibinfo{pages}{299} (\bibinfo{year}{1980}).

\bibitem[{\citenamefont{{Blanchet}}(1998{\natexlab{a}})}]{Blanchet98}
\bibinfo{author}{\bibfnamefont{L.}~\bibnamefont{{Blanchet}}},
  \bibinfo{journal}{Class.\ Quantum Grav.} \textbf{\bibinfo{volume}{15}},
  \bibinfo{pages}{1971} (\bibinfo{year}{1998}{\natexlab{a}}).

\bibitem[{\citenamefont{{Blanchet} and {Damour}}(1992)}]{Blanchet92}
\bibinfo{author}{\bibfnamefont{L.}~\bibnamefont{{Blanchet}}} \bibnamefont{and}
  \bibinfo{author}{\bibfnamefont{T.}~\bibnamefont{{Damour}}},
  \bibinfo{journal}{Phys.\ Rev.\ D} \textbf{\bibinfo{volume}{46}},
  \bibinfo{pages}{4304} (\bibinfo{year}{1992}).

\bibitem[{\citenamefont{{Blanchet}}(1998{\natexlab{b}})}]{Blanchet98a}
\bibinfo{author}{\bibfnamefont{L.}~\bibnamefont{{Blanchet}}},
  \bibinfo{journal}{Class.\ Quantum Grav.} \textbf{\bibinfo{volume}{15}},
  \bibinfo{pages}{89} (\bibinfo{year}{1998}{\natexlab{b}}).

\bibitem[{\citenamefont{{Blanchet}}(1998{\natexlab{c}})}]{Blanchet98b}
\bibinfo{author}{\bibfnamefont{L.}~\bibnamefont{{Blanchet}}},
  \bibinfo{journal}{Class.\ Quantum Grav.} \textbf{\bibinfo{volume}{15}},
  \bibinfo{pages}{113} (\bibinfo{year}{1998}{\natexlab{c}}),
  \bibinfo{note}{{\bf 22}, 3381(E) (2005)}.

\bibitem[{\citenamefont{{Cutler} et~al.}(1993)\citenamefont{{Cutler},
  {Apostolatos}, {Bildsten}, {Finn}, {Flanagan}, {Kennefick}, {Markovic},
  {Ori}, {Poisson}, {Sussman} et~al.}}]{cutler_etal93}
\bibinfo{author}{\bibfnamefont{C.}~\bibnamefont{{Cutler}}},
  \bibinfo{author}{\bibfnamefont{T.~A.} \bibnamefont{{Apostolatos}}},
  \bibinfo{author}{\bibfnamefont{L.}~\bibnamefont{{Bildsten}}},
  \bibinfo{author}{\bibfnamefont{L.~S.} \bibnamefont{{Finn}}},
  \bibinfo{author}{\bibfnamefont{E.~E.} \bibnamefont{{Flanagan}}},
  \bibinfo{author}{\bibfnamefont{D.}~\bibnamefont{{Kennefick}}},
  \bibinfo{author}{\bibfnamefont{D.~M.} \bibnamefont{{Markovic}}},
  \bibinfo{author}{\bibfnamefont{A.}~\bibnamefont{{Ori}}},
  \bibinfo{author}{\bibfnamefont{E.}~\bibnamefont{{Poisson}}},
  \bibinfo{author}{\bibfnamefont{G.~J.} \bibnamefont{{Sussman}}},
  \bibnamefont{et~al.}, \bibinfo{journal}{Phys.\ Rev.\ Lett.}
  \textbf{\bibinfo{volume}{70}}, \bibinfo{pages}{2984} (\bibinfo{year}{1993}).

\bibitem[{\citenamefont{{Blanchet}
  et~al.}(2002{\natexlab{a}})\citenamefont{{Blanchet}, {Iyer}, and
  {Joguet}}}]{Blanchet02}
\bibinfo{author}{\bibfnamefont{L.}~\bibnamefont{{Blanchet}}},
  \bibinfo{author}{\bibfnamefont{B.~R.} \bibnamefont{{Iyer}}},
  \bibnamefont{and} \bibinfo{author}{\bibfnamefont{B.}~\bibnamefont{{Joguet}}},
  \bibinfo{journal}{Phys.\ Rev.\ D} \textbf{\bibinfo{volume}{65}},
  \bibinfo{pages}{064005} (\bibinfo{year}{2002}{\natexlab{a}}),
  \bibinfo{note}{{\bf 71}, 129903(E) (2005)}.

\bibitem[{\citenamefont{{Blanchet}
  et~al.}(2002{\natexlab{b}})\citenamefont{{Blanchet}, {Faye}, {Iyer}, and
  {Joguet}}}]{Blanchet02a}
\bibinfo{author}{\bibfnamefont{L.}~\bibnamefont{{Blanchet}}},
  \bibinfo{author}{\bibfnamefont{G.}~\bibnamefont{{Faye}}},
  \bibinfo{author}{\bibfnamefont{B.~R.} \bibnamefont{{Iyer}}},
  \bibnamefont{and} \bibinfo{author}{\bibfnamefont{B.}~\bibnamefont{{Joguet}}},
  \bibinfo{journal}{Phys.\ Rev.\ D} \textbf{\bibinfo{volume}{65}},
  \bibinfo{pages}{061501(R)} (\bibinfo{year}{2002}{\natexlab{b}}),
  \bibinfo{note}{{\bf 71}, 129902(E) (2005)}.

\bibitem[{\citenamefont{{Blanchet} and {Iyer}}(2005)}]{Blanchet05a}
\bibinfo{author}{\bibfnamefont{L.}~\bibnamefont{{Blanchet}}} \bibnamefont{and}
  \bibinfo{author}{\bibfnamefont{B.~R.} \bibnamefont{{Iyer}}},
  \bibinfo{journal}{Phys.\ Rev.\ D} \textbf{\bibinfo{volume}{71}},
  \bibinfo{pages}{024004} (\bibinfo{year}{2005}).

\bibitem[{\citenamefont{{Blanchet}
  et~al.}(2004{\natexlab{b}})\citenamefont{{Blanchet}, {Damour},
  {Esposito-Far{\`e}se}, and {Iyer}}}]{Blanchet04a}
\bibinfo{author}{\bibfnamefont{L.}~\bibnamefont{{Blanchet}}},
  \bibinfo{author}{\bibfnamefont{T.}~\bibnamefont{{Damour}}},
  \bibinfo{author}{\bibfnamefont{G.}~\bibnamefont{{Esposito-Far{\`e}se}}},
  \bibnamefont{and} \bibinfo{author}{\bibfnamefont{B.~R.}
  \bibnamefont{{Iyer}}}, \bibinfo{journal}{Phys.\ Rev.\ Lett.}
  \textbf{\bibinfo{volume}{93}}, \bibinfo{pages}{091101}
  (\bibinfo{year}{2004}{\natexlab{b}}).

\bibitem[{\citenamefont{{Blanchet} et~al.}(2005)\citenamefont{{Blanchet},
  {Damour}, {Esposito-Far{\`e}se}, and {Iyer}}}]{Blanchet05}
\bibinfo{author}{\bibfnamefont{L.}~\bibnamefont{{Blanchet}}},
  \bibinfo{author}{\bibfnamefont{T.}~\bibnamefont{{Damour}}},
  \bibinfo{author}{\bibfnamefont{G.}~\bibnamefont{{Esposito-Far{\`e}se}}},
  \bibnamefont{and} \bibinfo{author}{\bibfnamefont{B.~R.}
  \bibnamefont{{Iyer}}}, \bibinfo{journal}{Phys.\ Rev.\ D}
  \textbf{\bibinfo{volume}{71}}, \bibinfo{pages}{124004}
  (\bibinfo{year}{2005}).

\bibitem[{\citenamefont{Arun et~al.}(2004)\citenamefont{Arun, Blanchet, Iyer,
  and Qusailah}}]{Arun2004}
\bibinfo{author}{\bibfnamefont{K.}~\bibnamefont{Arun}},
  \bibinfo{author}{\bibfnamefont{L.}~\bibnamefont{Blanchet}},
  \bibinfo{author}{\bibfnamefont{B.}~\bibnamefont{Iyer}}, \bibnamefont{and}
  \bibinfo{author}{\bibfnamefont{M.}~\bibnamefont{Qusailah}},
  \bibinfo{journal}{Class.\ Quantum Grav.} \textbf{\bibinfo{volume}{21}},
  \bibinfo{pages}{3771} (\bibinfo{year}{2004}), \bibinfo{note}{{\bf 22},
  3115--3117(E) (2005)}.

\bibitem[{\citenamefont{{Kidder} et~al.}(2007)\citenamefont{{Kidder},
  {Blanchet}, and {Iyer}}}]{Kidder07}
\bibinfo{author}{\bibfnamefont{L.~E.} \bibnamefont{{Kidder}}},
  \bibinfo{author}{\bibfnamefont{L.}~\bibnamefont{{Blanchet}}},
  \bibnamefont{and} \bibinfo{author}{\bibfnamefont{B.~R.}
  \bibnamefont{{Iyer}}}, \bibinfo{journal}{Class.\ Quantum Grav.}
  \textbf{\bibinfo{volume}{24}}, \bibinfo{pages}{5307} (\bibinfo{year}{2007}).

\bibitem[{\citenamefont{{Wiseman}}(1993)}]{Wiseman93}
\bibinfo{author}{\bibfnamefont{A.~G.} \bibnamefont{{Wiseman}}},
  \bibinfo{journal}{Phys.\ Rev.\ D} \textbf{\bibinfo{volume}{48}},
  \bibinfo{pages}{4757} (\bibinfo{year}{1993}).

\bibitem[{\citenamefont{{Blanchet} and {Sch{\"{a}}fer}}(1993)}]{Blanchet93}
\bibinfo{author}{\bibfnamefont{L.}~\bibnamefont{{Blanchet}}} \bibnamefont{and}
  \bibinfo{author}{\bibfnamefont{G.}~\bibnamefont{{Sch{\"{a}}fer}}},
  \bibinfo{journal}{Class. Quant. Grav.} \textbf{\bibinfo{volume}{10}},
  \bibinfo{pages}{2699} (\bibinfo{year}{1993}).

\bibitem[{\citenamefont{Blanchet}(1996)}]{Blanchet96}
\bibinfo{author}{\bibfnamefont{L.}~\bibnamefont{Blanchet}},
  \bibinfo{journal}{Phys. Rev. D} \textbf{\bibinfo{volume}{54}},
  \bibinfo{pages}{1417} (\bibinfo{year}{1996}).

\bibitem[{\citenamefont{{Blanchet}}(1997)}]{Blanchet97}
\bibinfo{author}{\bibfnamefont{L.}~\bibnamefont{{Blanchet}}},
  \bibinfo{journal}{Phys.\ Rev.\ D} \textbf{\bibinfo{volume}{55}},
  \bibinfo{pages}{714} (\bibinfo{year}{1997}).

\bibitem[{\citenamefont{Poisson}(2004)}]{Poisson2004}
\bibinfo{author}{\bibfnamefont{E.}~\bibnamefont{Poisson}},
  \bibinfo{journal}{Phys.\ Rev.\ D} \textbf{\bibinfo{volume}{70}},
  \bibinfo{pages}{084044} (\bibinfo{year}{2004}).

\bibitem[{\citenamefont{{Kr{\'o}lak} et~al.}(1995)\citenamefont{{Kr{\'o}lak},
  {Kokkotas}, and {Sch{\"a}fer}}}]{Krolak95}
\bibinfo{author}{\bibfnamefont{A.}~\bibnamefont{{Kr{\'o}lak}}},
  \bibinfo{author}{\bibfnamefont{K.~D.} \bibnamefont{{Kokkotas}}},
  \bibnamefont{and}
  \bibinfo{author}{\bibfnamefont{G.}~\bibnamefont{{Sch{\"a}fer}}},
  \bibinfo{journal}{Phys.\ Rev.\ D} \textbf{\bibinfo{volume}{52}},
  \bibinfo{pages}{2089} (\bibinfo{year}{1995}).

\bibitem[{\citenamefont{{Wahlquist}}(1987)}]{Wahlquist87}
\bibinfo{author}{\bibfnamefont{H.}~\bibnamefont{{Wahlquist}}},
  \bibinfo{journal}{General Relativity and Gravitation}
  \textbf{\bibinfo{volume}{19}}, \bibinfo{pages}{1101} (\bibinfo{year}{1987}).

\bibitem[{\citenamefont{Faye et~al.}(2006)\citenamefont{Faye, Blanchet, and
  Buonanno}}]{Faye-Blanchet-Buonanno:2006}
\bibinfo{author}{\bibfnamefont{G.}~\bibnamefont{Faye}},
  \bibinfo{author}{\bibfnamefont{L.}~\bibnamefont{Blanchet}}, \bibnamefont{and}
  \bibinfo{author}{\bibfnamefont{A.}~\bibnamefont{Buonanno}},
  \bibinfo{journal}{Phys.\ Rev.\ D} \textbf{\bibinfo{volume}{74}},
  \bibinfo{pages}{104033} (\bibinfo{year}{2006}).

\bibitem[{\citenamefont{Blanchet et~al.}(2006)\citenamefont{Blanchet, Buonanno,
  and Faye}}]{Blanchet-Buonanno-Faye:2006}
\bibinfo{author}{\bibfnamefont{L.}~\bibnamefont{Blanchet}},
  \bibinfo{author}{\bibfnamefont{A.}~\bibnamefont{Buonanno}}, \bibnamefont{and}
  \bibinfo{author}{\bibfnamefont{G.}~\bibnamefont{Faye}},
  \bibinfo{journal}{Phys.\ Rev.\ D} \textbf{\bibinfo{volume}{74}},
  \bibinfo{pages}{104034} (\bibinfo{year}{2006}), \bibinfo{note}{{\bf 75},
  049903(E) (2007)}.

\bibitem[{\citenamefont{Kidder}(1995)}]{kidder95}
\bibinfo{author}{\bibfnamefont{L.~E.} \bibnamefont{Kidder}},
  \bibinfo{journal}{Phys.\ Rev.\ D} \textbf{\bibinfo{volume}{52}},
  \bibinfo{pages}{821} (\bibinfo{year}{1995}).

\bibitem[{\citenamefont{{Will} and {Wiseman}}(1996)}]{Will96}
\bibinfo{author}{\bibfnamefont{C.~M.} \bibnamefont{{Will}}} \bibnamefont{and}
  \bibinfo{author}{\bibfnamefont{A.~G.} \bibnamefont{{Wiseman}}},
  \bibinfo{journal}{Phys.\ Rev.\ D} \textbf{\bibinfo{volume}{54}},
  \bibinfo{pages}{4813} (\bibinfo{year}{1996}).

\bibitem[{LAL()}]{LAL}
\emph{\bibinfo{title}{{\normalfont LSC Algorithm Library software packages
  {\scshape lal}, {\scshape lalwrapper}, and {\scshape lalapps}}}},
  \urlprefix\url{http://www.lsc-group.phys.uwm.edu/lal}.

\bibitem[{\citenamefont{Miller}(2004)}]{Miller2004}
\bibinfo{author}{\bibfnamefont{M.}~\bibnamefont{Miller}},
  \bibinfo{journal}{Phys.\ Rev.\ D} \textbf{\bibinfo{volume}{69}},
  \bibinfo{pages}{124013} (\bibinfo{year}{2004}).

\bibitem[{\citenamefont{Lovelace}(2007)}]{Thesis:Lovelace}
\bibinfo{author}{\bibfnamefont{G.~M.} \bibnamefont{Lovelace}}, Ph.D. thesis,
  \bibinfo{school}{California Institute of Technology} (\bibinfo{year}{2007}),
  \eprint{http://etd.caltech.edu/etd/available/etd-05232007-115433}.

\end{thebibliography}

\end{document}